\documentclass[twocolumn]{revtex4-2}
\usepackage{amsfonts}
\usepackage[dvipsnames]{xcolor}
\usepackage{amsmath}
\usepackage{amssymb}
\usepackage{amsthm}
\usepackage{mathtools}
\usepackage{graphicx}
\usepackage{enumitem}
\usepackage{graphicx}

\usepackage[top = 1. in,bottom = 1. in, left = 1 in, right=1 in]{geometry}

\newcommand{\arXiv}[1]{\href{http://www.arXiv.org/abs/#1}{arXiv:#1}}
\usepackage[colorlinks=true, linkcolor=blue, bookmarks=true]{hyperref}

\newcommand{\beq}{\begin{equation}}
\newcommand{\eeq}{\end{equation}}
\newcommand{\beqnn}{\begin{equation*}}
\newcommand{\eeqnn}{\end{equation*}}
\newcommand{\ber}{\begin{array}}
\newcommand{\eer}{\end{array}}

\newcommand{\Veff}{V}
\newcommand{\bfx}{\mathbf{x}}
\newcommand{\bfk}{\mathbf{n}}

\newcommand{\w}{\omega}

\begin{document}

	\title{Exact solutions for a coherent phenomenon of condensation\\ in conservative Hamiltonian systems \vspace{-2mm}}
	\author{Anxo Biasi
		\vspace{2mm}}
	\affiliation{Laboratoire de Physique de l'Ecole Normale Sup\'erieure ENS Universit\'e PSL, CNRS, Sorbonne Universit\'e, Universit\'e de Paris, F-75005 Paris, France}
	
	\email{anxo.biasi@gmail.com}

	\begin{abstract}
		\noindent While it is known that Hamiltonian systems may undergo a phenomenon of condensation akin to Bose-Einstein condensation, not all the manifestations of this phenomenon have been uncovered yet. In this work, we present a novel form of condensation in conservative Hamiltonian systems that stands out due to its evolution through highly coherent states. The result is based on a deterministic approach to obtain exact explicit solutions representing the dynamical formation of condensates in finite time. We reveal a dual-cascade behavior during the process, featuring inverse and direct transfer of conserved quantities across the spectrum. The direct cascade yields the excitation of arbitrarily high modes in finite time, being associated with the formation of a small-scale coherent structure. We provide a fully analytic description of the processes involved.
		\\
	\end{abstract}
	
	\maketitle

	
	\section{Introduction}
	\label{sec:Introduction}
	
	In the last decades, it has been observed that nonlinear waves far from equilibrium may undergo a phenomenon of condensation \cite{Condensation_2001,Condensation_2004,Condensation_2005,Condensation_2005_2,Condensation_2012,Condensation_2006,Condensation_2011,Condensation_2018,Condensation_2019,Condensation_2020,Escobedo,Velazques} akin to Bose-Einstein condensation in quantum systems \cite{BEC, BEC2}. It happens when the fundamental mode experiences a macroscopic occupation and ends dominating the spectrum. In conservative Hamiltonian systems, this process is understood as part of the phenomenon of self-organization that takes place in strongly non-equilibrium regimes \cite{Josserand}, consisting in the emergence of large-scale coherent structures from highly fluctuating configurations. The common notion is that these systems evolve from out-of-equilibrium states towards the ``most disordered" configuration, i.e., the thermal equilibrium state \cite{Zakharov_1988,Zakharov_1989}. In the presence of conserved quantities, however, the thermal state cannot just consist of highly disordered fluctuations but, in addition, it requires the formation of coherent structures to store the quantities \cite{Josserand, Rumpf,Rumpf2}. The phenomenon of condensation arises in this picture when the thermal state spectrum diverges at the fundamental mode, something that induces the dramatic occupation of that mode and results in the formation of a large-scale coherent structure, a condensate. On many occasions, this process has found characterization within the framework of wave turbulence \cite{ZakharovBook, NazarenkoBook}, which has proved fruitful in the study of the dynamical formation of condensates \cite{NazarenkoBook,Condensation_2001,Condensation_2005,Condensation_2005_2,Condensation_2006,Condensation_2011}. Through this framework the phenomenon has been widely studied in the Gross-Pitaevskii equation \cite{ZK_Optical_Turbulence}, finding great applicability in the understanding of Bose-Einstein condensation \cite{Condensation_2001,Condensation_2005_2,Self-Similar_2001,Self-Similar_2009,Self-Similar_2021,Condensation_2014,Condensation_2021}, and wave condensation \cite{Condensation_2005,Condensation_2006,Condensation_2011,Condensation_2012,Condensation_2018,Condensation_2019,Condensation_2020,ReviewOpticalTurbulence}. The latter found experimental confirmation in optical turbulence \cite{Condensation_2012,Condensation_2020}.

	In this work, we move out of non-equilibrium and thermal regimes to report a condensation phenomenon that takes place in the ``coherent" regime of a conservative Hamiltonian system. Specifically, we report {\em exact explicit solutions} representing the dynamical formation of condensates in an infinite-dimensional Hamiltonian. These solutions reveal that the system condenses in finite time evolving through {\em highly coherent states} (all modes have well-determined phases and amplitudes during the process), contrasting sharply with the incoherent evolution and randomness inherent to the condensation predicted by wave turbulence \cite{NazarenkoBook}. It leads to three notable results in this paper. First, the existence of a {\em coherent} phenomenon of condensation in conservative Hamiltonian systems. Second, the characterization of a direct cascade that excites arbitrarily high modes in finite time. Third, our fully analytic and explicit description of all the processes involved. This is particularly remarkable due to the extensive use of numerical methods to study the dynamical formation of condensates.

	
	\section{Setup}
	\label{sec:Setup}
	
	The coherent phenomenon of condensation uncovered in this work takes place in the following class of infinite-dimensional Hamiltonian systems 
	\beq
		i\frac{d\alpha_n}{dt} = \underset{n+m=i+j}{\underbrace{\sum_{m=0}^{\infty}\sum_{i=0}^{\infty}\sum_{j=0}^{\infty}}}C_{nmij} \bar{\alpha}_m\alpha_i\alpha_j,
	\label{eq:Resonant_Equation}
	\eeq
	where $t$ represents time, $\alpha_n(t)\in \mathbb{C}$ are complex variables labeled by $n\in\mathbb{N}$, the bar represents complex conjugation, and  $C_{nmij}\in\mathbb{R}$ are time-independent couplings with symmetries $C_{nmij}=C_{mnij}=C_{nmji}=C_{ijnm}$. Systems of this form and similar commonly arise via resonant approximations of weakly nonlinear waves in diverse fields of physics \cite{BBCE1,BBCE2,BEM1,CEM,BMP,CEL,CF,DR,MBox,BR,CEV1,CEV2,GGT,GHT,BEF,BMR,BCE,Balasubramanian,CE,E,E2}, including nonlinear optics and cold atoms \cite{BBCE1,BBCE2,BEM1,BMP,CEM}, nonlinear waves on the sphere \cite{CEL}, or general relativity \cite{BMR,BEF,MBox,CF,DR,CE}. For instance, the well-known Gross-Pitaevskii equation with the harmonic trap enjoys an effective system of the form (\ref{eq:Resonant_Equation}) in any number of dimensions under radial symmetry \cite{BMP}. In those contexts, $\alpha_n$ is the amplitude of the $n$-th normal mode of the system, while $n+m=i+j$ represents the resonance condition ($\w_n+\w_m=\w_i+\w_j$) between these modes when the linearized spectrum of frequencies is fully resonant ($\w_n = a n+b$) \cite{BBCE1,BEM2,CF,CEL,BR,E}. See Appendix~\ref{apx:Coherent_Regime} for further details on this kind of derivation and Ref.~\cite{DWT1} for examples of systems with different frequencies $\w_n$.	This kind of Hamiltonian structures have been useful to study a rich catalog of dynamics, including turbulent cascades \cite{BE,Xu,BMR,MBox,GG,Jalmuzna} (e.g., in the formation of small black holes \cite{BR,BMR}), time-periodic energy flows \cite{CF,BBCE1,BBCE2,CEL,BEF,E} (e.g., vortex precession in  2-dimensional Bose-Einstein condensates (BECs) \cite{BBCE1,BBCE2}, or breathing modes \cite{E}), Fermi-Pasta-Ulam-Tsingou recurrences (in BECs \cite{BEM1} and general relativity \cite{BCE}), or stationary dynamics \cite{CF,GHT,GGT,BBE2,Balasubramanian}. There are also systems in (\ref{eq:Resonant_Equation}) or similar with exceptional analytic structures, belonging to the class of integrable Hamiltonians \cite{BE,Xu,GG,CMEH}, or presenting multidimensional invariant manifolds \cite{BBE1}. The quantum version of (\ref{eq:Resonant_Equation}) has been explored in \cite{	EP,CE_2020,CEM}. 	
	
	Our motivation to search for condensation  processes in the  class of Hamiltonian systems (\ref{eq:Resonant_Equation}) was twofold. First, these systems arise in phase-sensitive (coherent) regimes of wave models, clearly differing from the typical scenarios where condensation phenomena have been observed. We then expected that if Hamiltonian systems (\ref{eq:Resonant_Equation}) exhibited condensation, it would present distinctive  features from previous observations. Our second motivation came from  the presence of systems (\ref{eq:Resonant_Equation}) in disparate areas of physics, something that endows their study with great interdisciplinary value. If one of these systems displays a certain kind of dynamics, then we expect that other systems in (\ref{eq:Resonant_Equation}) with different physical origins may exhibit the same behavior. An example of this idea is the series of works \cite{CF,CEL,BBCE1,BBCE2,BEF}, which were unified in \cite{BBE1}.
	
	Driven by these considerations, we developed a Hamiltonian system in (\ref{eq:Resonant_Equation}) that admitted exact solutions undergoing condensation in finite time. Our goal was to demonstrate the existence of  this phenomenon  in this kind of coherent regime, hoping it leads to future observations in various wave models of physics. Our system is  given by the couplings
	\beq
	C_{nmij} = \frac{1}{2}(n+m+2) \frac{f_n f_m f_i f_j}{f_{n+m}^2}
	\label{eq:C_nmij_equation}
	\eeq
	where $f_n=\sqrt{A_n}$, and $A_n = (2n+1)^{-1}\binom{2n+1}{n}$ are the Catalan numbers \cite{Catalan}. Details on its construction are provided in Appendix~\ref{apx:Origin}.  It belongs to the class of Hamiltonian systems with two extra conserved quantities since conserves the Hamiltonian 
	\beq
	\mathcal{H} = \frac{1}{2} \underset{n+m=i+j}{\underbrace{\sum_{n=0}^{\infty}\sum_{m=0}^{\infty}\sum_{i=0}^{\infty}\sum_{j=0}^{\infty}}}C_{nmij} \bar{\alpha}_n\bar{\alpha}_m\alpha_i\alpha_j,
	\label{eq:Hamiltonian}
	\eeq
	and two additional quantities
	\beq
	N = \sum_{n=0}^{\infty} |\alpha_n|^2, \quad \text{and} \quad E = \sum_{n=0}^{\infty} n |\alpha_n|^2,
	\label{eq:Conserved_equatities_N_E}
	\eeq 
	which receive the interpretation of the ``particle number" and the ``energy" when they are associated with the Gross-Pitaevskii equation \cite{BBCE2}.

	
	\section{Coherent condensation}
	\label{sec:Coherent condensation}
	
	We here show that our system (\ref{eq:Resonant_Equation})-(\ref{eq:C_nmij_equation}) admits exact solutions representing the phenomenon of condensation. The key element is the preservation of the following ansatz by the evolution
	\beq
	\alpha_0(t) = b(t), \qquad \alpha_{n\geq 1} = f_n c(t) p(t)^{n-1},
	\label{eq:invariant_manifold}
	\eeq
	where $b,c,p \in\mathbb{C}$ are unknowns and $f_n$ is the same time-independent function as that in (\ref{eq:C_nmij_equation}). The restriction $|p|^2<1/4$ is necessary to guarantee finite values for the conserved quantities $N$ and $E$ in (\ref{eq:Conserved_equatities_N_E}) due to the exponential growth of $f_{n\gg1}\sim 2^n n^{-3/4}$. The  advantage  of this ansatz is the reduction of the infinite-dimensional system (\ref{eq:Resonant_Equation})-(\ref{eq:C_nmij_equation}) to three equations for three unknowns 
	\begin{align}
		&i \dot{p} =  p \left(N +  \frac{b \bar{c} p}{x} F +\frac{\bar{b} c \bar{p}}{x}\right), \label{eq:pdot}\\
		&i \dot{b} =\left(N + \frac{2E}{F+1}\right) b+\frac{2 E F}{(F+1) x} c \bar{p}, \label{eq:bdot}\\
		&i \dot{c} = (3 N+2E)c + 2 E\frac{F-2}{F+1}(c-b p), \label{eq:cdot}
	\end{align}
	where $N$ and $E$ are given in (\ref{eq:Conserved_equatities_N_E}), $x = |p|^2$, and 
	\beq
	F(x) = \sum_{n=1}^{\infty} f_n^2 x^n = \frac{2 x}{1-2 x+\sqrt{1-4 x}}.
	\label{eq:F_function}
	\eeq
	Technical details about this and subsequent derivations are provided in Appendix~\ref{apx:Condensation}.
	
	The resulting system of equations is solvable. First, we combine Eq.~(\ref{eq:pdot}) and the quantities $N$, $E$, and $\mathcal{H}$ to write an  equation for $x$ in the form of zero-energy trajectories of a point particle
	\beq
	\dot{x}^2 + V(x) = 0,
	\label{eq:xdot}
	\eeq
	where $V(x)$ only depends on $x$, $N$, $E$, and $\mathcal{H}$. Once we obtain a trajectory $x(t)$, the expressions for $|b(t)|^2$ and $|c(t)|^2$ follow from the combination of the ansatz for $\alpha_n$ and the conserved quantities
	\beq
	|b(t)|^2 = N - \frac{E F(x(t))}{x(t) F'(x(t))}, \  |c(t)|^2 = \frac{E}{F'(x(t))},
	\label{eq:b_c_in_terms_of_x_N_E}
	\eeq
	where $F'=dF/dx= F/(x\sqrt{1-4x})$. One then obtains the evolution for $|\alpha_0(t)|^2=|b(t)|^2$ and $|\alpha_{n\geq1}(t)|^2= f_n^2 |c(t)|^2 x(t)^{n-1}$. The expressions for the phases come from the integration of (\ref{eq:pdot})-(\ref{eq:cdot}) once we have substituted $x(t)$, $|b(t)|$, and $|c(t)|$. Altogether, this constitutes an exact solution $\alpha_n(t)$ to our Hamiltonian system.

	For the sake of concreteness, we focus the discussion on a particular initial condition that undergoes condensation, while a family of conditions will be introduced at the end of the section. We work with $\alpha_0(0)=\alpha_1(0)=1$, $\alpha_{n\geq 2}(0)=0$, being equivalent to $p(0)=0$, $b(0)=c(0)=1$. Equation (\ref{eq:xdot}) reduces to
	\beq
	\dot{x}^2 = -2 (1-4x)\left(1-4x - \sqrt{1-4x}\right),
	\eeq 
	and is solved by
	\beq
	x(t) = \frac{1}{4}\sin^2\sqrt{2}t \left(1+\cos^2\sqrt{2}t\right).
	\label{eq:x_solution}
	\eeq
	Applying the above strategy, we have explicit formulas for the evolution of $|\alpha_n(t)|$, while the phases of $\alpha_n$ come from the phases of $p,\ b$, and $c$
		\begin{align}
		& \phi_p(t) = -\frac{\pi }{2}-\arctan\left(2^{-\frac{1}{2}}\tan{\sqrt{2} t}\right), \label{eq:phase_p}\\
		& \phi_b(t) = -2 t - \arctan\left(2^{\frac{1}{2}} \tan{\sqrt{2} t}\right), \label{eq:phase_b} \\
		& \phi_c(t) = -2 t -2 \arctan\left(2^{-\frac{1}{2}}\tan{\sqrt{2} t} \right). \label{eq:phase_c}
	\end{align} 
	Then, we have an exact explicit solution $\alpha_n(t)$ with well-determined amplitudes and phases in time. 
	\begin{figure}[t!]
		\centering
		\includegraphics[width=0.98\columnwidth]{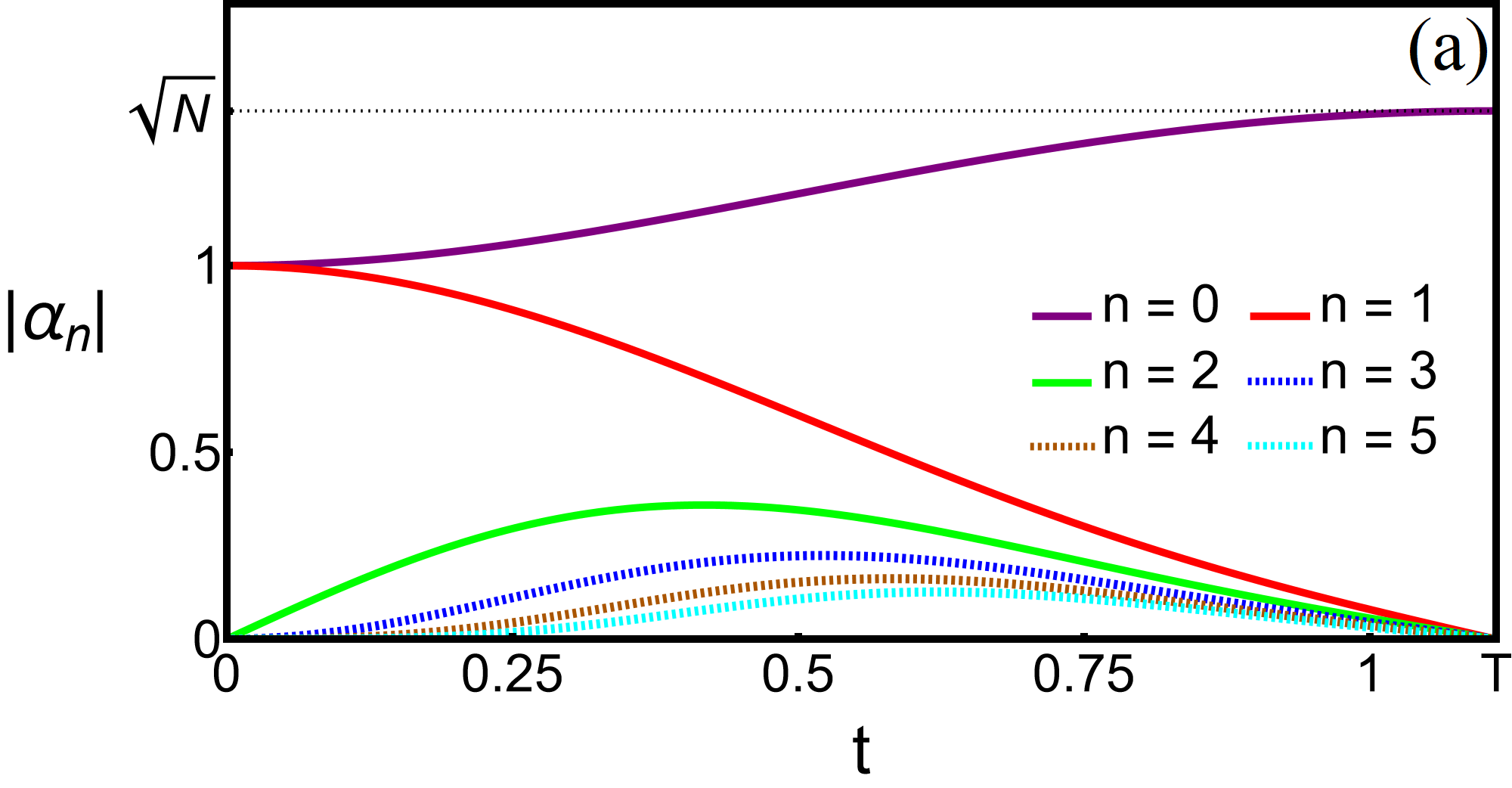}
		\includegraphics[width=\columnwidth]{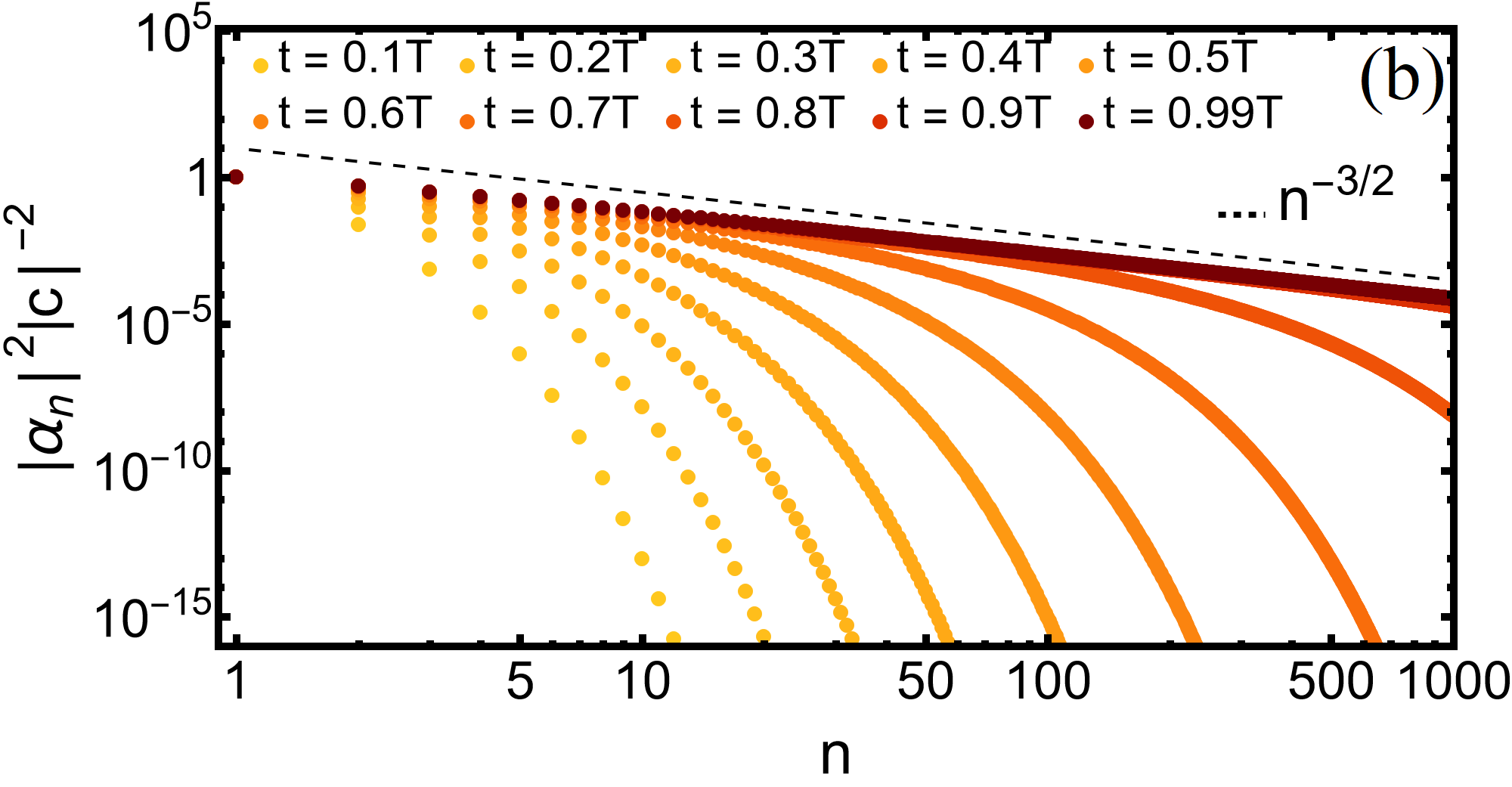}
		\caption{(a) Evolution of the first $|\alpha_n(t)|$. (b) Convergence  of the amplitude spectrum to the power-law $n^{-3/2}$ as $t$ approaches $T$ (visualized extracting $|c|^2$). The initial condition corresponds to the two-mode initial data described in the main text; $N=2$ and $E = 1$.}
		\label{fig:Time_Ev_alpha_n_main_text}
	\end{figure}
	As we now explain, it represents the dynamical formation of a condensate in finite time $T=\pi/(2\sqrt{2})$, being its most distinctive feature the evolution through {\em highly coherent states} (the phases of $\alpha_{n\geq 1}$ are always in a straight line (\ref{eq:invariant_manifold})). The evolution is illustrated in Fig.~\ref{fig:Time_Ev_alpha_n_main_text}. Initially, only the first two modes are excited since $x(0)=0$; at intermediate times, all modes get excited with an exponential decay for large $n$ since $0<x(t)<1/4$; and finally, the amplitude spectrum converges to the Kronecker delta distribution
	\beq
	|\alpha_n|^2\underset{t\to T}{\to}  N \delta_{0,n} ,
	\label{eq:wave_condensation_KroneckerDelta}
	\eeq
	storing the total amount of $N$ at the lowest mode while the rest of the spectrum goes to zero. This behavior is extracted from (\ref{eq:b_c_in_terms_of_x_N_E}) and the fact that $x(T) = 1/4$, $F(1/4)$ remains finite, but $F'(1/4)$ diverges. Then, the conservation of $N$ and $E$ leads to $|c|^2 \to0$, and $|b|^2\to N$ as $t\to T$.  
	
	Above, we have illustrated the process of coherent condensation through a particular example but, there is a family of exact solutions undergoing the same phenomenon. Their initial conditions  are written in terms of the conserved quantities $N$ and $E$ as follows
	\begin{align}
		&b(0) = \sqrt{\frac{N^3}{N^2 + 4 E^2}}, \quad p(0) = \frac{4 E^2 - N^2}{2 \left(N^2 + 4 E^2\right)}, \nonumber\\
		& \qquad c(0) = \frac{E (N + 2 E)^2}{N^2 + 4 E^2} \sqrt{\frac{N}{N^2 + 4 E^2}},
		\label{eq:family_conditions}
	\end{align}
	and all of them converge to the Kronecker delta distribution in finite time. 
	Note therefore that the condensation happens for any positive $E/N$, from arbitrarily small to arbitrarily large.  Technical details and explicit expressions are provided in Appendix~\ref{apx:Condensation}.
		

	\section{Dual cascade behavior}
	\label{sec:Dual cascade behavior} 
	
	We here demonstrate that the condensation process described by our family of solutions shares common features with turbulent systems, in particular, the transfer of conserved quantities across the spectrum \cite{Cascade_1,ZakharovBook,NazarenkoBook,Nz_2023}. As we shall see, our quantities $N$ and $E$ experience a separation in the spectrum as the condensation advances, being supported by modes with low and high $n$ respectively, similar to what happens in the condensation predicted by wave turbulence \cite{NazarenkoBook}. At the end of the process, $N$ and $E$ are entirely supported by the opposite ``edges" of the spectrum. We then say that our system undertook a dual-cascade behavior, in connection with the literature on wave turbulence. An inverse cascade transferred the entire amount of $N$ to the lowest mode, as we can see from (\ref{eq:wave_condensation_KroneckerDelta}), while a direct cascade transferred $E$ to high modes (modes with arbitrarily large $n$). The latter is indicated by the loss of the exponential suppression in the asymptotic spectrum when the condensate forms (i.e., $x\to 1/4$)
	\beq
	|\alpha_{n\gg1}|^2 \sim (4/\sqrt{\pi})|c|^2 n^{-\frac{3}{2}} (4x)^{n-1},
	\label{eq:powerlaw}
	\eeq
	as Fig.~\ref{fig:Time_Ev_alpha_n_main_text} shows. The development of the power-law $n^{-3/2}$ (the excitation of higher and higher modes) serves to ensure the conservation of $E$ when the spectrum approaches the Kronecker delta (\ref{eq:wave_condensation_KroneckerDelta}) because the decay of $|c|^2$ is compensated by the divergence of the series in (\ref{eq:Conserved_equatities_N_E}). To our knowledge, this is the first confirmation of the development of a power-law spectrum in finite time in the class of Hamiltonian systems (\ref{eq:Resonant_Equation}). Numerical evidence of this phenomenon has been reported in other systems \cite{BMR,MBox}, and confirmation of the development of a power-law spectrum in infinite time was achieved in \cite{BE,Xu}. In general, the dynamical excitation of the asymptotic spectrum is an important subject included in the reference list ``Problems in Hamiltonian PDE's" \cite{Bourgain}. In that context, the phenomenon is quantified through the growth of Sobolev norms
	 \cite{Staffilani2010,Bourgain2,Kuksin,Maspero,Hani,GG,BE,Xu,GG2,GGH},
	 \beq
	 	H^{\xi} = \left(\sum_{n=0}^{\infty} (n+1)^{2\xi} |\alpha_n|^2\right)^{1/2},
	 \eeq
	  which we have calculated in Appendix~\ref{apx:Sobolev_Norms}, confirming their growth to infinity as the spectrum converges to the Kronecker delta distribution (for $\xi>1/2$)
	  \beq
	  H^{\xi>1/2} \underset{t\sim T}{\sim} (T-t)^{2(1-2\xi)}.
	  \label{eq:Sobolev_Norms}
	  \eeq
	  The transition is at $\xi=1/2$ because that norm $H^{1/2} = \sqrt{N+E}$ remains finite due to the conservation of $E$ and $N$. 

	
	\section{Formation of a small-scale coherent structure}
	\label{sec:Formation of small- and large-scale structures} 
	
	\begin{figure}[t!]
		\centering
		\includegraphics[width=\columnwidth]{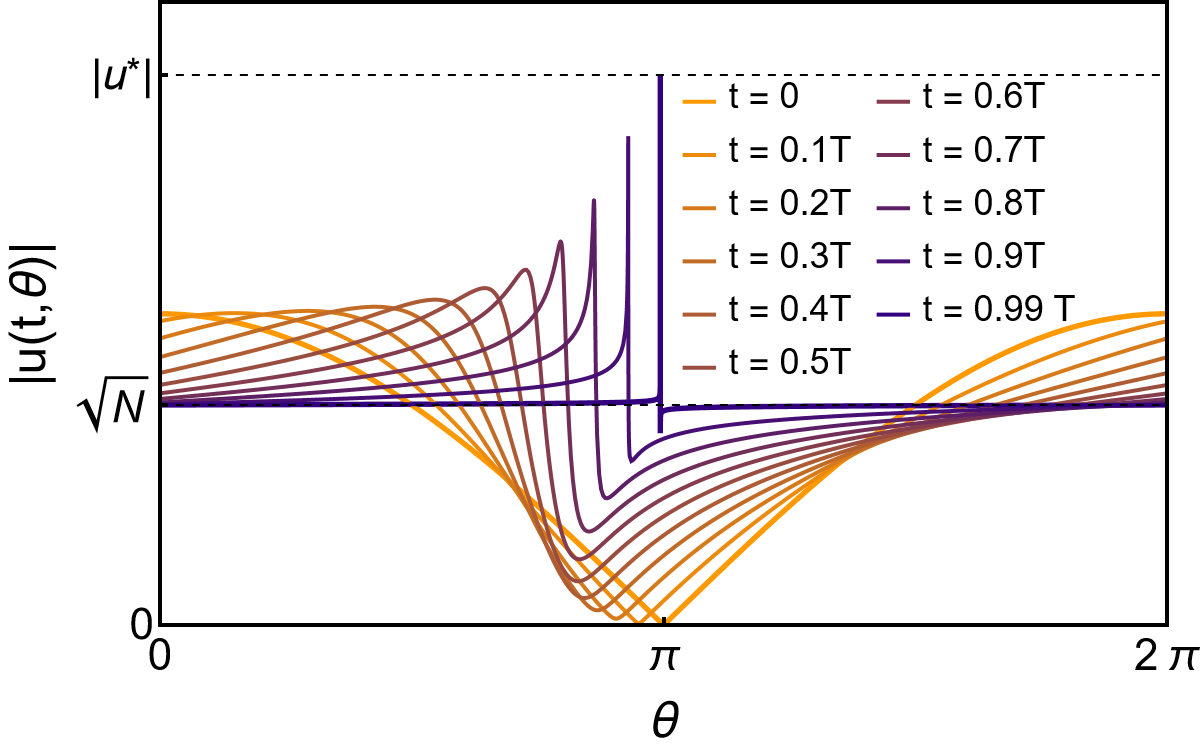}
		\caption{Position space representation of the coherent condensation process presented in Fig.~\ref{fig:Time_Ev_alpha_n_main_text}. Function $u(t,\theta)$ converges from the initial state to  the  fundamental mode $\sqrt{N}$ except at a single point, $\theta=\pi$.}
		\label{fig:DWC}
	\end{figure}
	
	We here demonstrate that the development of the power-law spectrum in Eq.~(\ref{eq:powerlaw}) represents a process of energy concentration in position space, which leads to the formation of a small-scale structure. This kind of process has been, for instance, associated with the formation of tiny black holes in general relativity \cite{BMR}, a result that supports the instability of anti-de Sitter spacetime \cite{BR}. To illustrate this point, we visualize the coherent condensation process  described by our solutions in a 1-dimensional box in Fig.~\ref{fig:DWC}. This is done by taking  $\{\alpha_n\}$ as the Fourier coefficients $u(t,\theta) = \sum_{n=0}^{\infty} \alpha_n(t) e^{i n\theta}$.	The figure shows the convergence of $|u(t,\theta)|$ from the initial state to the fundamental mode ($|u(T,\theta)|=\sqrt{N}$). In the process, a steep region is generated, approaching a discontinuity at a single point $|u(t,\pi)|\underset{t\to T}{\to}|u^*| = \sqrt{N+\Gamma(1/4)^2 E \sqrt{2/\pi}}$, where $\Gamma$ is the gamma-function; see Appendix~\ref{apx:Small-Scale_Structure}. It happens because function $u(t,\theta)$ translates the excitation of modes with large $n$ into high-frequency oscillations in $\theta$. Additionally, the phases of $\alpha_{n\geq 1}$ form a straight line in $n$, inducing the coherent superposition of high-frequency Fourier modes. Altogether, it results in the formation of a small-scale coherent structure, a discontinuity. This is a markedly different behavior from turbulent regimes. In those contexts, the excitation of high modes leads to a background of turbulent fluctuations consequence of the incoherent superposition of Fourier modes \cite{Josserand}.
 
 	
 	\section{Numerical Simulations}
 	\label{sec:Numerical_Simulations}

\begin{figure}[t!]
		\centering
		\includegraphics[width=0.97\columnwidth]{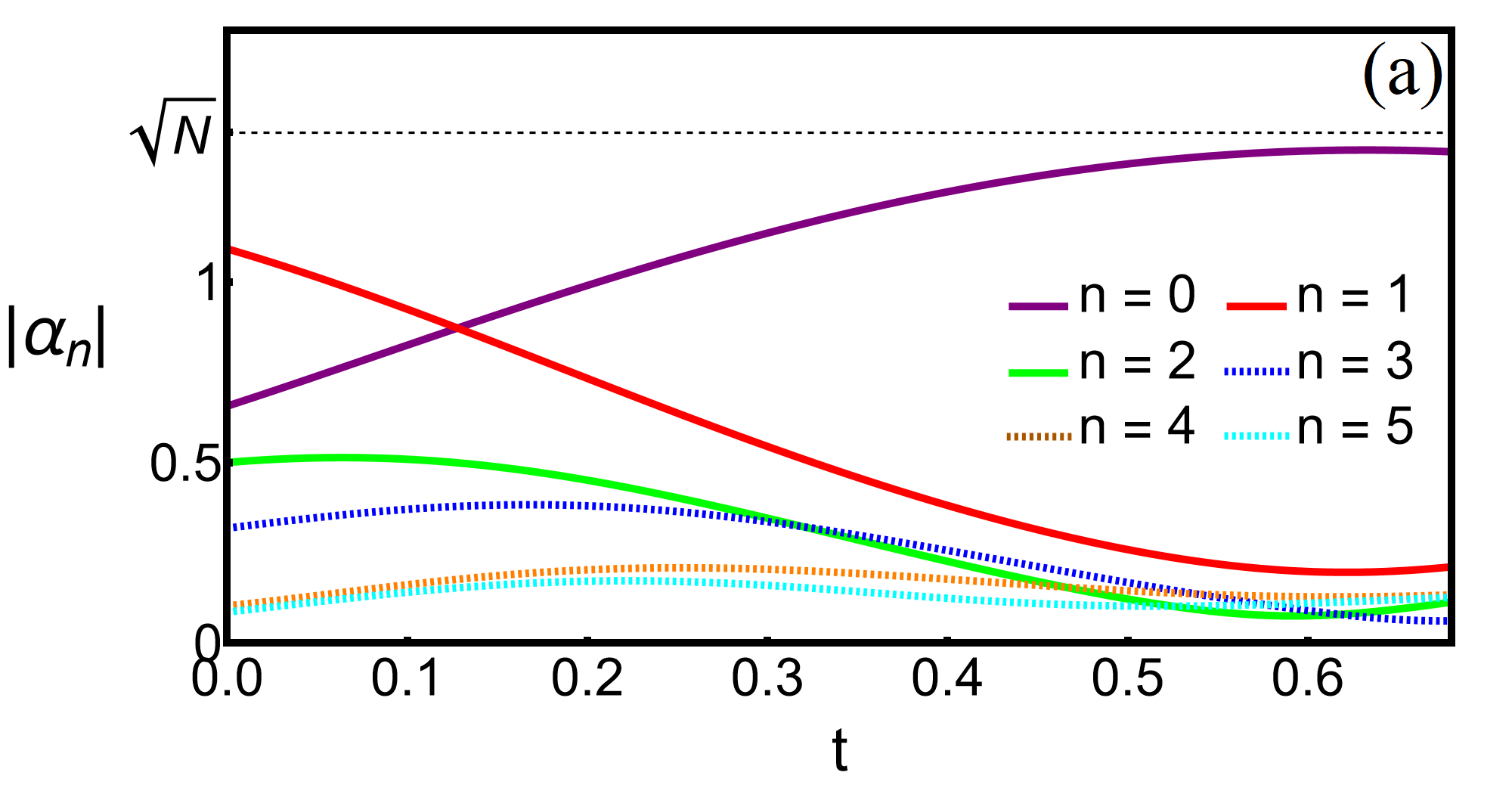}
		\includegraphics[width=\columnwidth]{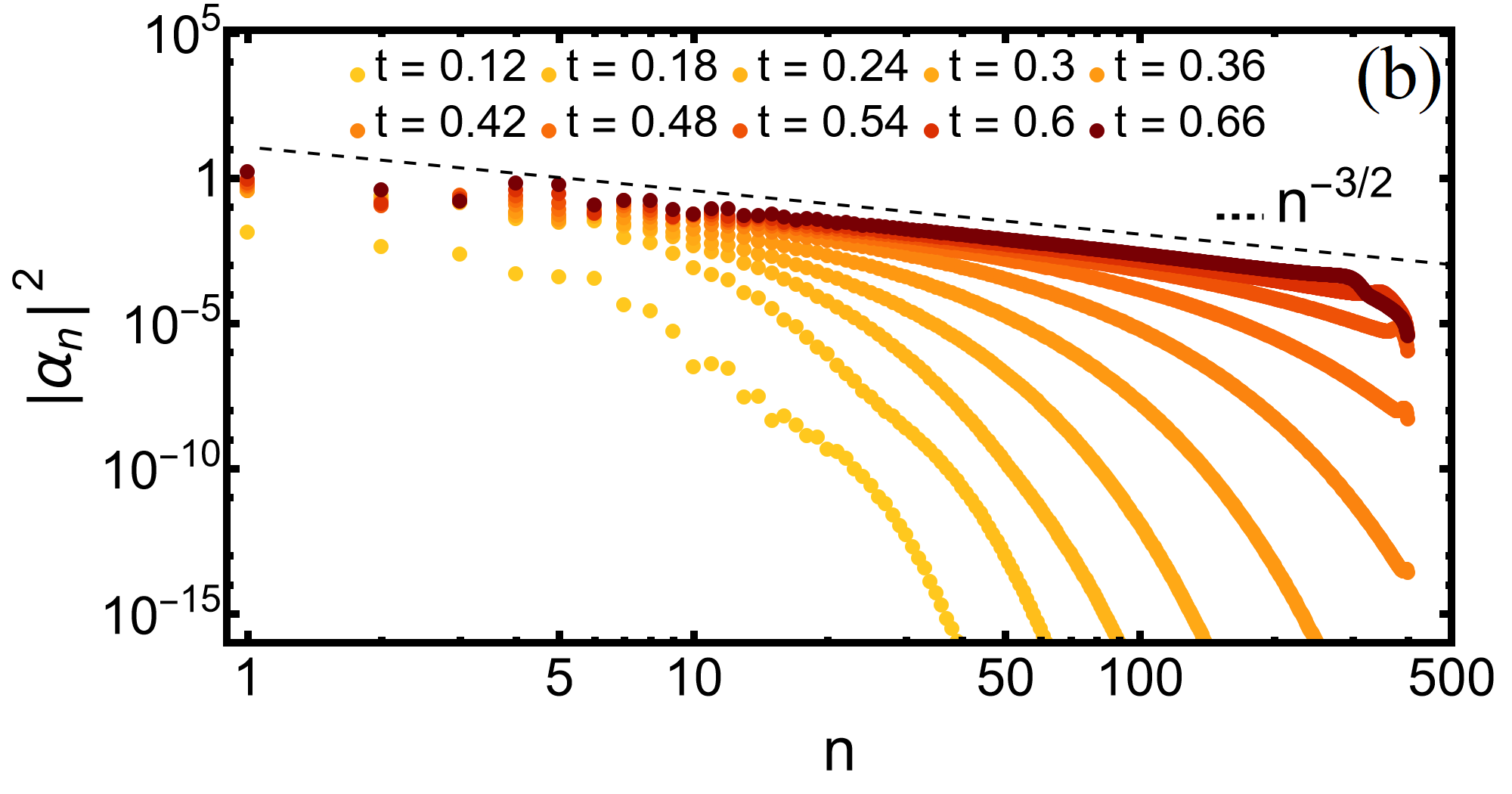}
		
		\includegraphics[width=0.49\columnwidth]{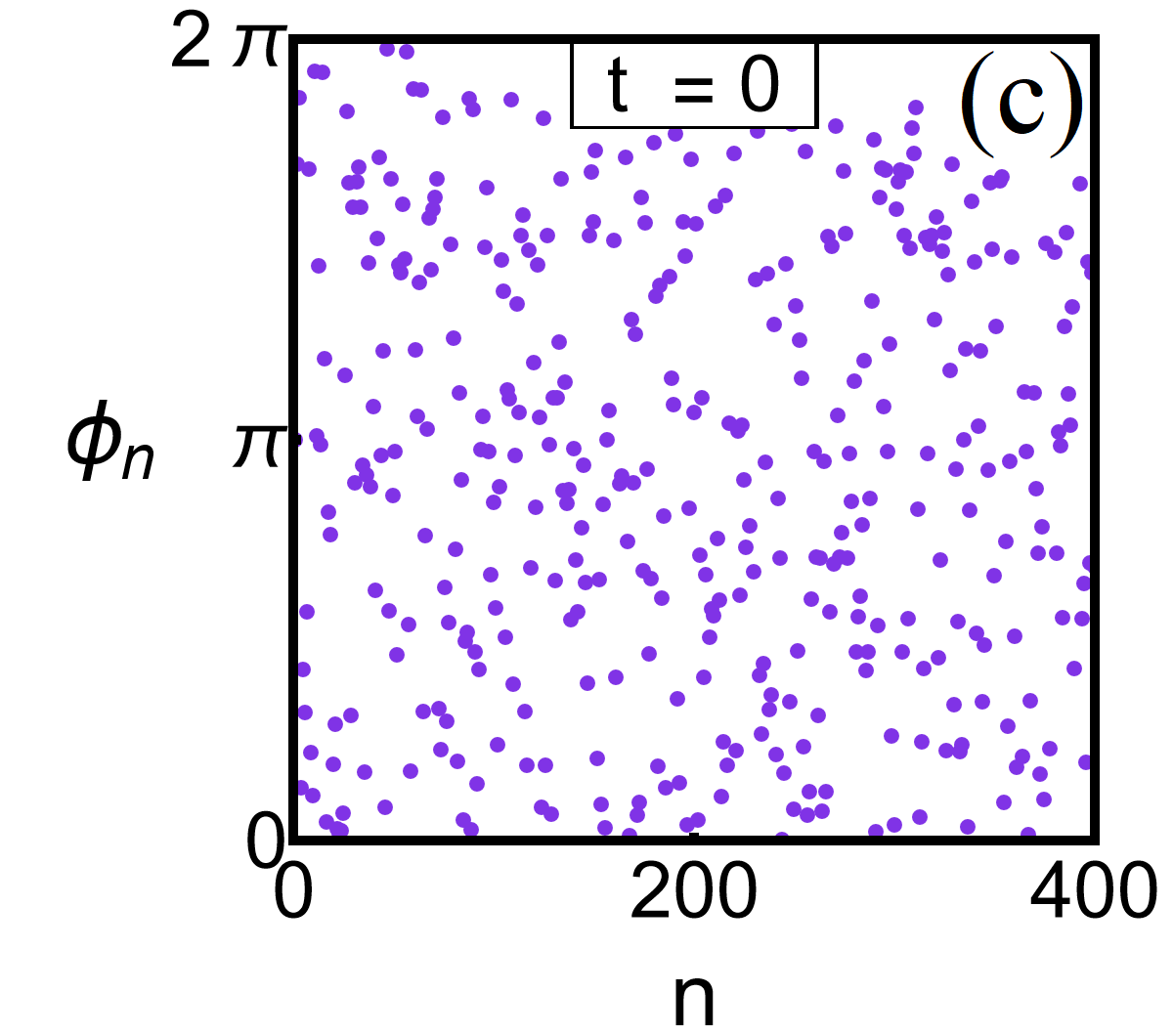}
		\includegraphics[width=0.49\columnwidth]{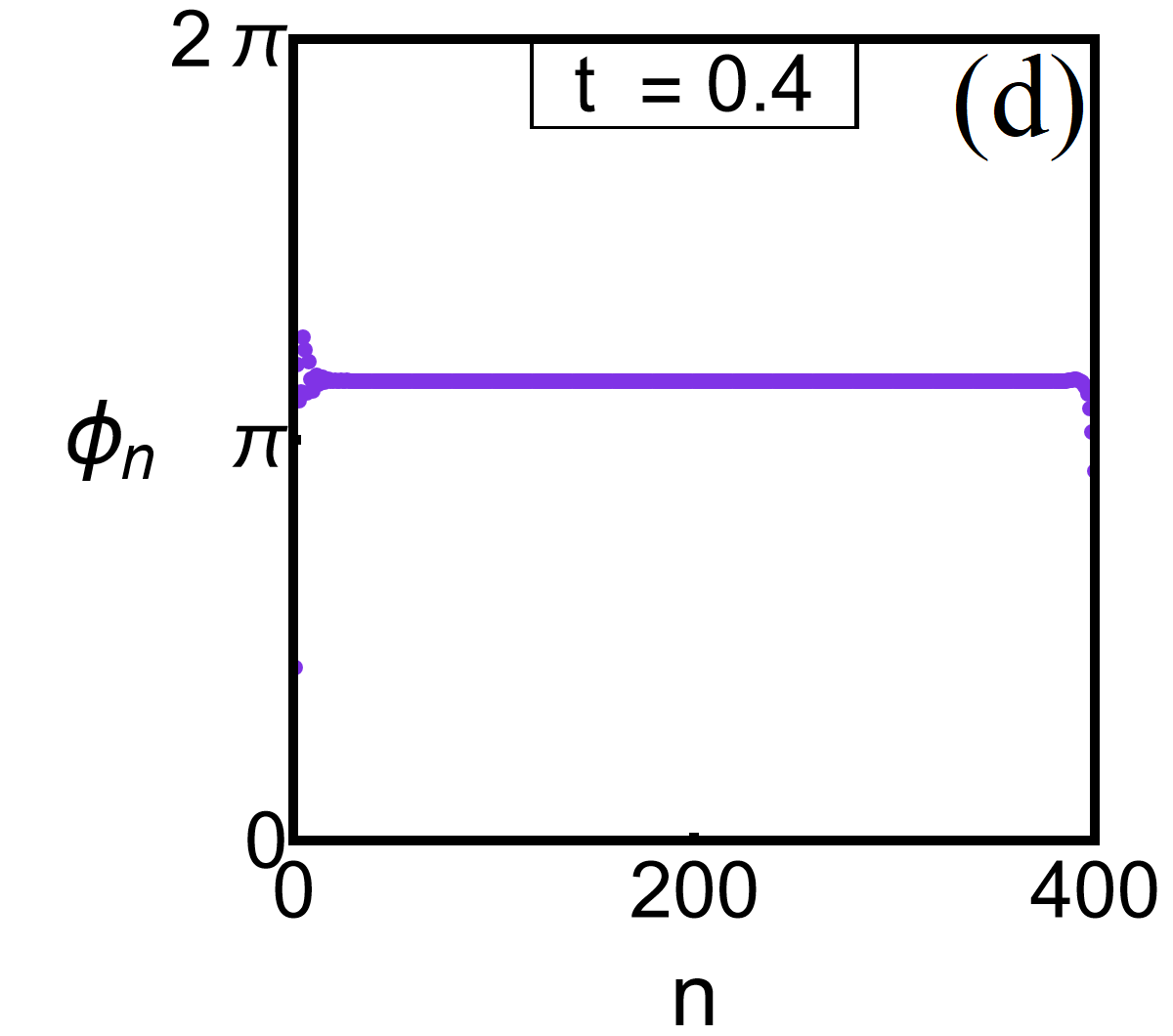}
		\caption{Numerical simulation of system (\ref{eq:Resonant_Equation})-(\ref{eq:C_nmij_equation}) with $400$ modes and an initial condition outside ansatz (\ref{eq:invariant_manifold}) with $N=2$ and $E=2.11$ (provided in Appendix~\ref{apx:Numerical_simulations}). A large fraction of $N$ accumulates  at  $|\alpha_0|^2$, while the amplitude spectrum approaches the power-law $n^{-3/2}$ (visualized compensating the spectrum decay). $\alpha_n$ transits from random phases (c) to a coherent state (d). Truncation effects are visible around $n\sim 400$.}
		\label{fig:Simulation_2}
	\end{figure}

\begin{figure*}[t]
		\centering
		
		\includegraphics[width=0.315\textwidth]{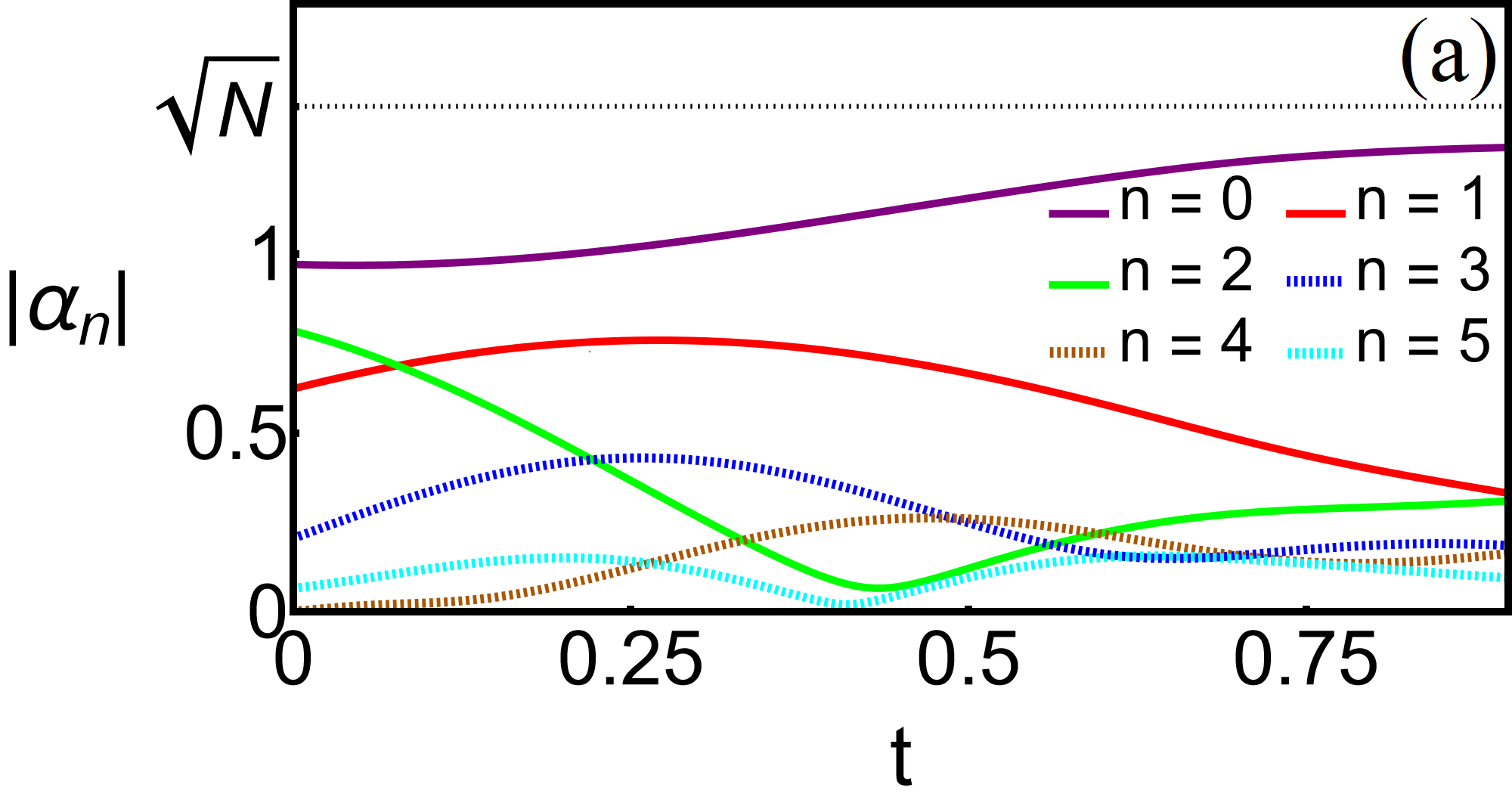} \hspace{0.15cm}
		\includegraphics[width=0.315\textwidth]{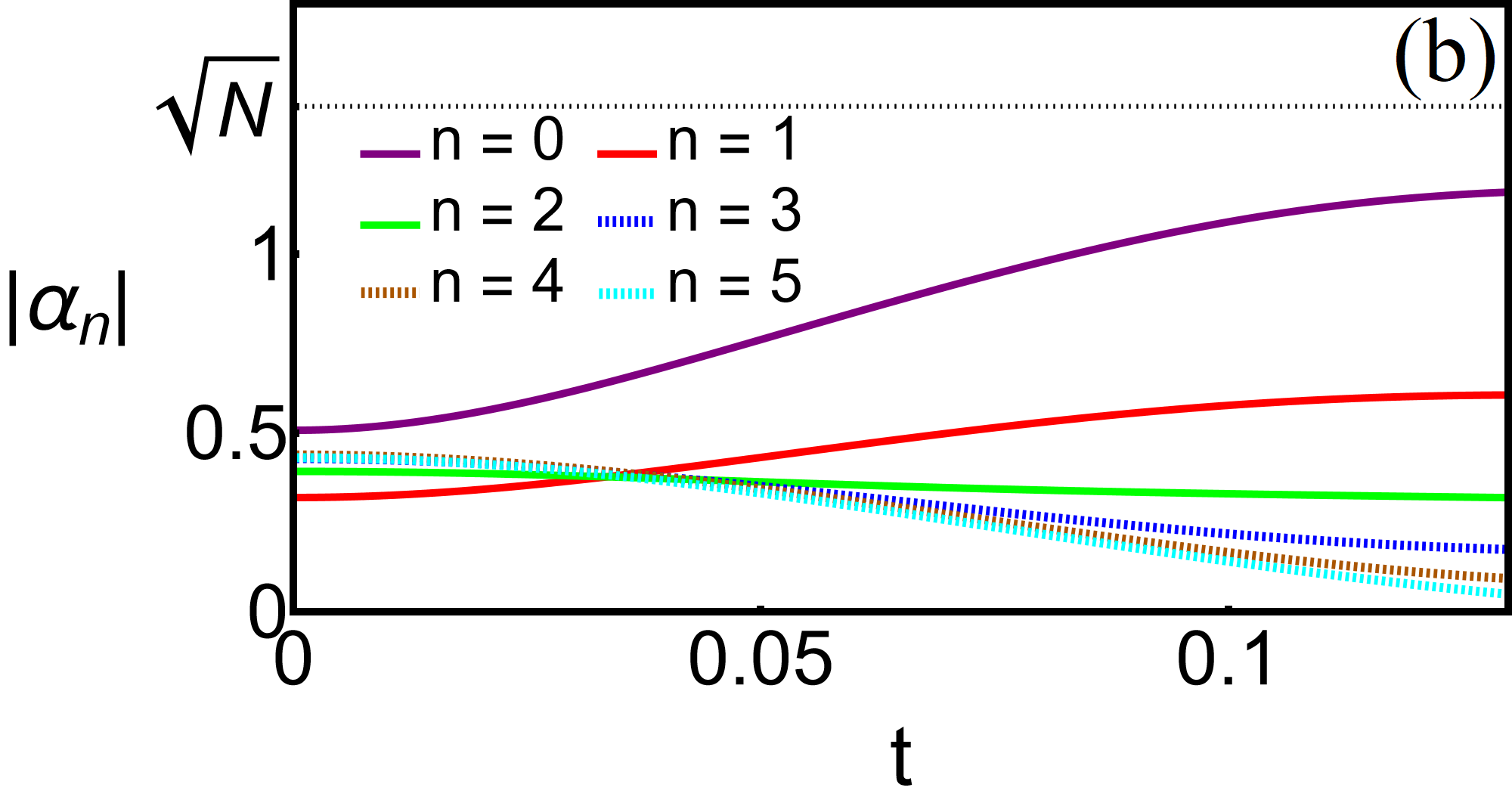} \hspace{0.15cm}
		\includegraphics[width=0.315\textwidth]{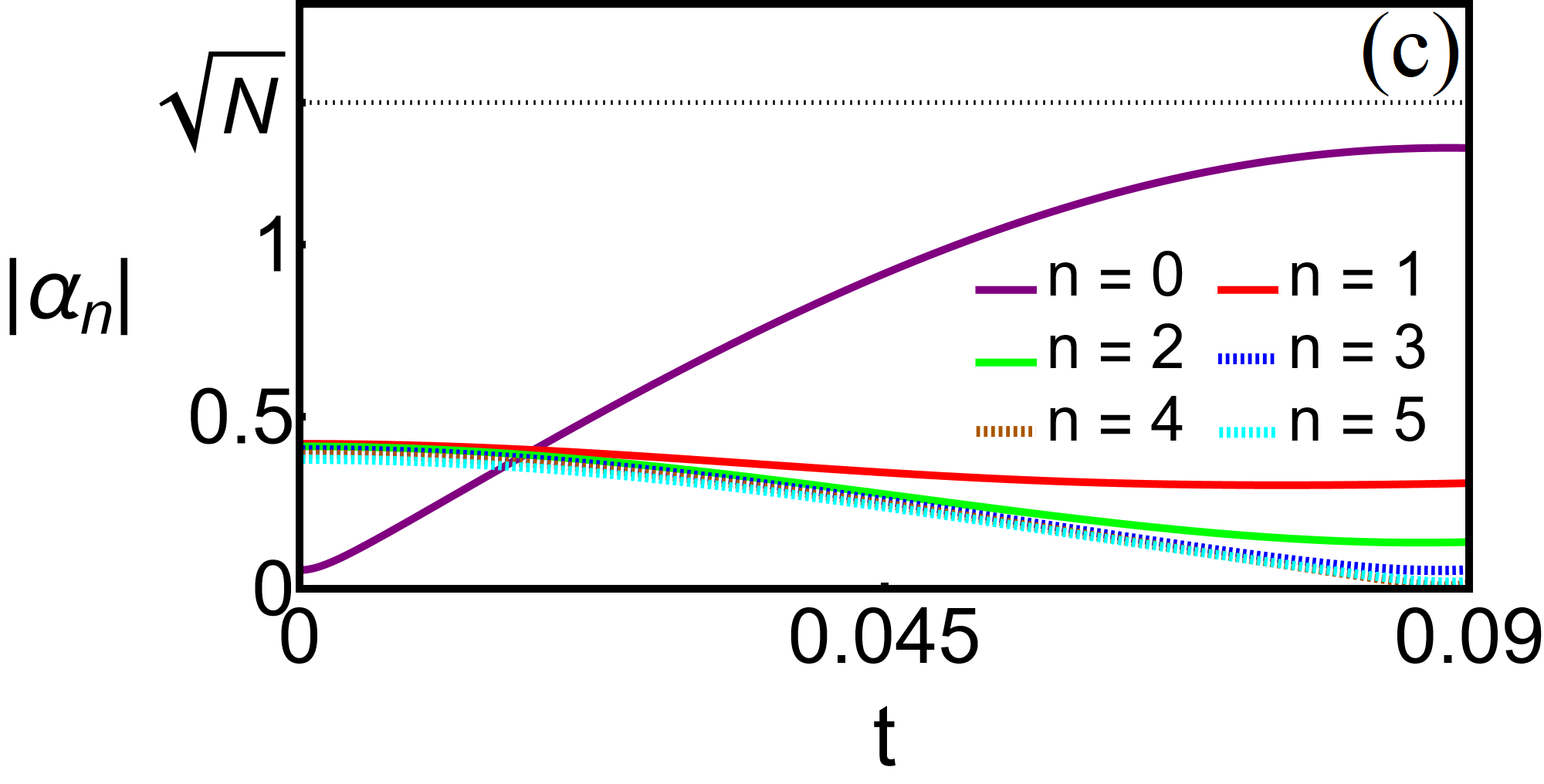}
		
		\includegraphics[width=0.325\textwidth]{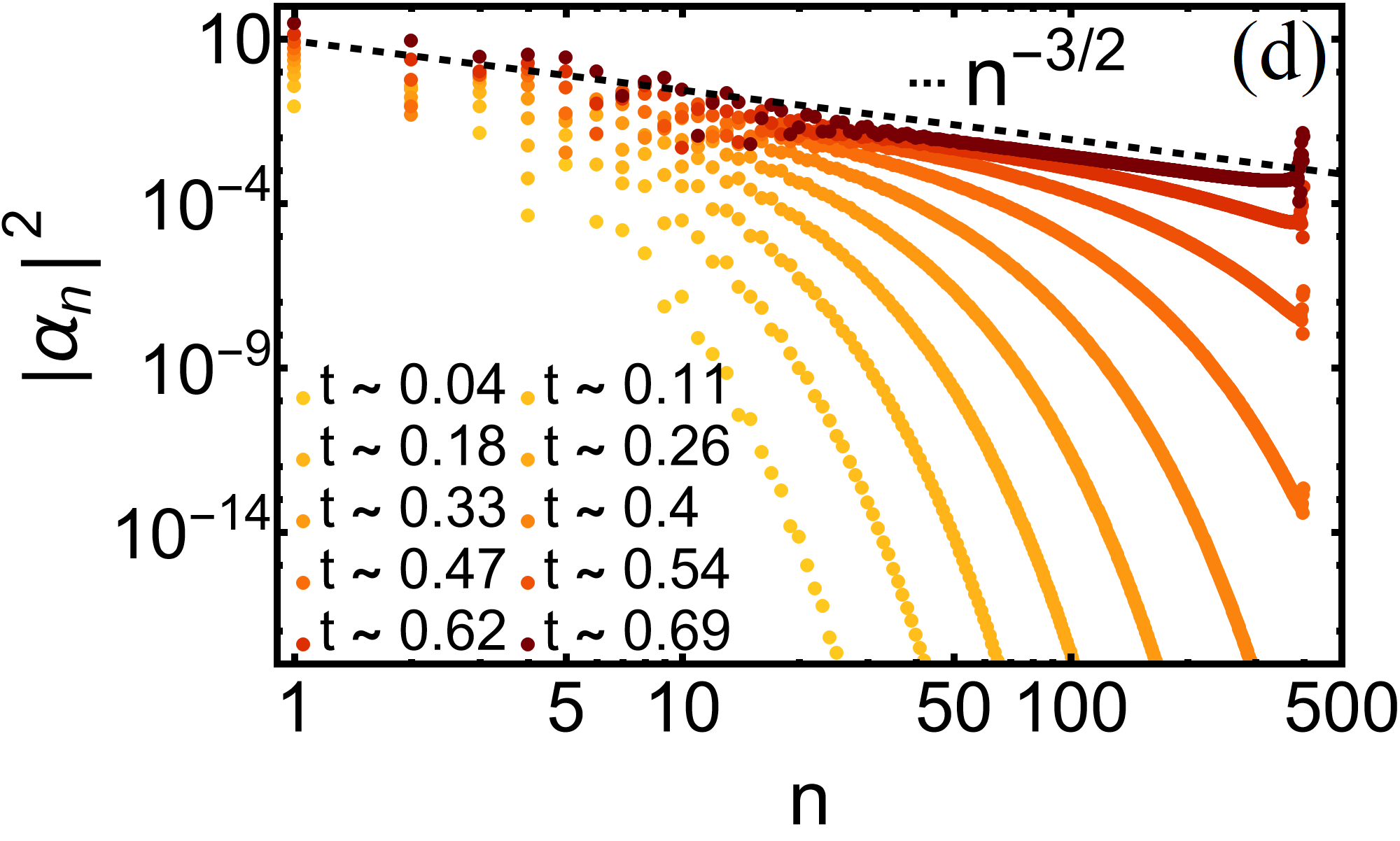}
		\includegraphics[width=0.325\textwidth]{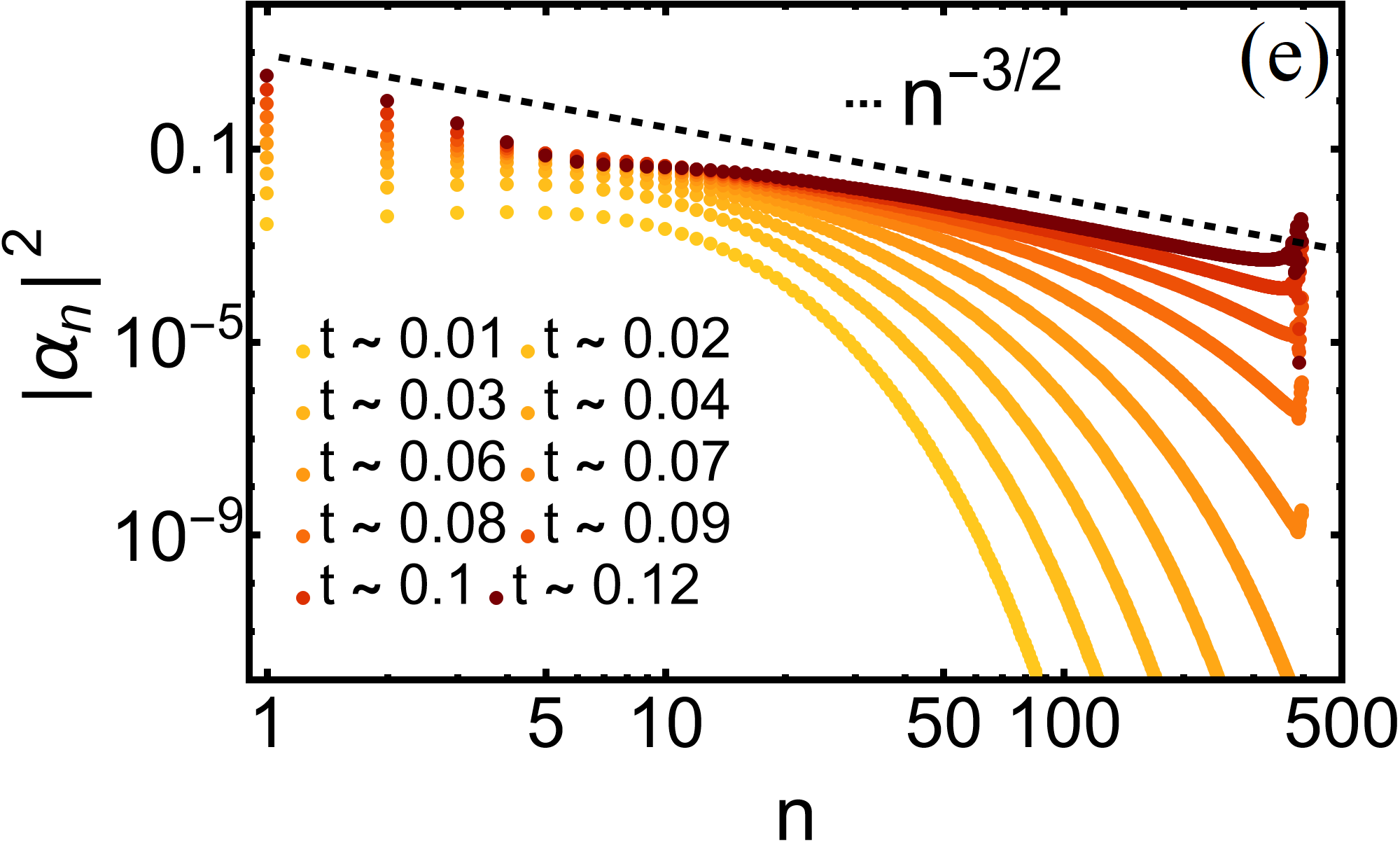}
		\includegraphics[width=0.325\textwidth]{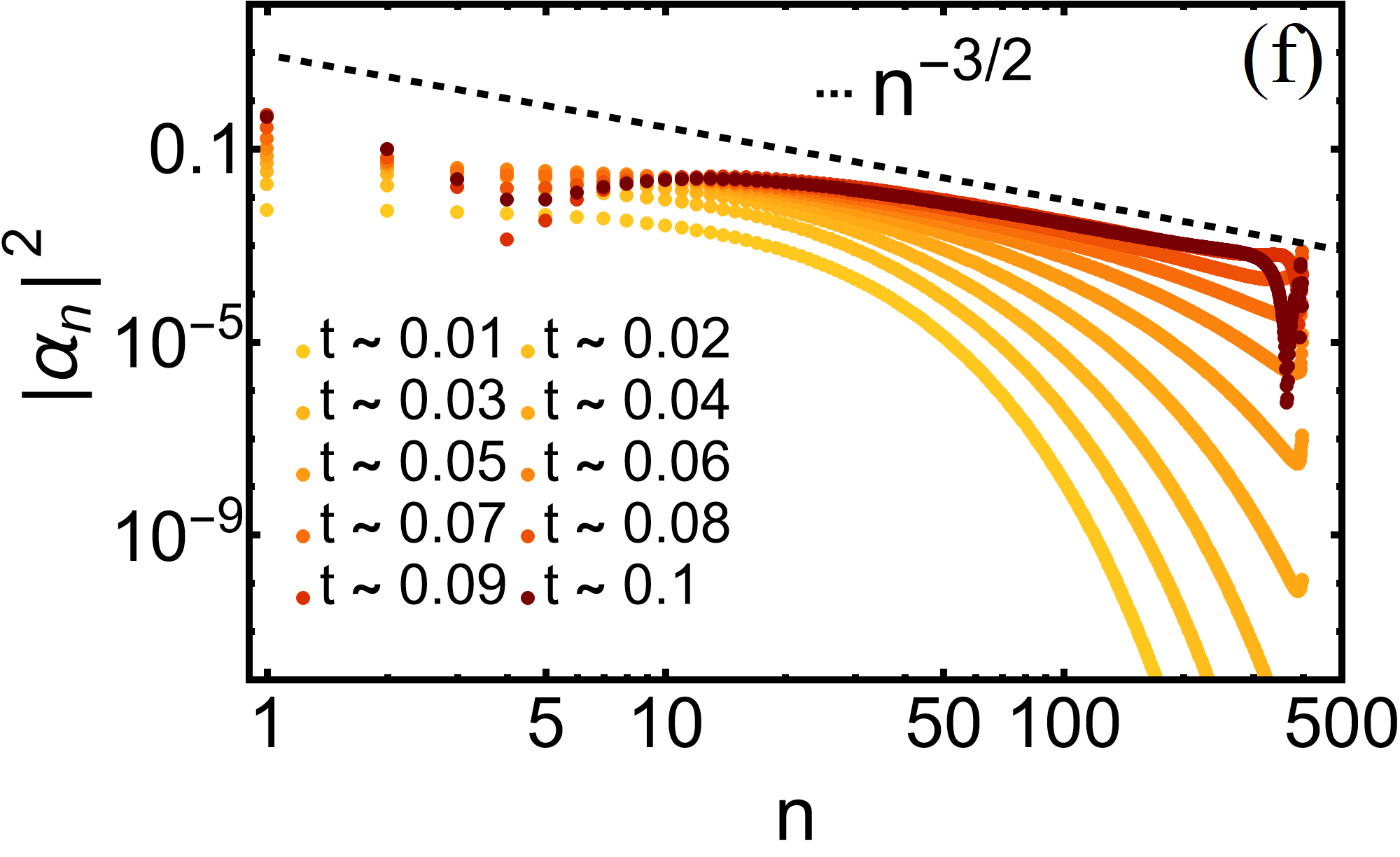}
		
		\caption{Numerical simulation of three initial conditions outside ansatz (\ref{eq:invariant_manifold}). First row: Evolution of the first mode amplitudes. Second row: Amplitude spectrum approaching the power-law $n^{-3/2}$ (visualized compensating the spectrum decay). Simulations have been performed using $400$ modes and truncation effects are visible at nearby modes.  (a, d) The first six modes with random phases (see Appendix~\ref{apx:Numerical_simulations}): $E=1.78$ and $N=2$. (b, e) $\alpha_0(0) = 0.51$, $\alpha_{n\geq 1}(0) = 0.2 (n+1) e^{-0.2n}$: $E=11.57$  and $N=2$. (c, f) $\alpha_0(0) = 0.05$, $\alpha_{n\geq 1}(0) = 0.46 (n+1)f_n e^{-0.77n}$: $E=18.24$ and $N=2$.}
		\label{fig:robustness_main_text}
	\end{figure*}

	Numerical simulations have been used to explore conditions more general than the ansatz in (\ref{eq:invariant_manifold}). Across different classes of conditions, these simulations revealed dynamics that exhibited characteristic features of our solutions for coherent  condensation. As Fig.~\ref{fig:Simulation_2} and Fig.~\ref{fig:robustness_main_text} illustrate when compared with Fig.~\ref{fig:Time_Ev_alpha_n_main_text}, we observe a strong concentration of $N$ at the lowest mode and the approach to the power-law $|\alpha_n|^2\sim n^{-3/2}$. Particularly interesting is the emergence of this kind of dynamics from highly incoherent states (random phases), as demonstrated in Fig.~\ref{fig:Simulation_2}. It indicates that a high degree of correlation between modes is not necessary for a close manifestation of coherent condensation, as it can arise during evolution. The initial condition in that figure has  been generated by adding random phases and amplitudes to our ansatz in the form $\alpha_n(0) = \mathcal{A}_n e^{i \mathcal{P}_n}\alpha_n^{(a)}(0)$, where $\alpha_n^{(a)}$ represents the ansatz and $\mathcal{A}_n$, and $\mathcal{P}_n$ are random numbers uniformly distributed on $[0,2]$ and $[0,2\pi)$, respectively. Similar  dynamics have been observed from other classes of initial conditions. For instance, several modes initially excited with random phases (Fig.~\ref{fig:robustness_main_text}a), or spectra with different asymptotic behaviors: $n^2$ (Fig.~\ref{fig:robustness_main_text}b),  and  $n^{1/2}$ (Fig.~\ref{fig:robustness_main_text}c). These  conditions revealed a preference of the system for spectra led by the power $n^{-3/2}$, as they approached it regardless of their initial spectrum. Further details are provided in Appendix~\ref{apx:Numerical_simulations}
	
\begin{figure}[t!]
		\centering
		\includegraphics[width=0.97\columnwidth]{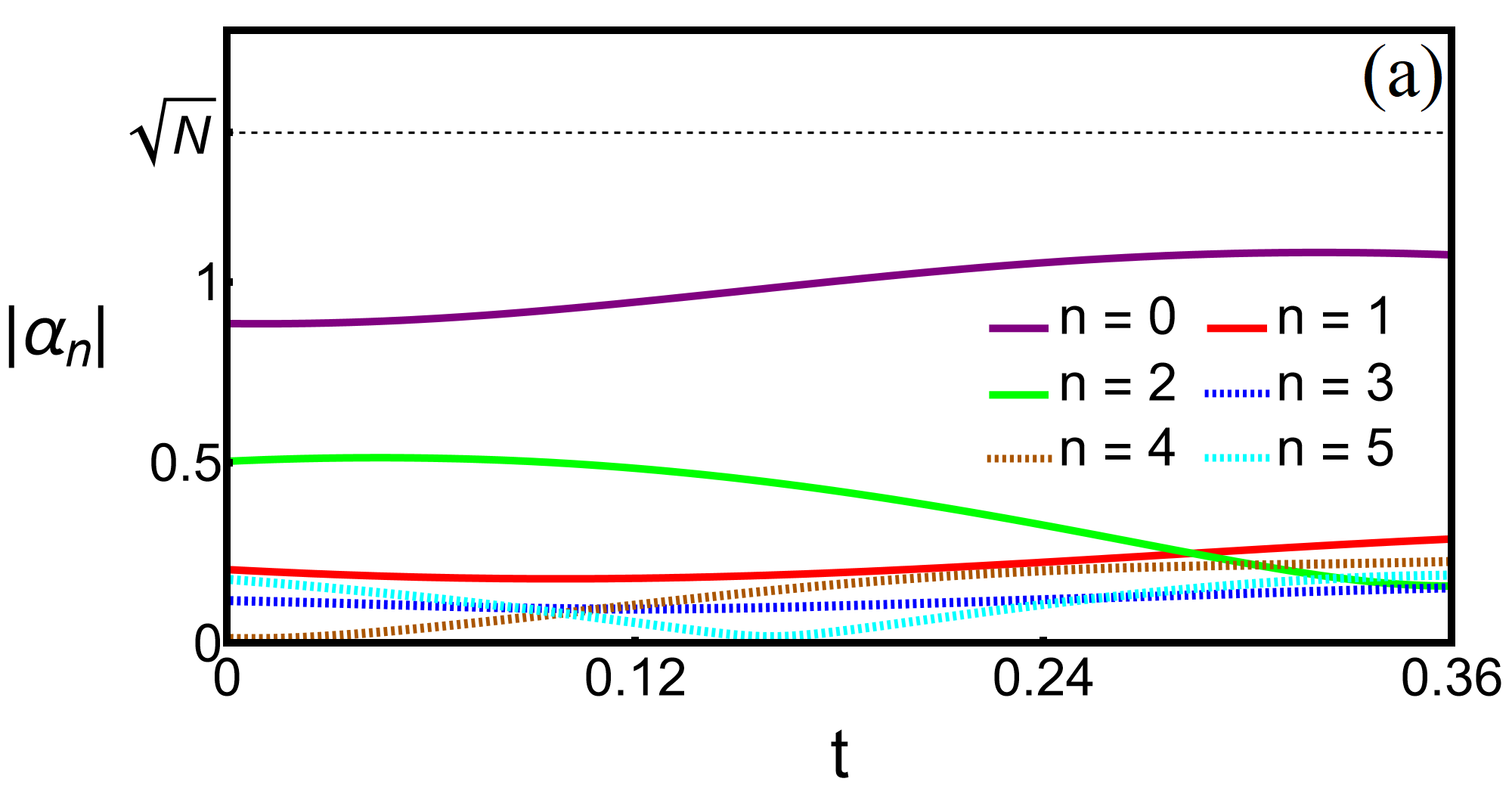}
		\includegraphics[width=\columnwidth]{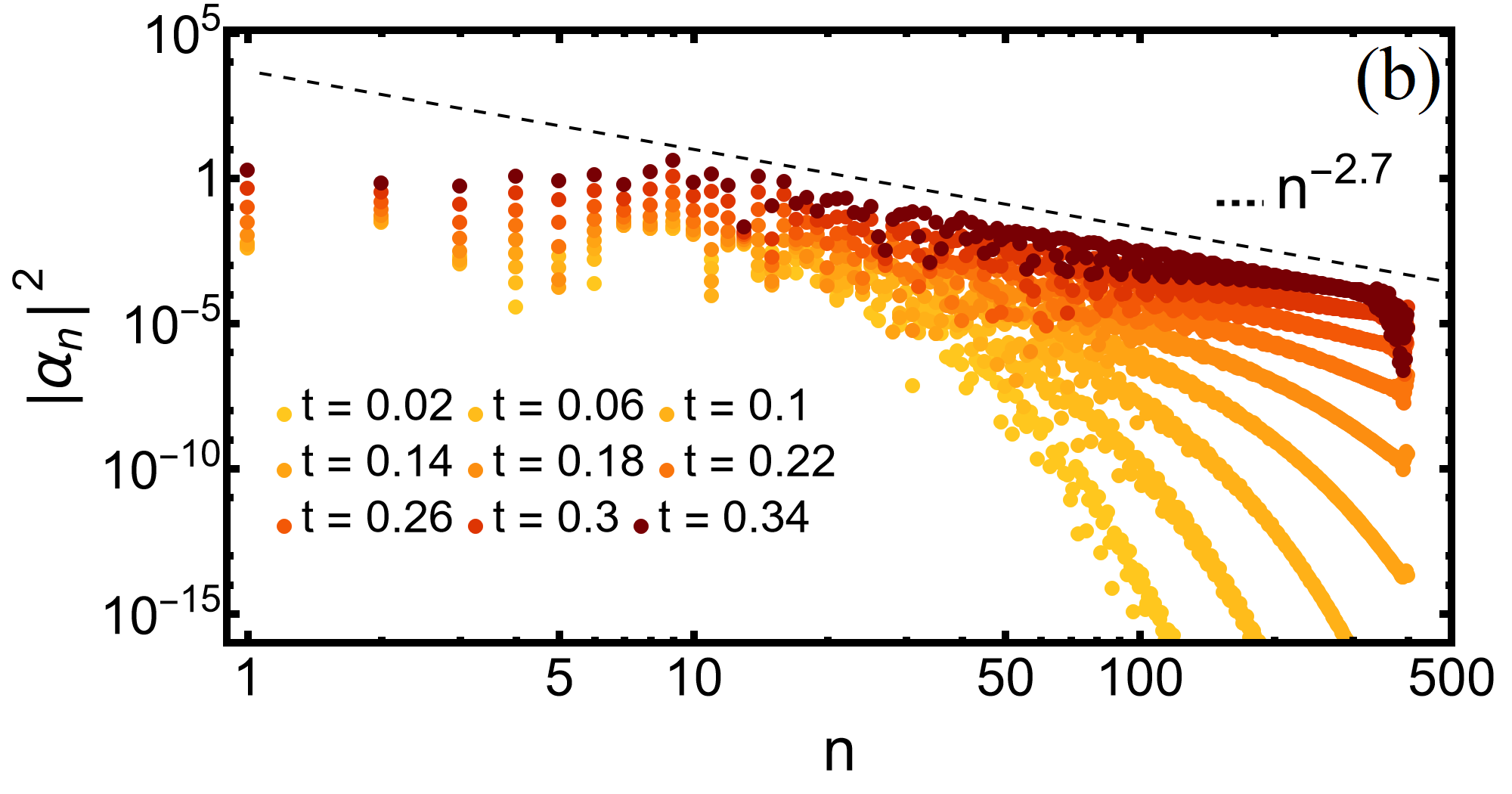}
		
		\includegraphics[width=0.49\columnwidth]{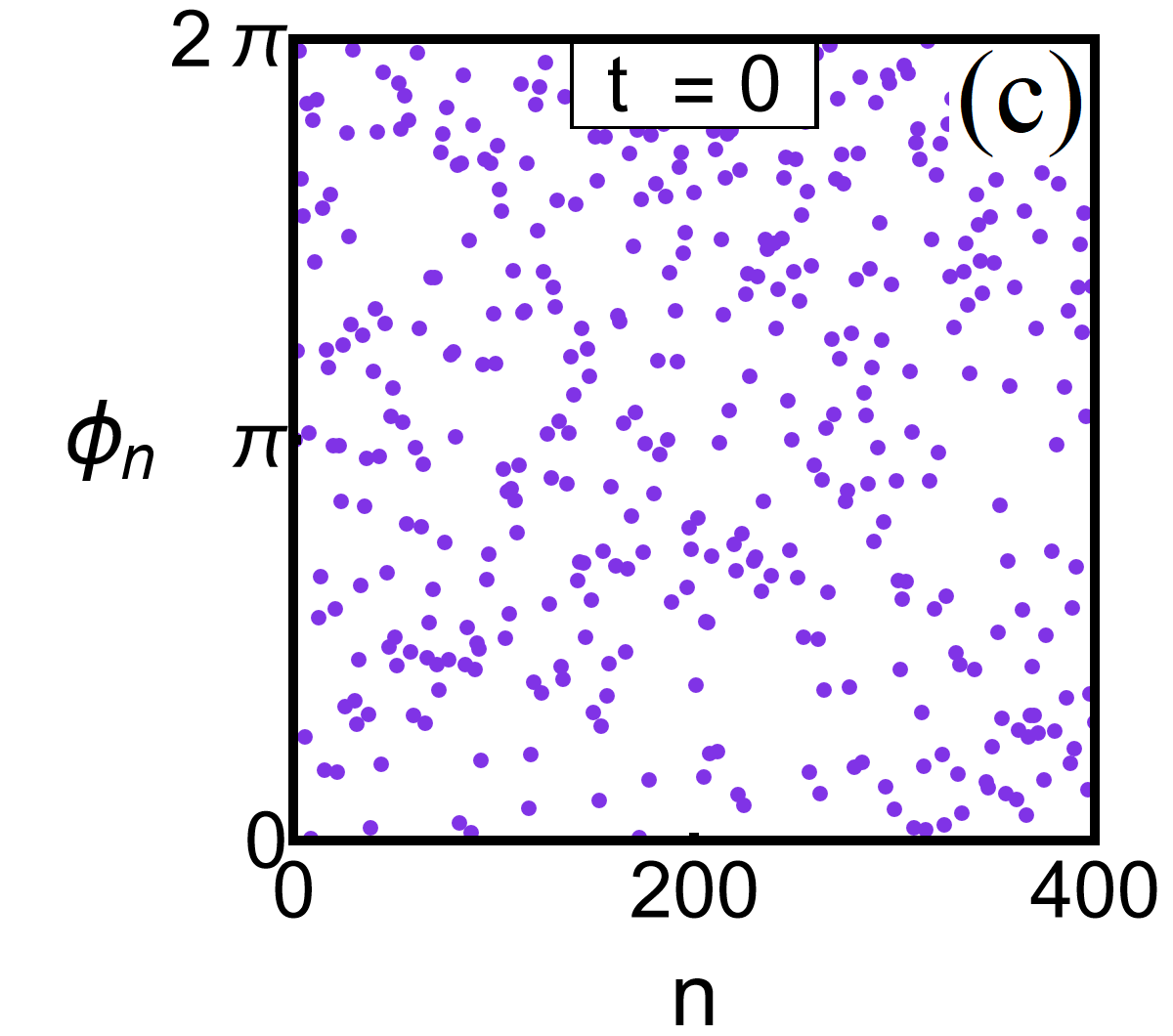}
		\includegraphics[width=0.49\columnwidth]{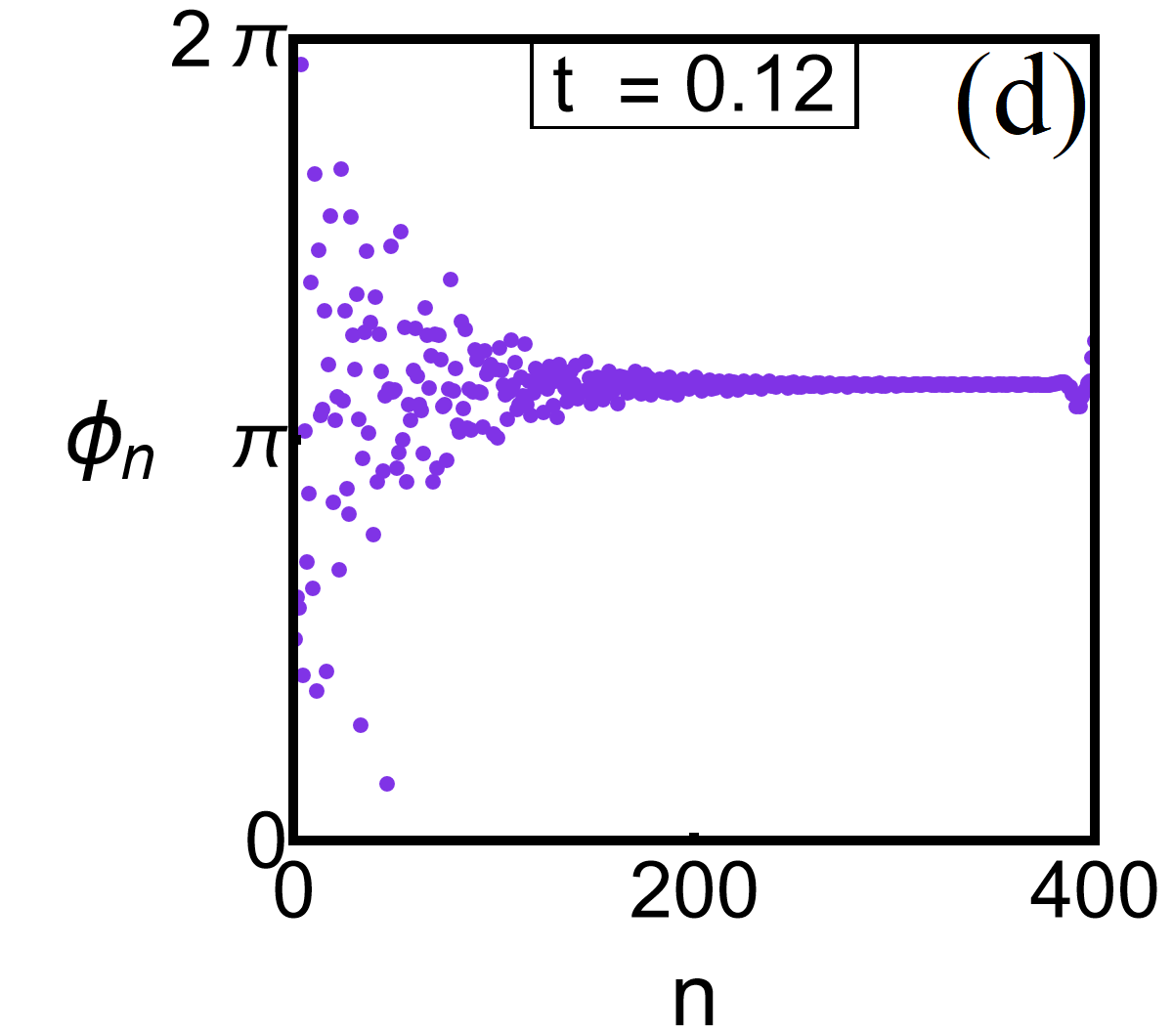}
		\caption{Numerical simulation starting from a fluctuating spectrum with random phases in the form:  $\alpha_n(0) = \mathcal{A}_n(n+1)e^{-0.2n + i\mathcal{P}_n}$ ($N=2$, $E=9.17$). High modes develop strong correlations, contrary to low ones. The amplitude spectrum  presents oscillations when the power-law is approached, contrary to Fig.~\ref{fig:Simulation_2} and Fig.~\ref{fig:robustness_main_text}.}
		\label{fig:Simulation_3}
	\end{figure}

	In general, we observe that our Hamiltonian system exhibits common features in the behavior of high modes: they get significantly  excited after some time and develop a high degree of correlation in their phases. We also find that an asymptotic spectrum of the form $|\alpha_{n\gg  1}|^2\sim n^{-3/2} e^{-\rho n}$ arises during the evolution, tending to approach the formation of a power-law ($\rho$ becomes small). Other powers have been occasionally observed when the  spectrum develops oscillations, as illustrated in Fig.~\ref{fig:Simulation_3}. Even in those cases, the power $n^{-3/2}$ dominated for part of the evolution, switching to a smaller one when the exponential suppression weakened. Regarding the low mode behavior, it has been more difficult to characterize. Contrary to high modes,  low ones did not exhibit common features independently of the initial conditions. For instance, low modes reached different degrees of correlation in the phases when starting from random data (Fig.~\ref{fig:Simulation_2} and  Fig.~\ref{fig:Simulation_3}), or they did not perfectly converge to the Kronecker delta distribution: $|\alpha_n|^2=N\delta_{n0}$ (Fig.~\ref{fig:robustness_main_text}). However, these observations come from simulations that span a few forward cascades in the system (i.e., a few times the spectrum increased) and are not applicable to longer times. As discussed in \cite{BCE} and Appendix~\ref{apx:Numerical_simulations}, the truncation in the number of modes introduces important limitations for the simulation of systems that exhibit strong cascades like ours. In particular, spurious effects associated with the truncation make the evolution unreliable when high modes get excited, something that quickly happens in our system, obstructing the execution of long-time simulations. Consequently, the simulated dynamics did not last long enough to display effects associated with relaxation or reach conclusions about  the role of condensation processes in the long-time behavior of the system. Addressing these questions requires enhancements in the numerical description of direct cascades in the Hamiltonian structures (\ref{eq:Resonant_Equation}). In this regard, our analytic solutions are well-suited for benchmarking, as they display a direct cascade in finite time and are exact and explicit.
	
	\section{Threshold in $E/N$}
	\label{sec:Threshold_Ec}
	
	A cutoff is commonly introduced in systems that exhibit energy concentration processes, such as the one discussed in Sec.~\ref{sec:Formation of small- and large-scale structures},  precisely to prevent energy leakage to arbitrarily small scales. This action is known to generate a threshold in the ``energy per particle" $(E/N)_c$ for condensation to occur in non-equilibrium setups  \cite{Condensation_2005,Condensation_2011}. This  fact is now discussed in our system. We first explain the emergence of a threshold within the ansatz in (\ref{eq:invariant_manifold}), where an exact expression can be derived analytically. Then, we  discuss this question for more general conditions. Before that, note the absence of a threshold without a cutoff, as we  have  obtained an exact solution undergoing  condensation for any positive $E/N$; see Eq.~(\ref{eq:family_conditions}).
	
	First, restricted  to the ansatz in (\ref{eq:invariant_manifold}), we use as a cutoff an upper bound for $x(t)< x^*\lesssim 1/4$. It indicates that, for physical reasons, the dynamics of the system remain reliable until the excitation of high modes reaches the admissible level $x(t)=x^*$. Under this restriction, a threshold follows from (\ref{eq:b_c_in_terms_of_x_N_E}):
	\beq
	\left(\frac{E}{N}\right)_{\hspace{-0.1cm} c} = \frac{1}{\sqrt{1-4 x^*}}, \quad \text{and} \quad \frac{|\alpha_0^*|^2}{N} = 1 - \frac{E}{E_c}.
	\label{eq:Threshold}
	\eeq
	The second expression is the fraction of $N$ stored at the lowest mode when the cutoff is reached $x(t)=x^*$. The origin of the threshold is quite simple in this case: it comes from the exclusion of trajectories with $\min[x(t)]\geq x^*$ and the interpretation of the  ones  with $\min[x(t)]< x^*  \leq  \max[x(t)]$ as the  same process of energy concentration. Interestingly, these processes now produce spectra with different fractions of $N$ stored at the lowest mode, contrary to the Kronecker delta distribution ($|\alpha_n|^2\to N \delta_{n0}$).	

	The presence  of a threshold is more subtle outside the  ansatz in (\ref{eq:invariant_manifold}). We illustrate it after making some assumptions  on the formation  of condensates, based on our analytic solutions  in Sec.~\ref{sec:Coherent condensation}  and numerical simulations in Sec.~\ref{sec:Numerical_Simulations}. We assume conditions that undergo condensation (i.e., $|\alpha_n|^2\to N \delta_{n0}$) accompanied by the power-law  $n^{-3/2}$. This is written in the form $|\alpha_{n\gg1}|^2 \sim (4/\sqrt{\pi})|c|^2 n^{-\frac{3}{2}} (4x)^{n-1}$ to define functions $x(t)$ and $c(t)$ outside the ansatz. They still need to satisfy $x\to 1/4$ and $|c|\to 0$ to represent condensation. Additionally, $|\alpha_{n\geq 1}|^2$ must vanish, and we assume this happens at the same rate as the asymptotic spectrum decays, or faster, guaranteeing that  all $|\beta_{n\geq  1}|^2 = |\alpha_{n}|^2 (x/|c|^2)$ remain bounded. For these conditions, the  following threshold can be derived when $0<1/4-x^*\ll 1$:
	\beq
		\left(\frac{E}{N}\right)_{\hspace{-0.1cm}  c} = \frac{1}{ B(x^*) \sqrt{1-4x^*}}, \ \ B(x^*) = \sum_{n=1}^{\infty} |\beta_n|^2\bigg{|}_{x=x^*}.
		\label{eq:Threshold_generalized}
	\eeq
	The new element $B(x^*)$ indicates that this expression depends on the specific way each mode approaches the Kronecker delta distribution. This dependency arises due to the power-law $n^{-3/2}$, which keeps $B(x^*)$ finite as $x^*\to 1/4$ and cannot be computed asymptotically. Higher powers would enable such computation, decoupling the  threshold from the low mode behavior; however, these have not been observed in our numerical simulations. Consequently, Eq.~(\ref{eq:Threshold_generalized}) may exhibit a significant dependence on the initial conditions through $B(x^*)$, potentially compromising  the existence of a global threshold  $(E/N)_c$. For it to exist, our system should exhibit a finite number of approaches, or a generic one, to the formation of a condensate (i.e., for given values of the conserved quantities, $\{\beta_n\}$ should tend to a common behavior as $x^*\to 1/4$). This issue thus demands an understanding of the various mechanisms in the system to approach the Kronecker delta distribution. Addressing the problem of asymptotic relaxation in Hamiltonian structures (\ref{eq:Resonant_Equation}) could shed light on this matter. The convergence to thermal equilibrium is precisely a mechanism behind the condensation in non-equilibrium setups \cite{Condensation_2005}. From a numerical perspective, the problem demands the exploration of long-time dynamics to observe the condensation  from larger classes  of  initial conditions and accurate characterization of its  formation. Specifically, the  expression  for $(E/N)_c$ in  Eq.~(\ref{eq:Threshold_generalized}) is valid asymptotically, indicating that simulations must describe the dynamics for very  small $1/4-x(t)$.  In  that regime,  determining $|\beta_n|$  becomes challenging as this is the quotient of two quantities approaching zero, and extracting the values of $x(t)$ and $|c(t)|^2$ from numerical data becomes particularly inaccurate as observed in Appendix~\ref{apx:Numerical_simulations}. Overall, enhancements in describing the direct cascade accompanying condensation are  necessary  to  clarify  the  existence of  a threshold $(E/N)_c$.

	
	 \section{Discussion and conclusions}
	\label{sec:Discussion_conclusions}
	
	In this work, we have uncovered a novel form of condensation in conservative Hamiltonian systems. Our fully analytic description of the process reveals features fundamentally different from the condensation observed in turbulent scenarios, being the most remarkable the evolution through {\em highly coherent states}. However, it shares common features with those scenarios as well, such as a dual-cascade behavior that transfers the conserved quantities towards the opposite parts of the spectrum, or the excitation of arbitrarily high modes in finite time.
	
	From a broader perspective, this paper sets the precedent that a novel form of condensation exists in a class of Hamiltonian equations associated with weakly nonlinear waves. Consequently, our result holds great potential for the observation of this coherent form of condensation in those scenarios, both theoretically and experimentally. The explicit solutions uncovered in this work will be valuable allies in that research, as they provide a full description of the development of the phenomenon and serve for benchmarking in numerical simulations. From an experimental viewpoint, we believe that nonlinear optics is a promising ground to initiate the search for coherent condensation. Hamiltonian systems similar to ours naturally arise in that field \cite{BBCE2,BEM2}, and it already has the necessary technology for the observation of wave condensation in kinetic regimes \cite{Condensation_2012,Condensation_2020}.


\section*{Acknowledgments}
I am indebted to the Simons Collaboration on Wave Turbulence for allowing me to participate in its meetings where the first ideas for this work arose. I am grateful to Patrick G\'erard for several meetings and discussions, and to Oleg Evnin, Sergey Nazarenko, and Miguel Onorato for useful discussions. This work has been supported by the LabEx ENS-ICFP: ANR-10-LABX-0010/ANR-10-IDEX-0001-02 PSL*.


\appendix


\section{The coherent regime where systems (\ref{eq:Resonant_Equation}) emerge as effective equations}
\label{apx:Coherent_Regime}

	We here illustrate the emergence of the Hamiltonian structures in Eq.~(\ref{eq:Resonant_Equation}) from a phase-sensitive regime of nonlinear waves. We use as an example the Gross-Pitaevskii equation (GPE)
	\beq
	i \partial_t \Psi = - \frac{1}{2}\Delta \Psi + V(\bfx) \Psi + g |\Psi|^2 \psi,
	\label{eq:GPE}
	\eeq
	where $\Delta$ is the Laplacian, $V(\bfx)$ a trapping potential (e.g., the harmonic trap), $\bfx\in\mathbb{R}^{d}$ the spatial coordinates, and $g\in \mathbb{R}$ the nonlinear coupling. It conserves the ``particle number" and the energy,
	\beq
		N = \int |\Psi|^2 d^{d}\bfx,
		\label{eq:GPE_conserved_quantities_N}
	\eeq
	\beq
	 	H = \int \left(\frac{1}{2} |\nabla\Psi|^2 + V |\Psi|^2 + \frac{g}{2}  |\Psi|^4\right) d^{d}\bfx.
	\label{eq:GPE_conserved_quantities_H}
	\eeq

	To derive the Hamiltonian structures in Eq.~(\ref{eq:Resonant_Equation}), one first decomposes the solution into the normal modes of the linearized equation ($g=0$),  $\Psi(t,\bfx) = \sum_{\bfk} \alpha_{\bfk}(t) \psi_{\bfk}(\bfx) e^{-i \omega_{\bfk} t}$. In this expression, $\bfk$ is a discrete $d$-wave vector, $\omega_\bfk$ is the normal frequency associated with mode $\psi_{\bfk}(\bfx)$, and $\alpha_{\bfk}(t)\in\mathbb{C}$ is the amplitude of that mode. The GPE  is then written as an infinite-dimensional system for $\alpha_{\bfk}$ 
	\beq
	i\frac{d\alpha_{\bfk}}{dt} = g \sum_{\mathbf{m}}\sum_{\mathbf{i}}\sum_{\mathbf{j}} C_{\mathbf{n}\mathbf{m}\mathbf{i}\mathbf{j}} \bar{\alpha}_{\mathbf{m}}\alpha_{\mathbf{i}}\alpha_{\mathbf{j}}\ e^{i \left(\omega_{\bfk} + \omega_{\mathbf{m}} - \omega_{\mathbf{i}} - \omega_\mathbf{j}\right) t}.
	\label{eq:qqq}
	\eeq	
	The couplings are $C_{\mathbf{n}\mathbf{m}\mathbf{i}\mathbf{j}} = \int \psi_{\bfk}\ \psi_{\mathbf{m}}\ \psi_{\mathbf{i}}\ \psi_{\mathbf{j}}\ d^d\bfx$, 
	and the symmetries $C_{\mathbf{n}\mathbf{m}\mathbf{i}\mathbf{j}}=C_{\mathbf{n}\mathbf{m}\mathbf{j}\mathbf{i}}=C_{\mathbf{m}\mathbf{n}\mathbf{i}\mathbf{j}}=C_{\mathbf{i}\mathbf{j}\mathbf{n}\mathbf{m}}$ naturally arise. This dynamical system is exactly the GPE in (\ref{eq:GPE}), we did not perform any approximation yet. It is now when the assumption of weak nonlinearities $|g|\ll 1$ is introduced, allowing us to apply perturbation techniques as the multiple scale method or time-averaging \cite{Murdock} to describe the evolution. The dynamics of the system split into timescales and relevant nonlinear effects happen at times of order $1/g$, which are captured by the following effective Hamiltonian system
	\beq
	i\frac{d\alpha_{\bfk}}{dt} = \underset{\omega_{\bfk} + \omega_{\mathbf{m}} = \omega_{\mathbf{i}} + \omega_\mathbf{j}}{\underbrace{\sum_{\mathbf{m}}\sum_{\mathbf{i}}\sum_{\mathbf{j}}}} C_{\mathbf{n}\mathbf{m}\mathbf{i}\mathbf{j}} \bar{\alpha}_{\mathbf{m}}\alpha_{\mathbf{i}}\alpha_{\mathbf{j}},
	\label{eq:Resonant_Eq_General}
	\eeq	
	where $g$ has been compensated by scaling $t$. This system receives the name of {\em resonant approximation} because only resonant terms  ($\omega_{\bfk} + \omega_{\mathbf{m}} = \omega_{\mathbf{i}} + \omega_\mathbf{j}$) remain, while others are neglected. Note that the derivation does not rely on random phases or amplitudes of $\alpha_{\bfk}$; just weak nonlinearities $|g|\ll 1$. Then, the Hamiltonian system in Eq.~(\ref{eq:Resonant_Eq_General}) provides a deterministic description of phase-sensitive (coherent) dynamics displayed by the GPE. 

	The Hamiltonian structure considered in the main text comes from Eq.~(\ref{eq:Resonant_Equation}) under two considerations. First, we reduce the d-dimensional space of modes $\bfk$ to a 1-dimensional one $n\in\mathbb{N}$. This is achieved by considering symmetries in the original equation, such as the radial one. Second, we considered a fully resonant/equidistant spectrum (i.e., $\omega_n = an+b$), which reduces the resonance condition $\omega_{n} + \omega_{m} = \omega_{i} + \omega_j$ to the simple expression $n+m=i+j$. As an example, the radially symmetric GPE (\ref{eq:GPE}) with the harmonic potential $V(\bfx) = |\bfx|^2/2$ belongs to this class of systems in any number of spatial dimensions (e.g., $\omega_n = 2n+d/2$ for $d\geq2$). This example is particularly convenient to visualize the origin of the conserved quantities $N$, $E$, and $\mathcal{H}$ in (\ref{eq:Conserved_equatities_N_E})-(\ref{eq:Hamiltonian}). They come from the conservation laws in (\ref{eq:GPE_conserved_quantities_N})-(\ref{eq:GPE_conserved_quantities_H}) getting the interpretation of the ``particle number", and the linear and nonlinear energies, respectively. Finally, the Hamiltonian systems in Eq.~(\ref{eq:Resonant_Equation}) enjoy scaling ($\alpha_n(t) \to \epsilon \alpha_n(\epsilon^2 t)$), phase-shift ($\alpha_n \to e^{i(\phi+n\theta)} \alpha_n$ with $\phi,\theta\in\mathbb{R}$), and time-reversal ($ \alpha_n \underset{t\to-t}{\longrightarrow} \bar{\alpha}_n$) symmetries.


\section{Catalan numbers}
\label{apx:Catalan}

Some properties of the Catalan numbers are listed  here \cite{Catalan}. They have the expression
\beq
	A_n = \frac{1}{2n+1}\binom{2n+1}{n},
\eeq 
with $n\in\mathbb{N}$, and the asymptotic behavior  $A_{n\gg1} \sim \frac{n^{-3/2}}{\sqrt{\pi}} 4^n$. They satisfy the identity on the left, which has been used as written on the right
\beq
\sum_{n=0}^{M} A_{M-i}A_{i} = A_{M+1}, \ \
\sum_{n=1}^{M-1} A_{M-i}A_{i} = 2\frac{M-1}{M+2} A_M
\label{eq:identity_catalan_APPENDIX}
\eeq
with $M = 2,3,4,...$ The generating function of the Catalan numbers may be rewritten as
\beq
 	F(x) = \sum_{n=1}^{\infty} A_n x^n = \frac{2 x}{1-2 x+\sqrt{1-4 x}}.
\label{eq:generating_function_APPENDIX}
\eeq
and its derivative has the form
\beq
F'(x) = \sum_{n=1}^{\infty} n A_n x^{n-1} =  \frac{F(x)}{x\sqrt{1-4x}}.
\label{eq:derivative_generating_function_APPENDIX}
\eeq
 $F(x)$ is finite at $x=1/4$, while $F'(x)$ diverges; a property we use in Sec.~\ref{sec:Coherent condensation} and Appendix~\ref{apx:Condensation}.


\section{The origin of system (\ref{eq:C_nmij_equation})}
\label{apx:Origin}
	For the construction of our system (\ref{eq:C_nmij_equation}), we first proposed a simple form for $\alpha_n(t)$ that could capture a condensation process. We realized that the ansatz presented in Eq.~(\ref{eq:invariant_manifold}) - i.e., $\alpha_0 = b,\ \alpha_{n\geq 1} = f_n c p^{n-1}$ -  could accomplish that. The idea behind that choice was to endow mode $\alpha_0$ with more freedom than the rest of modes, which would evolve following the same law. After, we constructed the couplings $C_{nmij}$ in (\ref{eq:C_nmij_equation}) by imposing that the ansatz was a solution (i.e., we substituted the ansatz in (\ref{eq:Resonant_Equation}) and searched for the couplings that reduced the system to 3-equations for the 3-unknowns). We did not have an algorithmic method for this search, but followed a trial-and-error process guided by the intuition acquired from our previous works \cite{BBE1,BE} on similar constructions. Catalan numbers arose due to their summation identity in (\ref{eq:identity_catalan_APPENDIX}), which allowed us to compute various sums once we realized how to combine them in (\ref{eq:C_nmij_equation}).


\section{Condensation in finite time}
\label{apx:Condensation}
Technical details omitted in Sec.~\ref{sec:Coherent condensation} are presented here. For the derivation of the system of equations in (\ref{eq:pdot})-(\ref{eq:cdot}), one plugs the ansatz ($\alpha_0=b$ and $\alpha_{n\geq 1} = f_n c p^{n-1}$) into the equations, divides by $f_n p^{n-1}$ on both sides, and uses the identities for the Catalan numbers in (\ref{eq:identity_catalan_APPENDIX})-(\ref{eq:derivative_generating_function_APPENDIX}) to replace the summations. It results in an expression linear in $n$ on both sides of the equations. From the equation for $\dot{\alpha}_0$ we obtain an equation for $\dot{b}$, while  the equations for $\dot{c}$ and $\dot{p}$ come from $\dot{\alpha}_{n\geq 1}$ after equating the coefficients that accompany $n$ on the one hand, and the independent terms on the other hand. The equation for $\dot{x} = \bar{p}\dot{p} + p \dot{\bar{p}}$ has the form 
\beq
\dot{x} = \sqrt{1-4 x}\frac{1+\sqrt{1-4 x}}{1-2 x+\sqrt{1-4 x}} (\bar{b}\bar{p}c - b p \bar{c}),
\eeq
which may be written as follows
\beq
\dot{x}^2 + V(x) = 0
\label{eq:xdot_apx}
\eeq
after using the relation $(\bar{b}\bar{p}c - b p \bar{c})^2 = (\bar{b}\bar{p}c + b p \bar{c})^2 - 4 |b|^2 |c|^2 x$ and writing the Hamiltonian as $\mathcal{H} = \frac{1}{2}\left(N^2 + E S\right)$ to work with the quantity
\begin{multline}
S = \frac{8}{1+\sqrt{1-4x}} \left(b\bar{c}p + \bar{b}c\bar{p}\right)\\ +  2 N-4E(1-4 x) + 2N \sqrt{1 - 4 x}.
\label{eq:conserved_quantity_S}
\end{multline}
The potential has the form
\begin{widetext}
	\begin{multline}
	V(x) = \frac{1}{64} \left(1-4 x+\sqrt{1-4 x}\right)^2 \Bigg{[}\left(4 N-\frac{2 S}{1+\sqrt{1-4
			x}}\right)^2 
\\	-8 E \frac{ \left(1-4 x+\sqrt{1-4
			x}\right) }{1-2 x+\sqrt{1-4 x}} \left(4 N-\frac{S \left(1-4 x+\sqrt{1-4 x}\right)}{1-2 x+\sqrt{1-4 x}}\right) 
		 + 16 E^2	\frac{ \left(1-4 x+\sqrt{1-4 x}\right)^2}{\left(1-2 x+\sqrt{1-4 x}\right)^2}\Bigg{]}. \label{eq:Veff}
	\end{multline}
\end{widetext}

The construction of solutions representing condensation is reduced to a standard study of this potential. As we have seen in the main text, the condensate forms as $x\to1/4$, namely, when $F'(x)$ diverges. Therefore, we had to find initial conditions for which $x(t)$ reached that value. Equation (\ref{eq:xdot_apx}) indicates  that  it only happens when the potential approaches $V(1/4)=0$ from negative values. Then, an exploration of the potential close to that point
\begin{multline}
V(x\sim1/4) \sim \frac{1}{16} (S-2 N)^2 (1 - 4 x)\\ - 
\frac{1}{4}  N (4 E + S - 2 N) (1 - 4 x)^{3/4}
\end{multline}
reveals that $x=1/4$ is only reached by initial conditions satisfying 
\beq
S=2N.
\label{eq:condition_condensation_S_2N}
\eeq
Namely, in  our ansatz  (\ref{eq:invariant_manifold}), only the  initial  conditions that satisfy  this  property develop a condensate. Any other conditions display a periodic or stationary motion.
Focusing on the ones that undergo condensation, we substitute $S$ and $N$ by its expressions in terms of $b$, $c$, and $p$, writing (\ref{eq:condition_condensation_S_2N}) in the form
\beq
\frac{F (F+1)^2}{x} |c|^2 = \Big{|}(1-F) b +\frac{2 F}{x} c \bar{p}\Big{|}^2,
\eeq
which is solved by
\beq
c = b p\frac{-2 F+e^{i \lambda}  (F+1) \sqrt{F} }{(F-1) F},
\label{eq:condensation_condition_c}
\eeq
where $\lambda\in [0,2\pi]$. Then, this is the family of initial conditions that undergo condensation in finite time. To see that, we substitute (\ref{eq:condition_condensation_S_2N})  in the potential $V(x)$, which is greatly simplified
\beq
V(x) = \frac{1}{4} (1 - 4 x) ((4 E^2 + N^2) (1 - 4 x) - 4 E N \sqrt{1 - 4 x}).
\label{eq:Veff_condensation}
\eeq
Eq.~(\ref{eq:xdot_apx}) is now solved by
\begin{multline}
x(t) = \frac{1}{4 (F_0+1)^2} \left(2+(F_0-1) \sin ^2(\Omega t )\right) \\
\times \left(2 F_0-(F_0-1) \sin ^2(\Omega t )\right)
\label{eq:x_solution}	
\end{multline}
with	
\beq
F_0 = \left(\frac{N-2E}{N+2E}\right)^2 \hspace{-0.2cm}, \ \ \ \Omega = \frac{N+2E}{2}\sqrt{\frac{F_0+1}{2}}.
\label{eq:Omega}
\eeq
From these expressions we find $x(T)=1/4$ at time $T = \pi/(2\Omega)$. Then, as $t$ approaches $T$  the function $F(x)$ remains  finite but  $F'(x)$  diverges,  as can be seen from (\ref{eq:generating_function_APPENDIX})  and (\ref{eq:derivative_generating_function_APPENDIX}).  The conservation of $N$ and $E$ in (\ref{eq:b_c_in_terms_of_x_N_E}) leads to  $|b|^2\to  N$  and  $|c|^2\to  0$. With this, the spectrum converges to the lowest mode
\beq
|\alpha_n|^2 \to N\delta_{n0},
\label{eq:Kronecker_delta}
\eeq
where $\delta_{n0}$ is the Kronecker delta. All these solutions develop the same power-law
	\beq
	|\alpha_{n\gg 1}|^2 \underset{t\sim T}{\sim} \frac{4}{\sqrt{\pi}}|c|^2 n^{-3/2},
	\label{eq:power-law}
	\eeq
	as extracted from the asymptotic behavior of Catalan numbers in Appendix~\ref{apx:Catalan}. 
	
	The expressions for the phases of $\alpha_n(t)$ follow from the phases of $b$, $c$,  and $p$. They are obtained by integrating the equations for $\dot{b}$, $\dot{c}$, and $\dot{p}$ in (\ref{eq:cdot}). First, we decompose these complex functions into modulus and phases $b(t) = |b(t)|e^{i\phi_b(t)}$, $c(t) = |c(t)|e^{i\phi_c(t)}$, and  $p(t) = |p(t)|e^{i\phi_p(t)}$, to obtain differential equations for the phases. Then, we substitute the explicit expressions for the modulus previously obtained, together with the constraint in the combination $\phi_b+\phi_p-\phi_c$ coming from the conserved quantity $S$ in (\ref{eq:conserved_quantity_S}). The integration of each equation leads to
\begin{widetext}
\beq
	\phi_b(t) = \phi_b(0) -\frac{1}{2} (2 E+N) t -\arctan\left(\frac{\sqrt{4 E^2+N^2}}{N} \tan(\Omega t)\right),
\eeq
\beq
	\phi_c(t) = \phi_c(0) + \frac{1}{2}\left(2E - 3 N\right) t -2 \arctan\left(\frac{\sqrt{4 E^2+N^2}}{2 E+N} \tan(\Omega t)\right),
\eeq
\beq
	\phi_p(t) = \phi_p(0) + (2 E -N) t + \arctan\left(\frac{\sqrt{4 E^2+N^2}}{N-2 E} \tan(\Omega t)\right)-\arctan\left(\frac{\sqrt{4
				E^2+N^2}}{2 E+N} \tan(\Omega t)\right).
\eeq
\end{widetext}
The case $N=2E$ is the two-mode initial data provided in the main text.

We conclude by providing an initial condition that  undergoes condensation for any non-vanishing values of the quantities $E$ and $N$:
	\beqnn
	b(0) = \sqrt{\frac{N^3}{N^2 + 4 E^2}}, \quad p(0) = \frac{4 E^2 - N^2}{2 \left(N^2 + 4 E^2\right)},
	\eeqnn
	\beq
		c(0) = \frac{E (N + 2 E)^2}{N^2 + 4 E^2} \sqrt{\frac{N}{N^2 + 4 E^2}}.
	\eeq
	These expressions guarantee that the conserved quantities $N$ and $E$ have the values there inserted, the condensation condition $S=2N$ in (\ref{eq:condition_condensation_S_2N}) is satisfied, and $x(0) < 1/4$ for any $N$ and $E$ greater than zero. Then, their evolution leads to the formation of condensates in finite time. We obtained these expressions by first fixing $x(0)$ to the lowest root of $\Veff(x)$ in (\ref{eq:Veff_condensation}). We then used that value to write $|b(0)|^2$ and $|c(0)|^2$ in terms of $N$ and $E$ from (\ref{eq:b_c_in_terms_of_x_N_E}). After, we required that these expressions satisfied the condensation condition in (\ref{eq:condensation_condition_c}), resulting in the values $\lambda = 0$ and $\pi$. Finally, we obtained $b(0)$, and $p(0)$, from $|b(0)|^2$ and $x(0)$, and after $c(0)$ from (\ref{eq:condensation_condition_c}).


\section{Growth of Sobolev norms}
\label{apx:Sobolev_Norms}

We here compute the growth of Sobolev norms presented in Sec.~\ref{sec:Dual cascade behavior}. Recall they have the form
\beq
H^{\xi} = \left(\sum_{n=0}^{\infty} (n+1)^{2\xi} |\alpha_n|^2\right)^{1/2}.
\eeq
Once we introduce the ansatz for $\alpha_n$ given in (\ref{eq:invariant_manifold}) they take the form
\beq
H^{\xi} = \left(|b|^2 + \frac{|c|^2}{x} \sum_{n=1}^{\infty} (n+1)^{2\xi} f_n^2 x^n\right)^{1/2}.
\eeq
In order to study their behavior close to the formation of the condensate ($x\sim 1/4$ and $t\to T$), we use the leading behavior of the following functions
\beq
|c|^2 = \frac{E}{F'(x)} \underset{x\sim 1/4}{\sim} \frac{E}{4} \sqrt{1 - 4 x}, \ \text{and} \	 	f^2_{n\gg1} \sim \frac{n^{-3/2}}{\sqrt{\pi}} 4^n,
\label{eq:dominat_terms}
\eeq
which comes from the expressions in Appendix~\ref{apx:Catalan}. The leading term is
\beq
H^{\xi}  \underset{x\sim 1/4}{\sim} \left(\frac{E}{\sqrt{\pi}} \sqrt{1 - 4 x} \sum_{n=1}^{\infty} n^{2\xi-\frac{3}{2}} (4x)^n\right)^{1/2}.
\eeq
Using  then the expression for $x(t)$ in (\ref{eq:x_solution}), and the leading contribution of the series \cite{Polylogarithms}
\beq
\sum_{n=1}^{\infty} n^{a} z^n \underset{z\sim 1^{-}}{\sim} \frac{\Gamma(1+a)}{(1-z)^{1+a}} \quad \text{for } a>-1,
\label{eq:Li_asymptotics}
\eeq
one obtains the leading contribution
\beq
H^{\xi>1/2}\underset{x\sim 1/4}{\sim} C_x (1-4x)^{\frac{1-2\xi}{2}} \underset{t\sim T}{\sim}  C_t (T-t)^{2(1-2\xi)},
\label{eq:Sobolev_Norms}
\eeq
where $C_x$ and $C_t$ are constants that depend on $E$ and $N$. Then, Sobolev norms with $\xi>1/2$ blow up at time $T$. The norm $H^{1/2} = \sqrt{N+E}$ remains finite due to the conservation of $E$ and $N$.  


\section{Condensation in position space}
\label{apx:Small-Scale_Structure}

We here present the calculation that led to the formation of a small-scale structure in Sec.~\ref{sec:Formation of small- and large-scale structures}. We first use the ansatz in (\ref{eq:invariant_manifold}) to write function $u(t,\theta) = \sum_{n=0}^{\infty} \alpha_n(t) e^{i n\theta}$ in the form
\begin{multline}
|u(t,\theta)|^2 = |b|^2 \\ + \frac{|c|^2}{x} \left(\sum_{n=1}^{\infty} f_n (p e^{i \theta})^n\right)\left(\sum_{n=1}^{\infty} f_n (\bar{p} e^{-i \theta})^n\right) \\ + \frac{\bar{b} c \bar{p}}{x} \left(\sum_{n=1}^{\infty} f_n (p e^{i \theta})^n\right) + \frac{b\bar{c}p}{x} \left(\sum_{n=1}^{\infty} f_n (\bar{p} e^{-i \theta})^n\right). \label{eq:expression_mod+u}
\end{multline}
To compute this expression as $t\to T$, we inspect the competition between the decay of $|c|$ to zero and the possible divergence in the summations. The only contributions will come from $|b|^2$ and diverging series that are compensated by $|c|$. To find the diverging series, we use the expansion $f_{n\gg1} \sim 2^n (n^{-3/4}/\pi^{1/4} - 9 n^{-7/4}/(16 \pi^{1/4}) + ...)$, observing that only the first term may produce a divergence. We then study a series of the form 
\beq
c \sum_{n=1}^{\infty} n^{-3/4} \left(2 \sqrt{x} e^{i (\theta + \phi_p)}\right)^n = 	c \ \text{Li}_{3/4}\left(2 \sqrt{x} e^{i(\theta+\phi_p)}\right)
\label{eq:Li_p_exp}
\eeq
where we have decomposed $p = \sqrt{x} e^{i\phi_p}$, and $\text{Li}_a(z)$ is the polylogarithm function \cite{Polylogarithms}. When $x\to1/4$ the polylogarithm function converges to $ \text{Li}_{3/4}\left(e^{i(\theta+\phi_p)}\right)$ which is finite except when the argument is $1$, that diverges, i.e., when $\theta = - \phi_p$. Due to the factor $c$, the expression in (\ref{eq:Li_p_exp}) and its complex conjugation vanish for all $\theta$ except at $\theta = -\phi_p$, leading to $|u(t\to T,\theta\neq-\phi_p)| \to \sqrt{N}$ in (\ref{eq:expression_mod+u}).

To analyze the point $\theta = -\phi_p$ in (\ref{eq:Li_p_exp}), we study the behavior of $c\ \text{Li}_{3/4}(2\sqrt{x})$. Close to $x=1/4$ we combine the property of the series in (\ref{eq:Li_asymptotics}) with the leading term of $|c|$ in (\ref{eq:dominat_terms}) to see that the second term in (\ref{eq:expression_mod+u}) is $\Gamma\left(1/4\right)^2 E \sqrt{2/\pi}  $,
while the third term vanishes. The latter observation required the conservation of $S$ in (\ref{eq:conserved_quantity_S}) to see that the phases of $b$, $c$, and $p$ satisfy $\cos(\phi_b+\phi_p-\phi_c) \to 0$ when $x\to 1/4$.
To sum up, we obtained that, when $t\to T$,  $|u(t,\theta)| \to \sqrt{N}$ for $\theta \in [0,2\pi) - \{-\phi_p\}$, and $|u(t,\theta = -\phi_p)|  \to \sqrt{ N + E\ \Gamma\left(1/4\right)^2 \sqrt{2/\pi}  }$. For the solution presented in Fig.~\ref{fig:DWC}, the phase of $p$ is $\phi_p \underset{t\to T}{\to} -\pi$, showing agreement between the visual representation and the calculation.


\section{Numerical  Simulations}
\label{apx:Numerical_simulations}

\begin{figure}[t!]
	\centering
	
	\includegraphics[width=0.88\columnwidth]{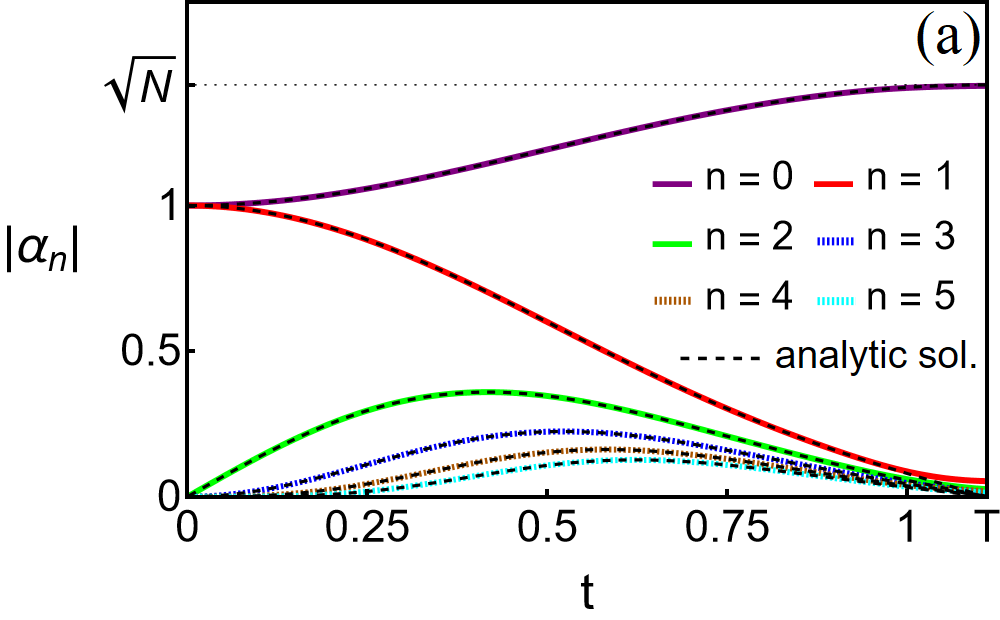}
	\includegraphics[width=0.90\columnwidth]{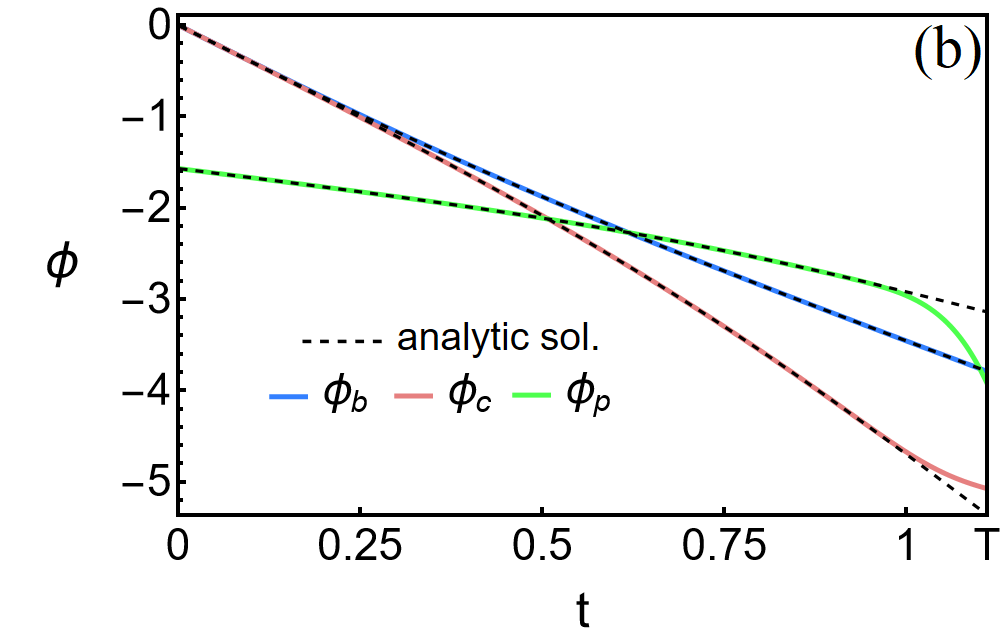}
	\includegraphics[width=0.98\columnwidth]{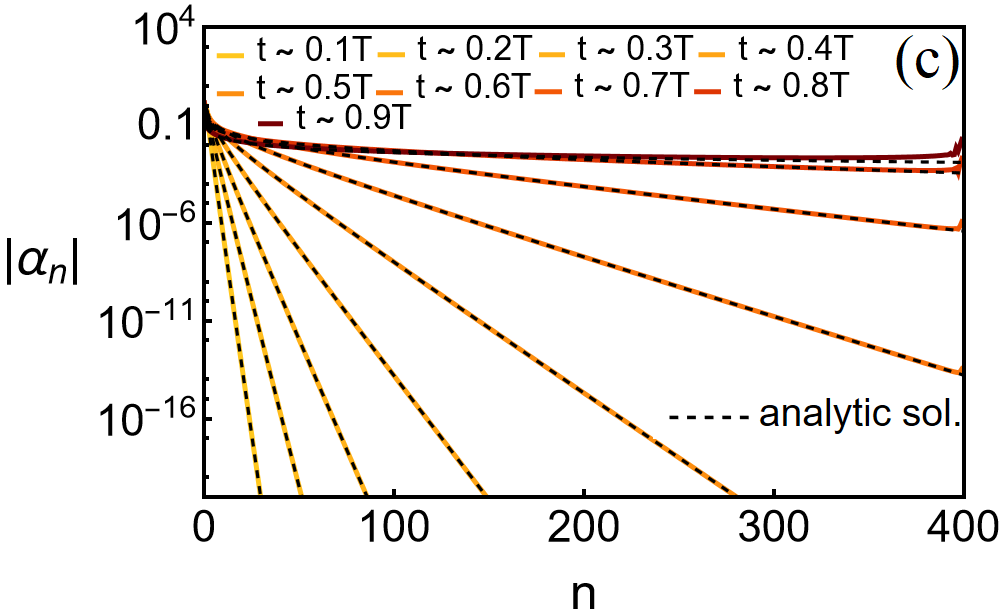}
	
	\caption{Coherent condensation process that concludes at time $T$. Comparison between our analytic solution presented in the main text (black dashed lines in each plot) and a numerical simulation of the same process using the Hamiltonian system (\ref{eq:Resonant_Equation}) truncated to $400$ modes (colored (gray) lines). (a) Time-evolution of the first modes, (b) time-evolution of the phases of $b$, $c$, and $p$, and (c) the amplitude spectrum $|\alpha_n|$ at different times. Spurious effects associated with the truncation in the number of modes are visible in (b) and in the zoomed-in sections provided in Fig.~\ref{fig:numerics_2}.}
	\label{fig:numerics}
\end{figure}

\begin{figure}[t!]
	\centering
	
	\includegraphics[width=0.88\columnwidth]{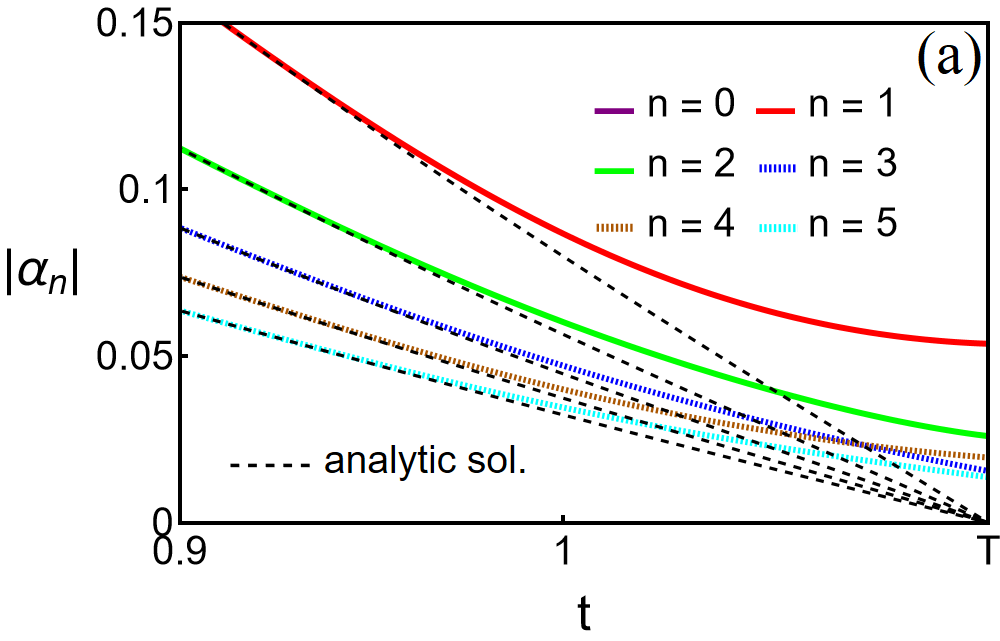}
	\includegraphics[width=0.90\columnwidth]{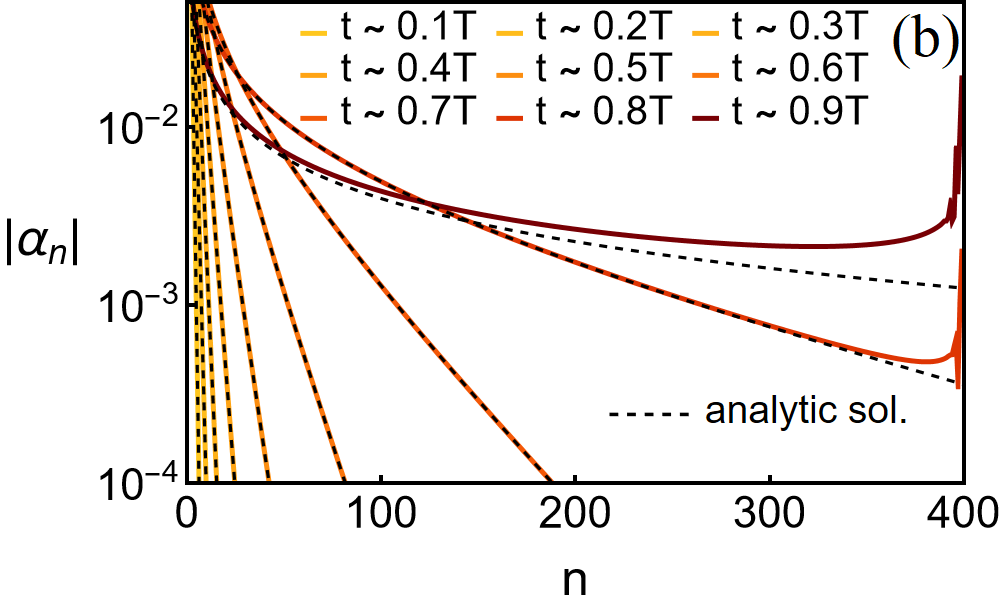}

	\caption{Zoom-in sections for Fig.~\ref{fig:numerics} (a) and (c) where deviations  from the exact solution (black dashed lines)  are  appreciated.}
	\label{fig:numerics_2}
\end{figure}

{\bf Comparison between an analytic solution and numerical simulations:} We here illustrate the difficulties in the numerical description of coherent condensation. To do so, we compare an analytic solution and a numerical simulation of the Hamiltonian system (\ref{eq:Resonant_Equation}) with the same initial condition. As Fig.~\ref{fig:numerics} and Fig.~\ref{fig:numerics_2} show, there is excellent agreement between them, excluding times close to the formation of the condensate ($t\sim T$). Deviations from the analytic prediction are associated with the necessary truncation in the number of modes to simulate infinite-dimensional Hamiltonian systems like ours. Spurious effects associated with this truncation arise as the asymptotic part of the spectrum develops the power-law $|\alpha_n(t\sim T)|^2\sim n^{-3/2}$ (\ref{eq:power-law}). It happens because contributions from the omitted modes become more relevant as the exponential decay of the spectrum weakens. Consequently, truncating the number of modes has a dramatic effect near $T$, causing serious difficulties for simulations in describing the final part of the evolution. A larger  number of modes should provide better resolution of the dynamics close to $T$; however, the computational cost is high and the improvement rather small. This is because the numerical problem (the truncated Hamiltonian system) scales cubically with the number of modes and $|\alpha_n|$ decays slowly with $n$ when time is close to $T$ (modes significantly far from each other are of practically the same order $|\alpha_{10 n}|/|\alpha_n| \sim 0.18$).

\begin{figure}[t!]
	\centering
	
	\includegraphics[width=0.95\columnwidth]{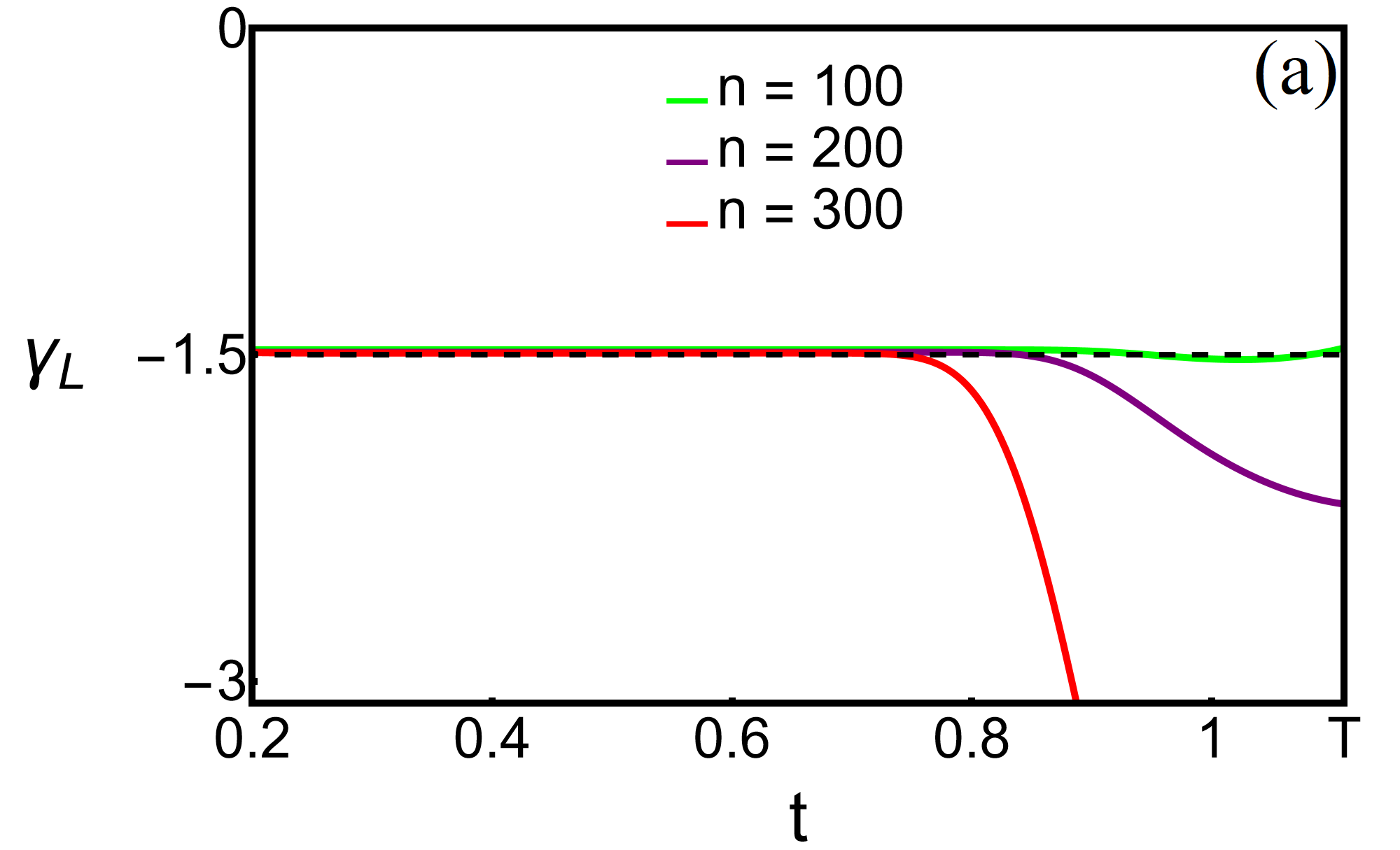}
	\includegraphics[width=0.95\columnwidth]{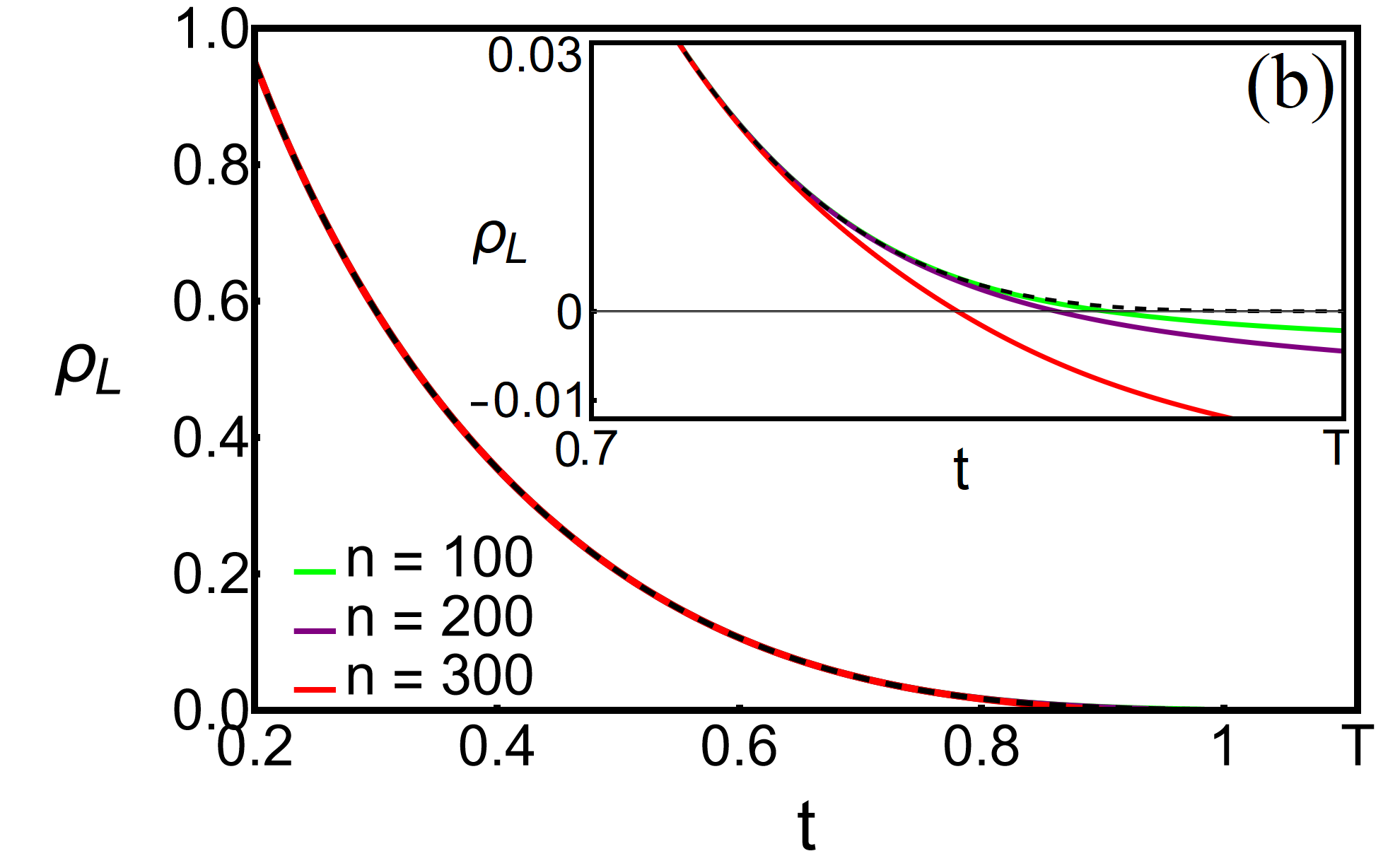}
	\includegraphics[width=0.95\columnwidth]{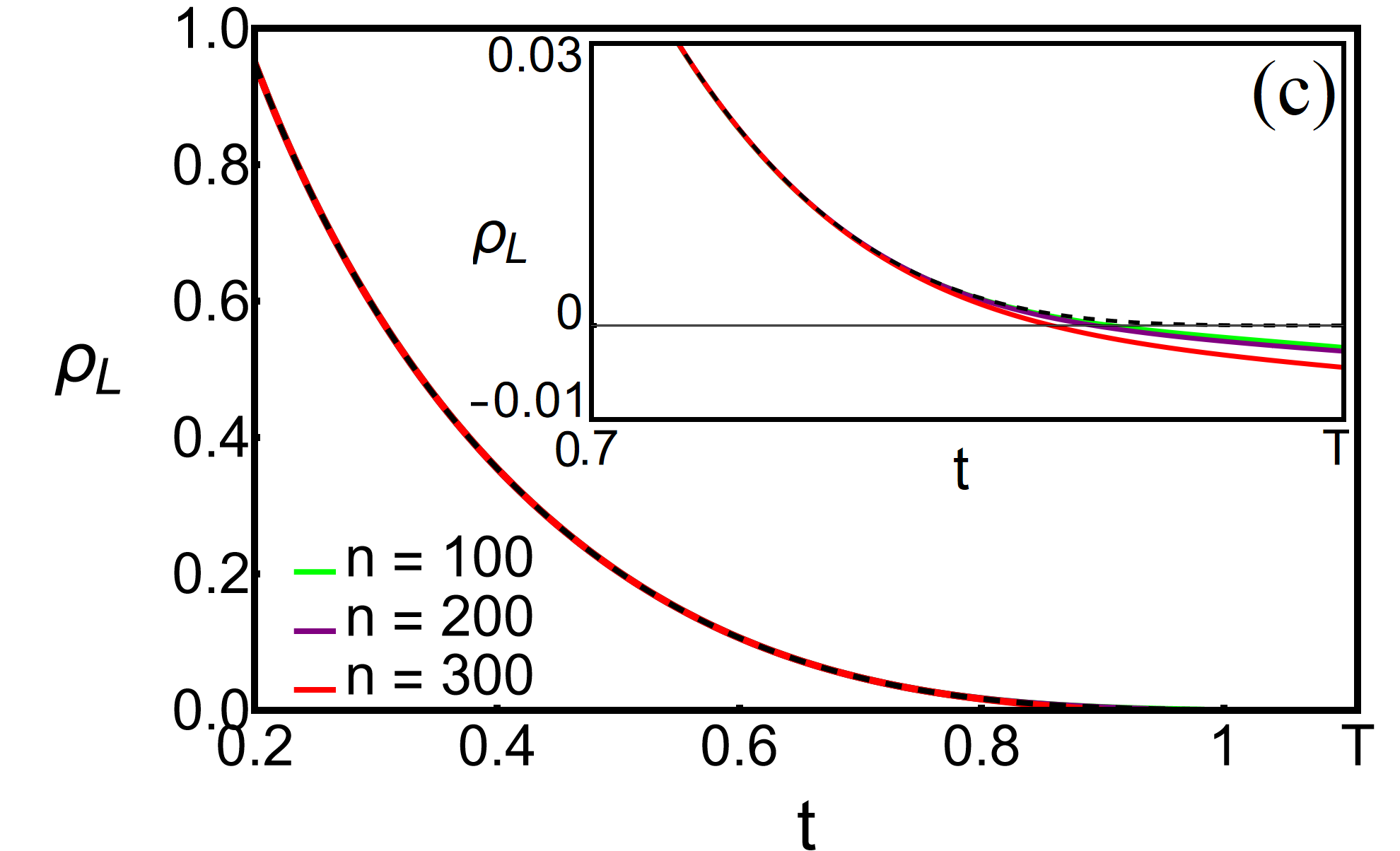}
	
	\caption{Time evolution of the local exponents $\gamma_L$ and $\rho_L$  (\ref{eq:local_exponents_gamma})-(\ref{eq:local_exponents_rho}) calculated at fixed $n$ for the simulation presented in Fig.~\ref{fig:numerics}. Black dashed lines represent the values coming from the  exact solution (e.g., $\gamma = -3/2$ and $\rho(t\to T)\to 0$). In (c) $\rho_L$ has been calculated fixing $\gamma_L=-3/2$.}
	\label{fig:analyticity_strip_method_analytic_solution}
\end{figure}
\begin{figure}[t!]
	\centering

	\includegraphics[width=0.99\columnwidth]{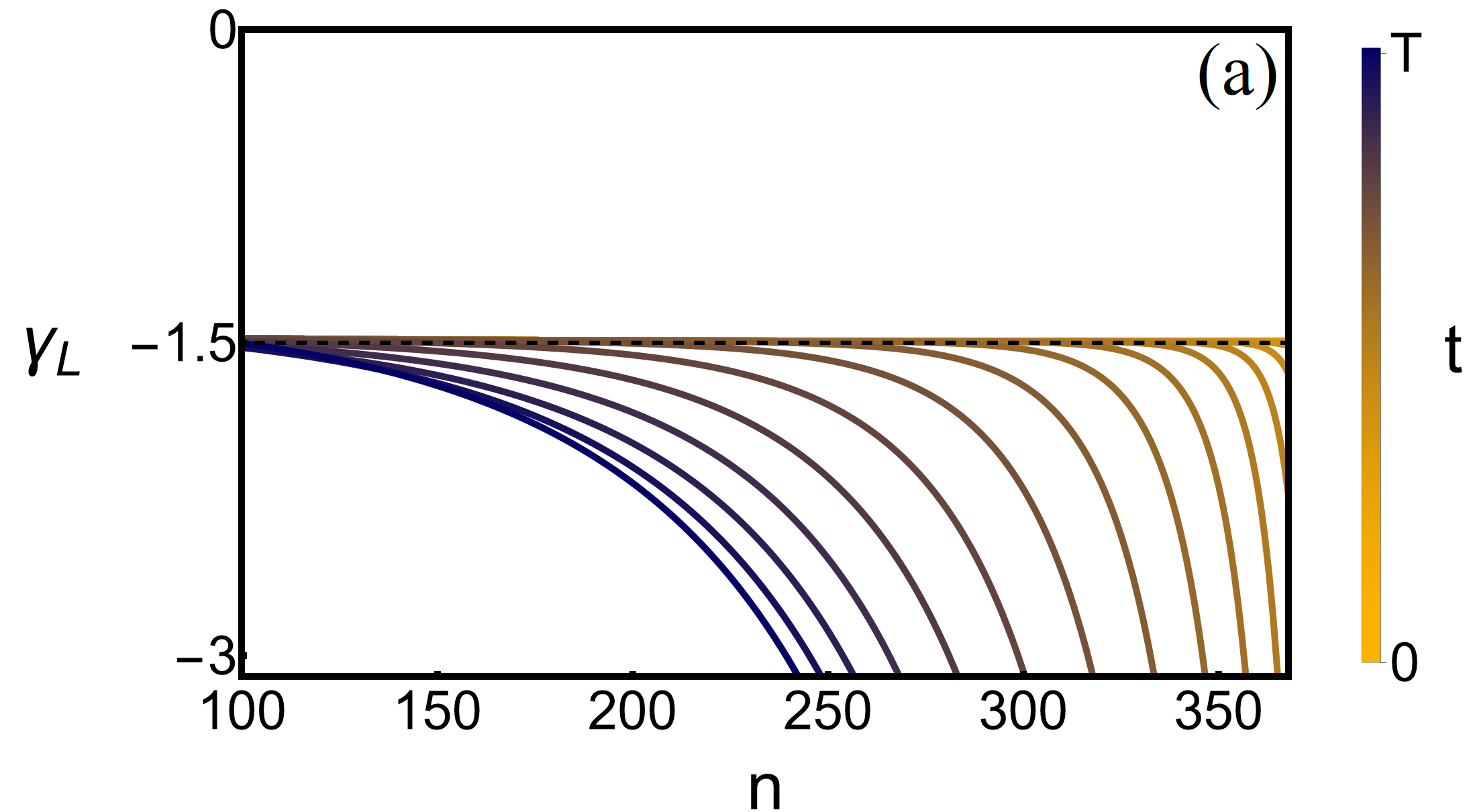}
	\includegraphics[width=0.99\columnwidth]{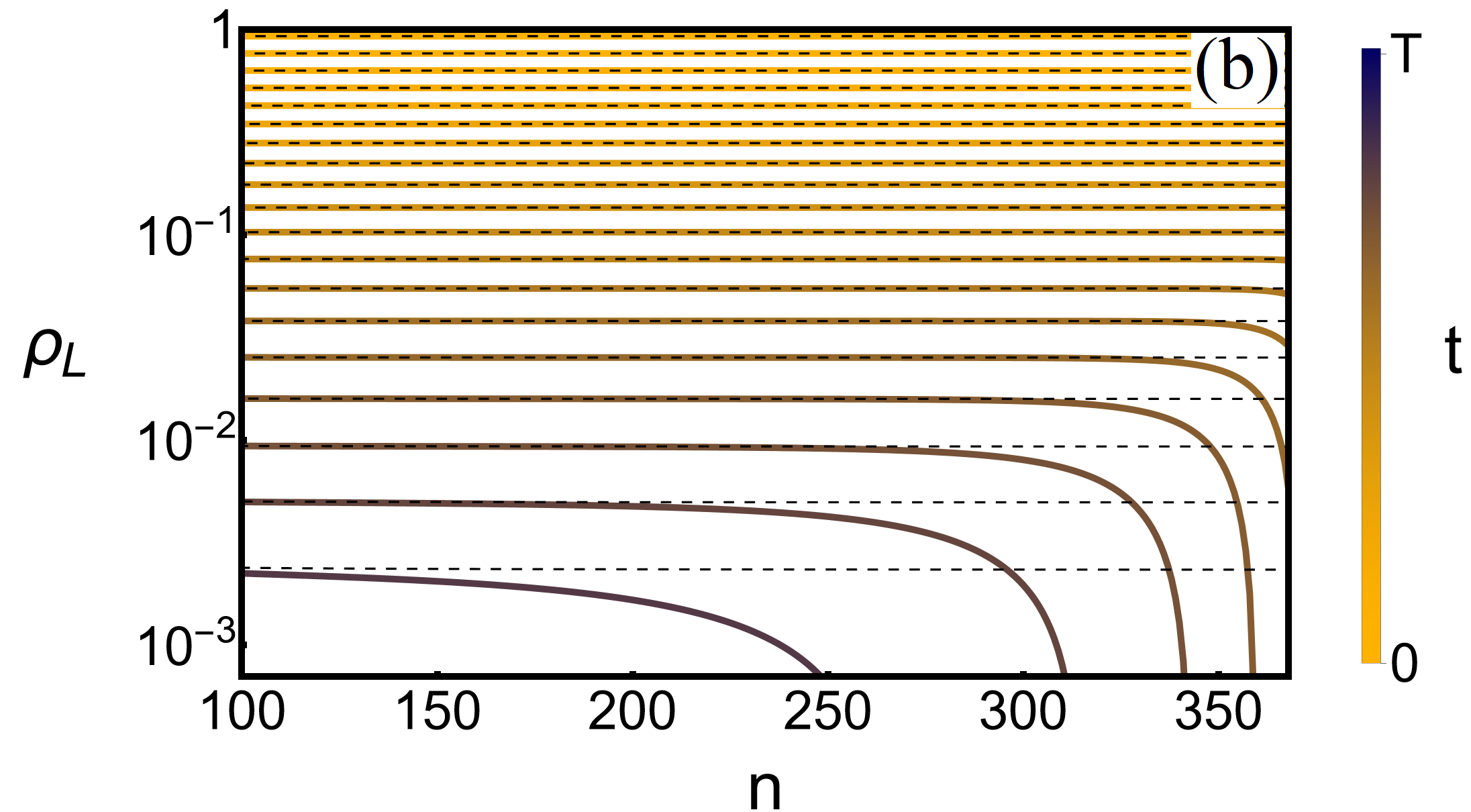} 
	\includegraphics[width=0.99\columnwidth]{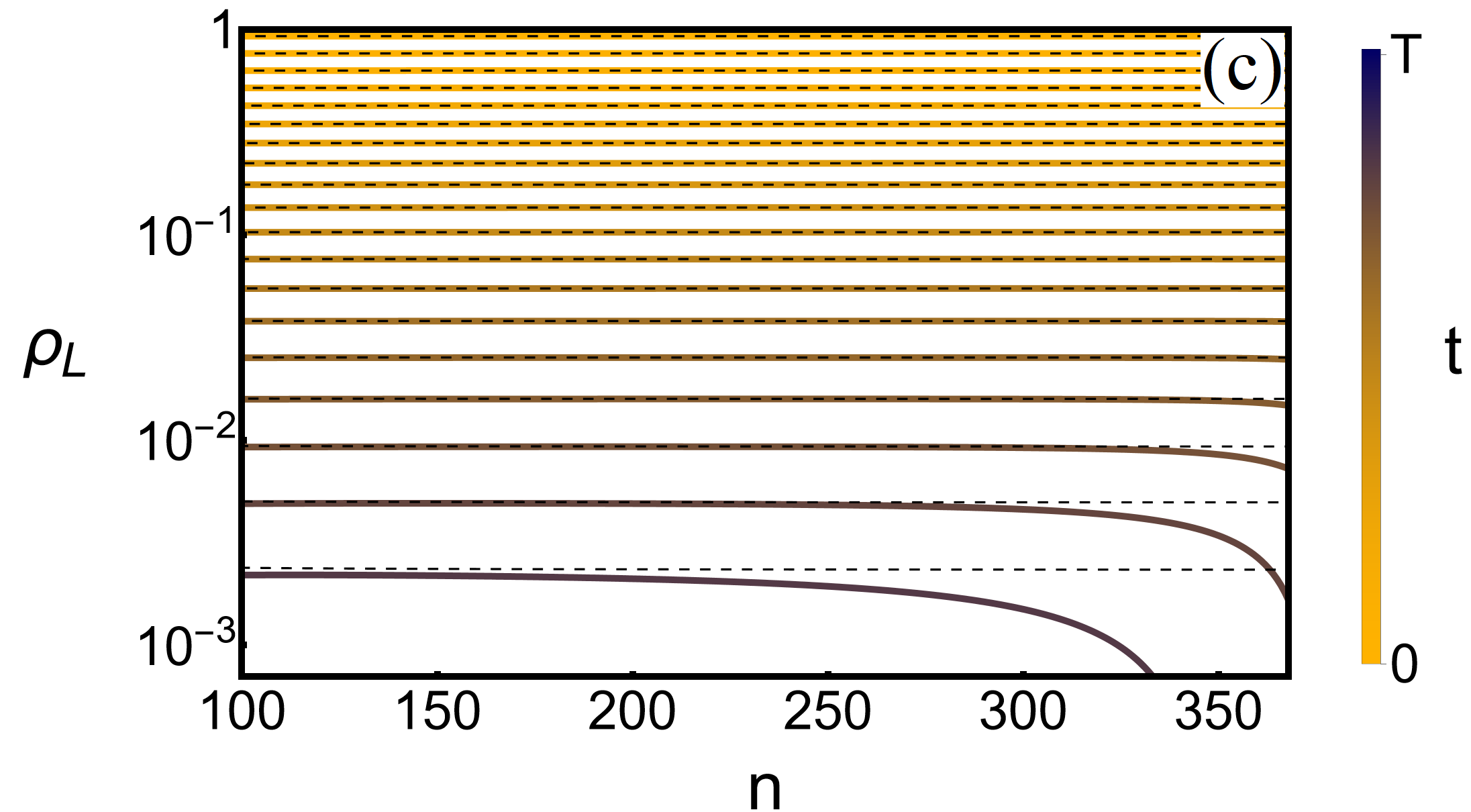}
	\caption{$\gamma_L$ and $\rho_L$ at different times as a function of the mode number. In (c), $\rho_L$ has been calculated fixing $\gamma_L=-3/2$. Black dashed lines represent the values coming from the  exact solution (e.g., $\gamma = -3/2$).}
	\label{fig:analyticity_strip_method_analytic_solution_v2}
\end{figure}

{\bf The analyticity strip method:} We here present the {\em analyticity strip method} \cite{Sulem}, which is used to characterize direct cascades in Hamiltonian systems (\ref{eq:Resonant_Equation}) from numerical simulations \cite{Jalmuzna,BMR}. We used this method to study the behavior of initial conditions outside the ansatz for $\alpha_n(t)$ in (\ref{eq:invariant_manifold}). Its basic idea is to assume the following asymptotic form for the spectrum
	\beq
	|\alpha_{n\gg1}(t)| \sim n^{\gamma(t)/2} e^{-\rho(t)n},
	\label{eq:analyticity_strip_method}
	\eeq
	and extract the behavior of the parameters $\gamma(t)$ and $\rho(t)$ from the numerical data. If $\rho(t)$ becomes zero, then it indicates the formation of a power-law. We calculated these parameters locally by using \cite{Shelley}
	\beq
	\gamma_{L}(t) = 2\frac{\log\left(\frac{|\alpha_{n-1}||\alpha_{n+1}|}{|\alpha_n|^2}\right)}{\log\left(\frac{(n-1)(n+1)}{n^2}\right)},
	\label{eq:local_exponents_gamma}
	\eeq
	\beq
		 \rho_L(t) =  -\frac{1}{n}\left( \log\left(\frac{|\alpha_n|}{|\alpha_{n-1}|}\right) - \frac{\gamma_L(t)}{2} \log\left(\frac{n}{n-1}\right)\right).
	\label{eq:local_exponents_rho}
	\eeq
	(a fit is more appropriate when the spectrum presents oscillations). In these expressions, $\gamma_L(t)$ and $\rho_L(t)$ depend on the mode number $n$. The advantage is to have multiple observations of two parameters that should be practically constant for large enough $n$,  allowing us to identify spurious effects associated with the truncation in the spectrum. In order to gain intuition and understanding, we  applied this procedure to the simulation presented above in Fig.~\ref{fig:numerics}, which underwent condensation and we had the analytic solution to compare. Fig.~\ref{fig:analyticity_strip_method_analytic_solution} and Fig.~\ref{fig:analyticity_strip_method_analytic_solution_v2} summarize the results, providing valuable observations. First, the method captures the analytic prediction $\gamma = -3/2$ except close to $T$, it can accurately reach an exponent smaller than $\rho = 0.01$, and hints at $\rho(t)$ becoming zero. Second, deviations from the exact values for the parameters $\gamma$ and $\rho$ originate at the truncation of modes and propagate toward low modes. Third, $\rho_L$ is more robust than $\gamma_L$ since its deviation from the exact value happens at later times. Finally, the robustness of $\rho_L(t)$ is  further improved by replacing in its expression the parameter $\gamma_L(t)$ by $-3/2$ (the value before spurious effects arose).
	
\begin{figure}[t!]
	\centering
	
	\includegraphics[width=0.9\columnwidth]{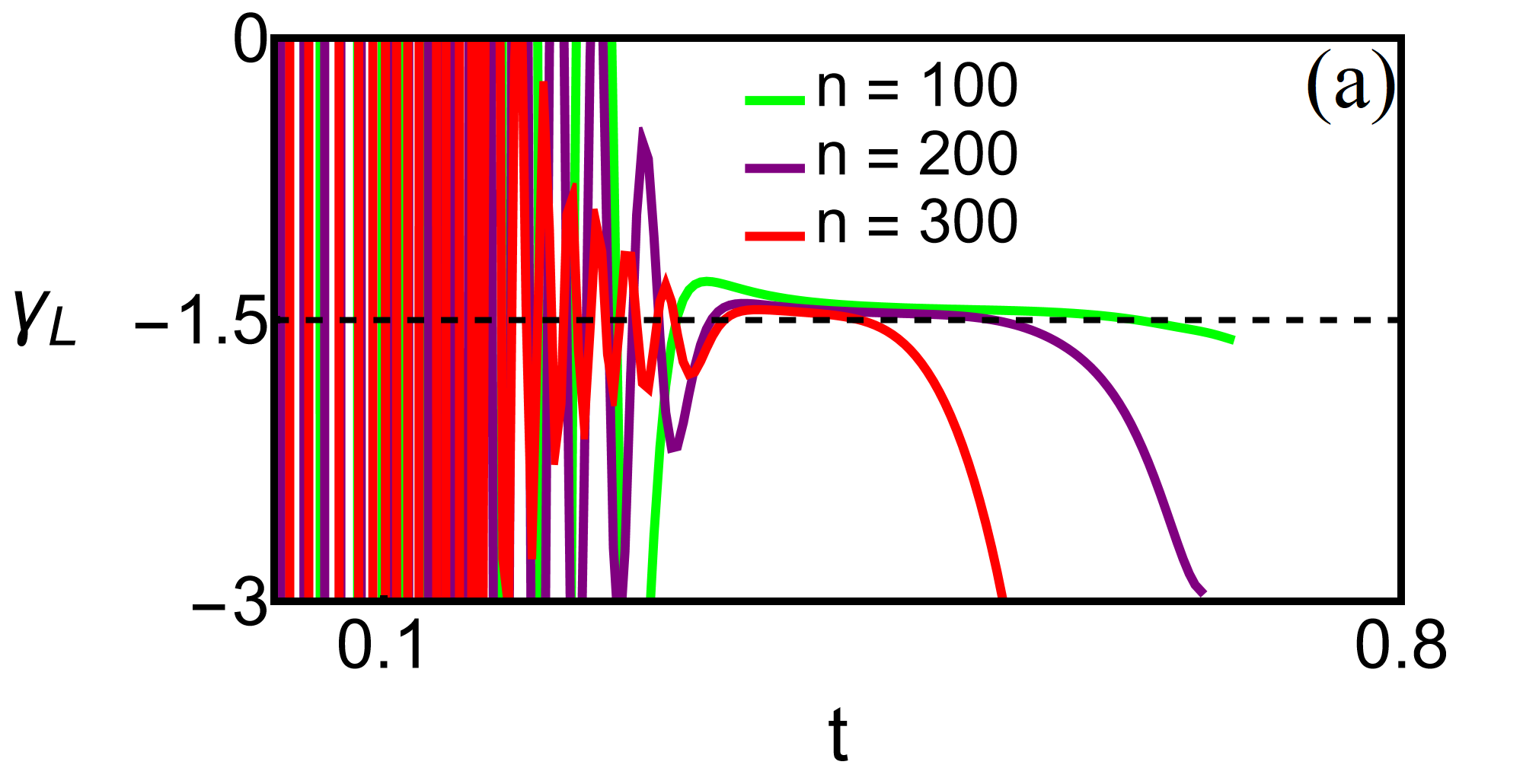}
	
	\includegraphics[width=0.9\columnwidth]{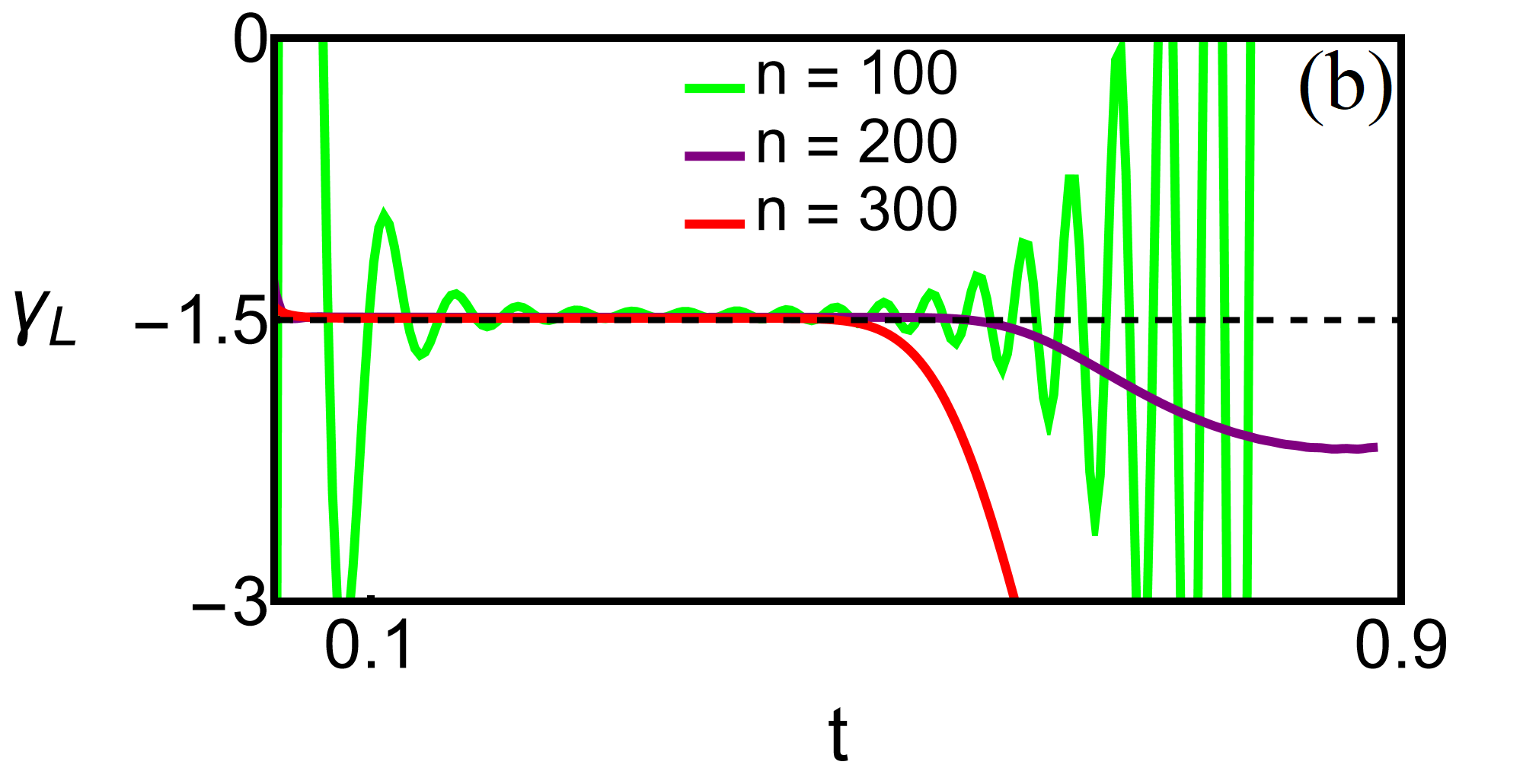}
	
	\includegraphics[width=0.9\columnwidth]{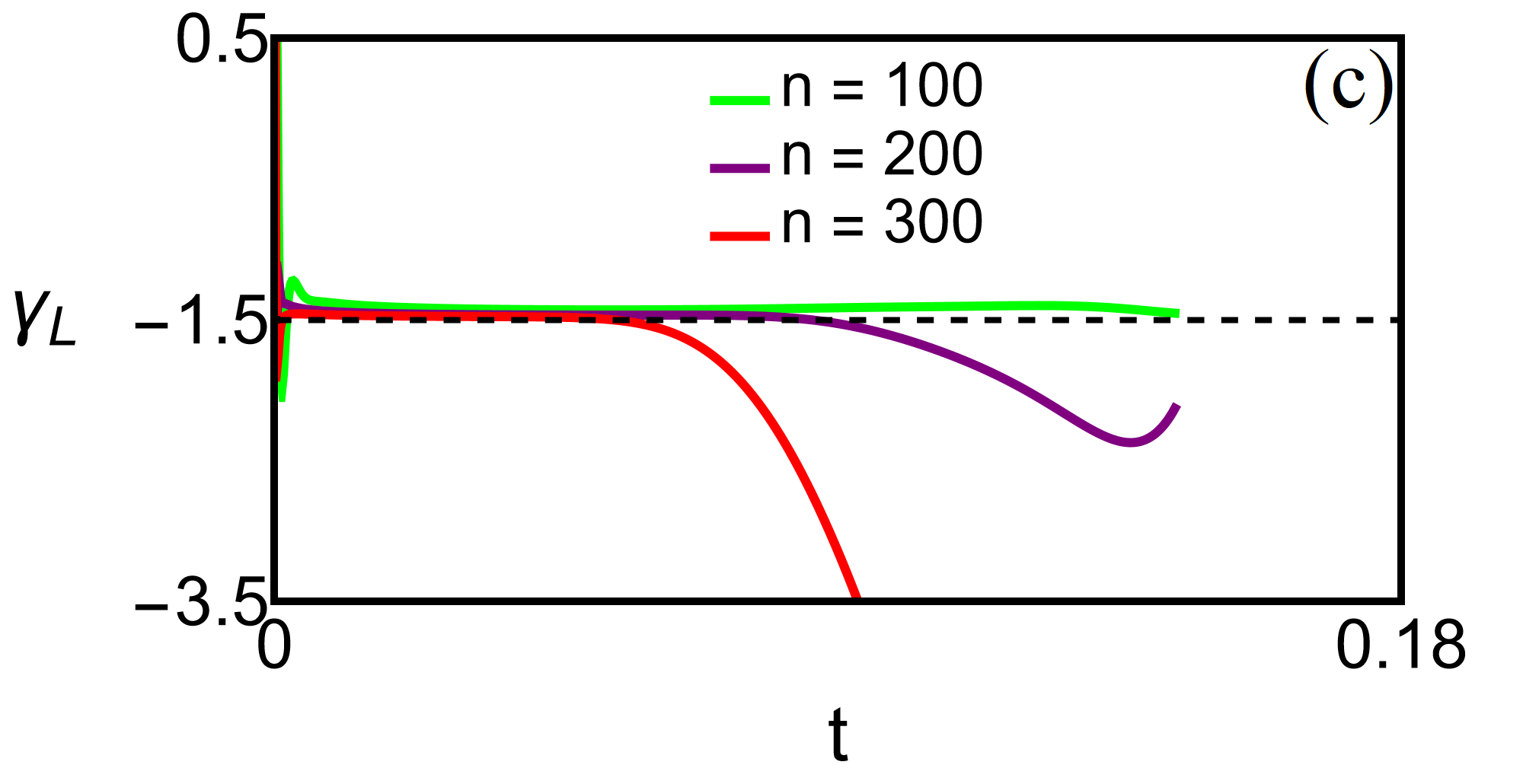}
	
	\includegraphics[width=0.9\columnwidth]{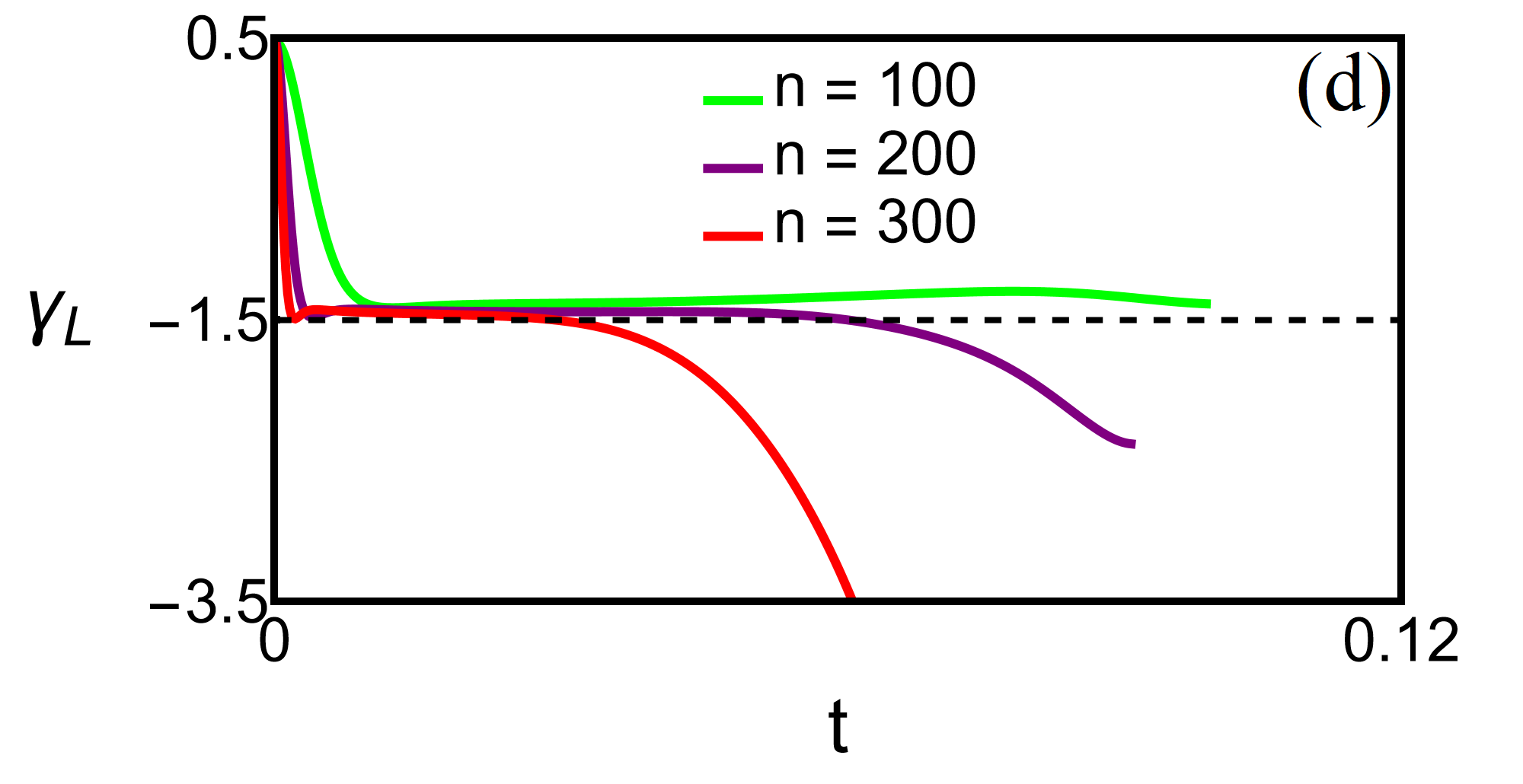}
	
	\caption{Local exponent $\gamma_L$ ($|\alpha_{n\gg1}|^2\sim n^{\gamma_L}e^{-2\rho_L n}$) calculated at different $n$ for the simulations presented in Fig.~\ref{fig:Simulation_2} and Fig.~\ref{fig:robustness_main_text}a-c, from (a)  to (d). Oscillations observed in (a) and (b) are associated with fluctuations in the spectra during those times.}
	\label{fig:robustness}
\end{figure}

\begin{figure}[t!]
	\centering
	
	\includegraphics[width=0.9\columnwidth]{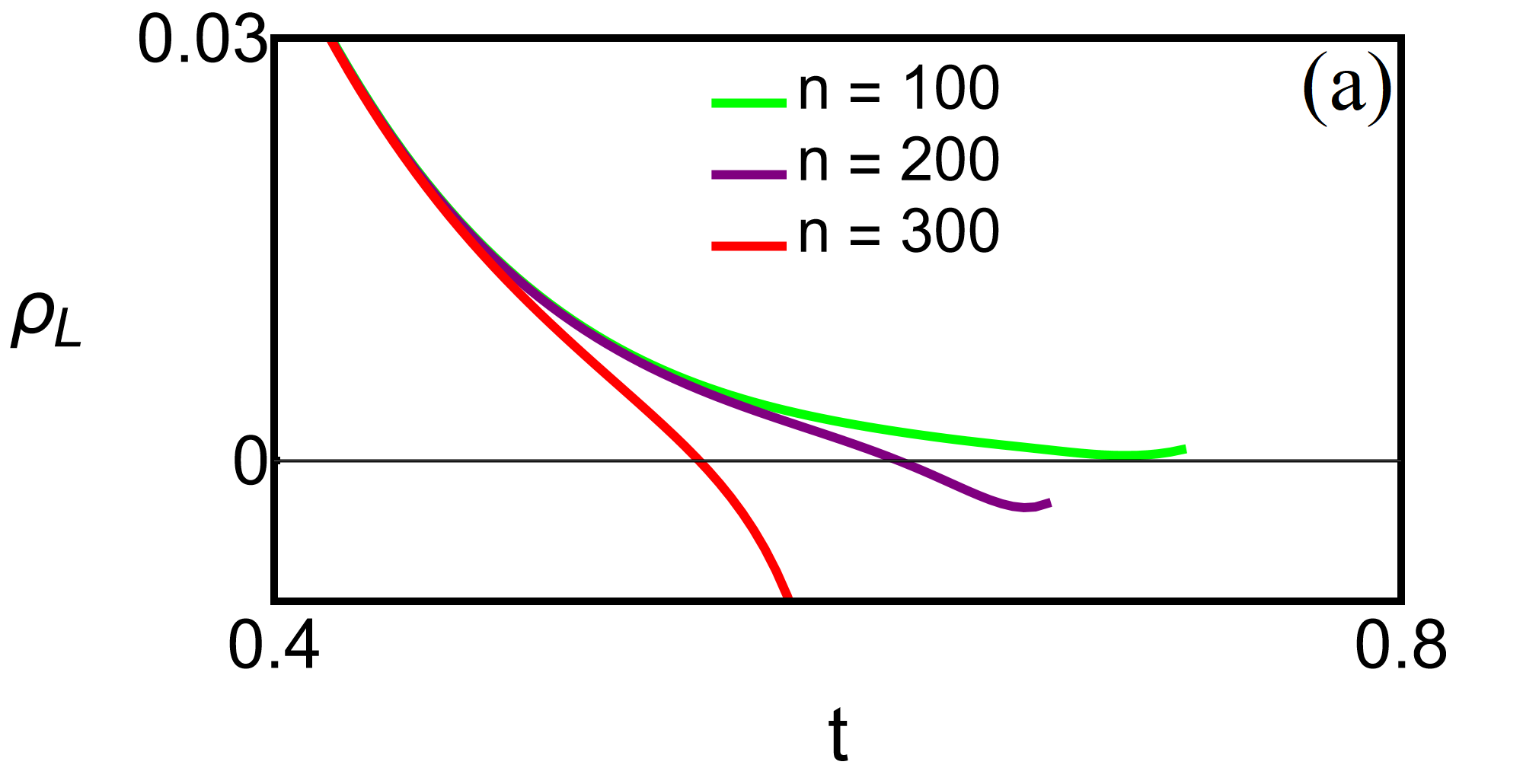}
	
	\includegraphics[width=0.9\columnwidth]{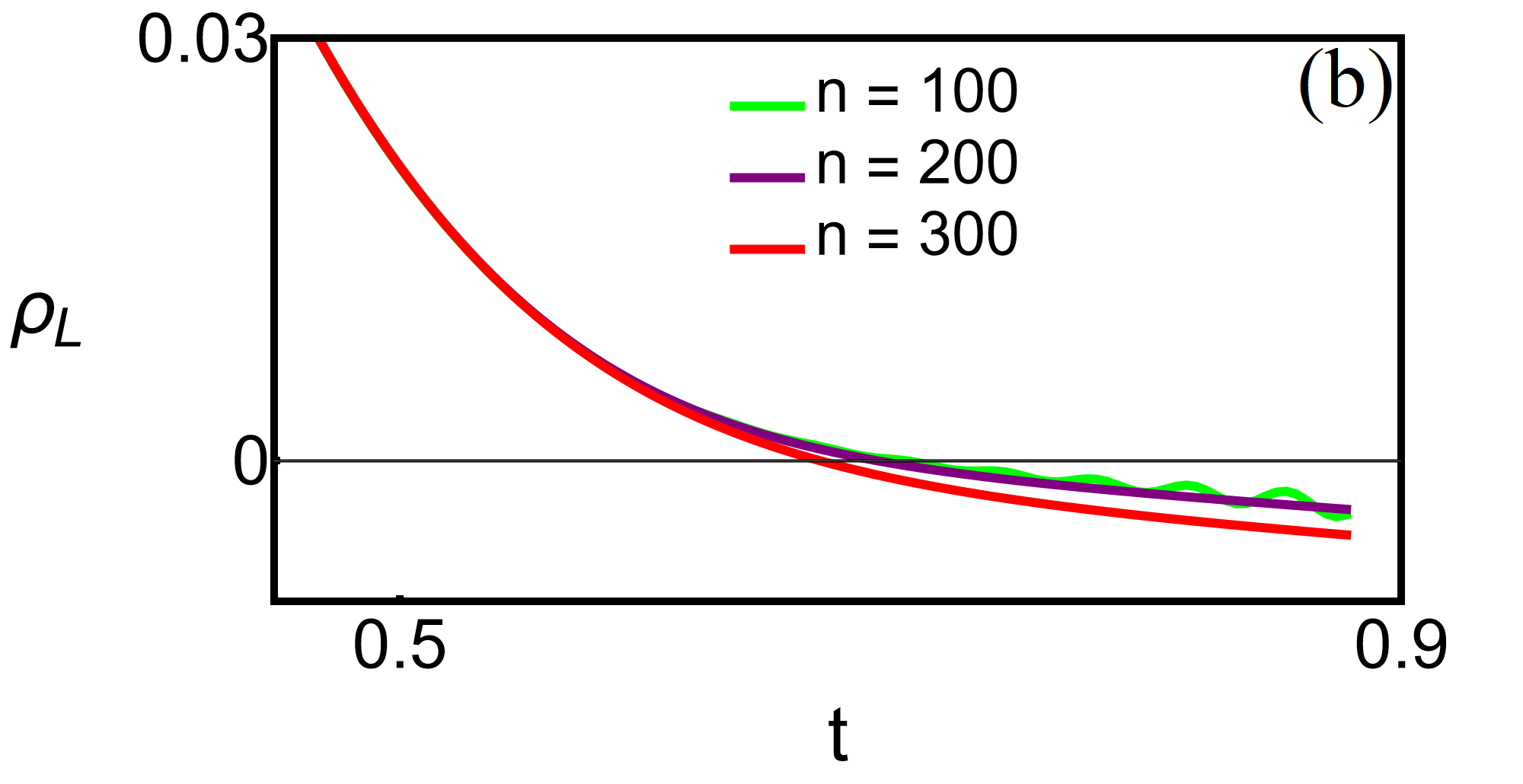}
	
	\includegraphics[width=0.9\columnwidth]{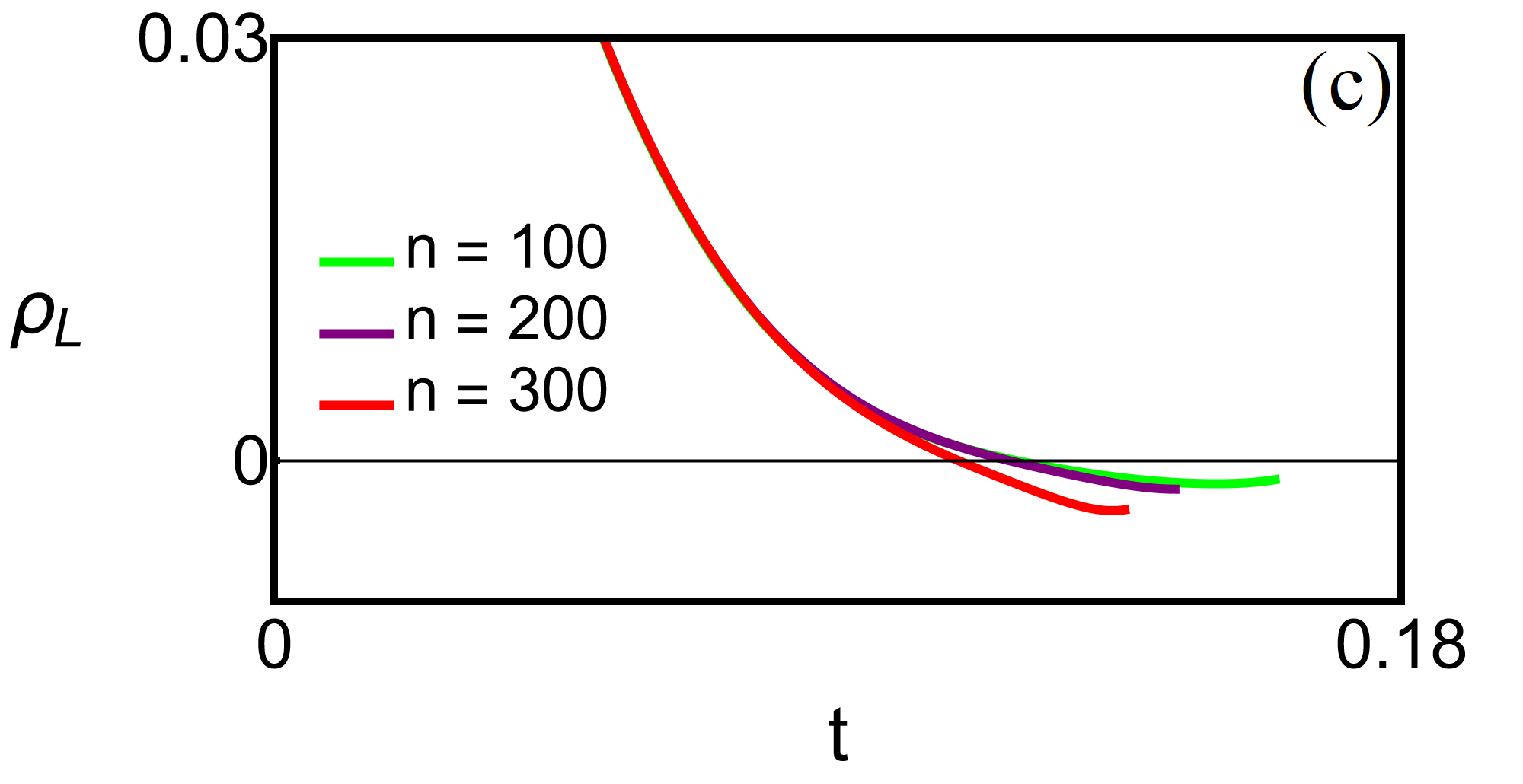}
	
	\includegraphics[width=0.9\columnwidth]{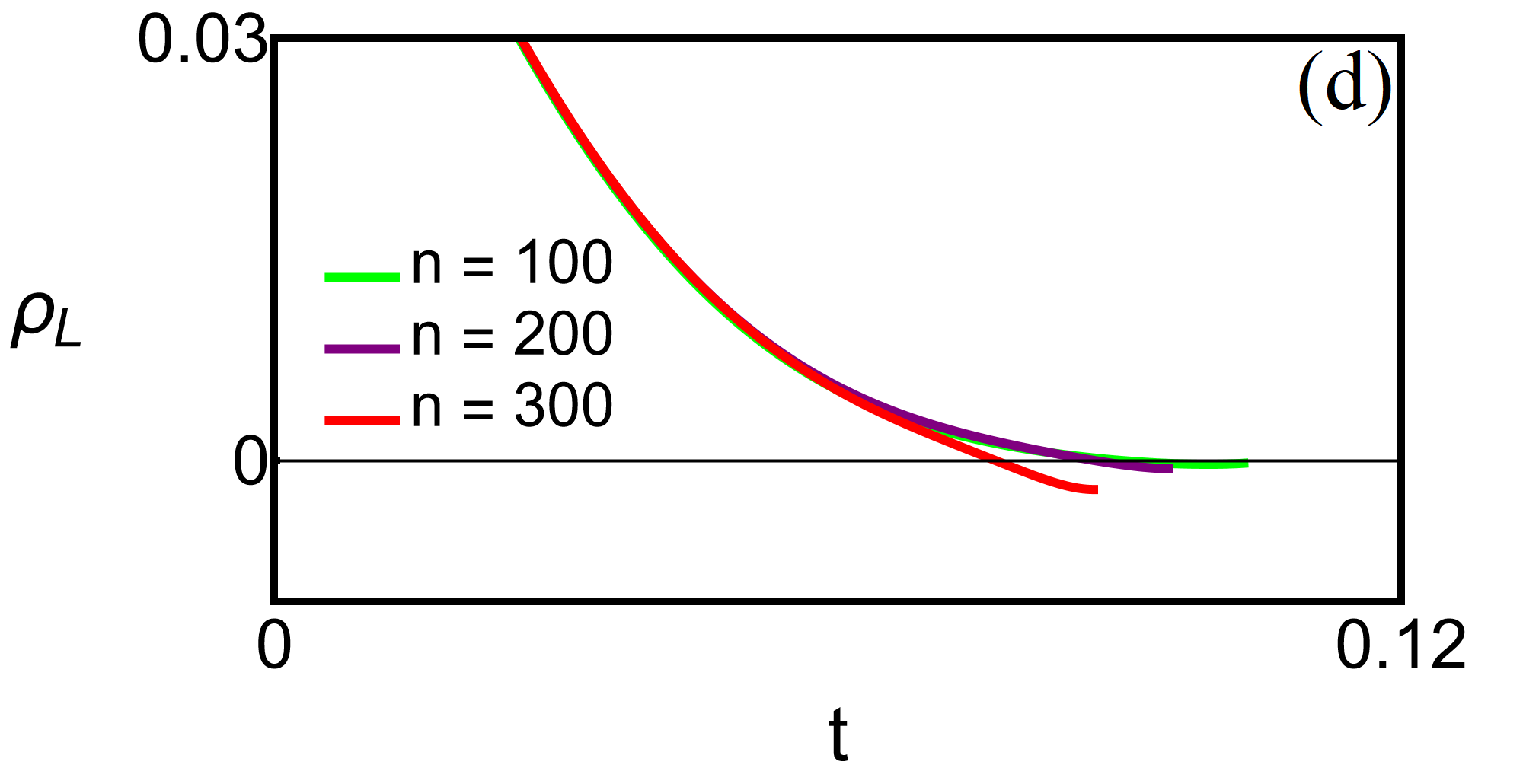}
	
	\caption{Local exponent $\rho_L$ ($|\alpha_{n\gg1}|^2\sim n^{\gamma_L}e^{-2\rho_L n}$) calculated at different $n$ for the simulations presented in Fig.~\ref{fig:Simulation_2} and Fig.~\ref{fig:robustness_main_text}a-c, from (a)  to (d). $\rho_L$ has been calculated using $\gamma_L= -3/2$ in (b)-(d).}
	\label{fig:robustness_2}
\end{figure}

	Fig.~\ref{fig:robustness} and Fig.~\ref{fig:robustness_2} provide the local exponents for the simulations presented in Fig.~\ref{fig:Simulation_2} and Fig.~\ref{fig:robustness_main_text}. We observe that $\rho_L$ deceases and $\gamma_L$ remains close to $-3/2$, indicating the approach  to the power-law $|\alpha_{n\gg 1}|^2 \sim n^{-3/2}$.

{\bf Initial conditions in the figures.} Fig.~\ref{fig:Simulation_2}:  $\alpha_0 \approx 0.6568,\ \alpha_1 \approx -0.6047 + 0.9099 i,\ \alpha_2 \approx -0.4481 + 0.223 i,\ \alpha_3 \approx -0.2922 - 
	0.1266 i,\ \alpha_4 \approx -0.0236 - 0.1007 i,\ \alpha_5 \approx -0.0465 - 0.0724 i,\ \alpha_6 \approx 0.0134 - 
	0.0594 i,\ \alpha_7 \approx -0.0138 + 0.0227 i,\ \alpha_8 \approx -0.0145 - 0.012 i,\ \alpha_9 \approx -0.0096 - 0.0035 i$, etc.  Fig.~\ref{fig:robustness_main_text}a: $\alpha_0  = 0.78-0.58 i$, $\alpha_1 = 0.53+0.33 i$, $\alpha_2 = -0.77-0.16 i$, $\alpha_3  = -0.20-0.05i$, $\alpha_4 = 0.002 + 0.003 i$, $\alpha_5  = 0.028 - 0.06 i$, $\alpha_{n\geq 6}=0$. The conditions in Fig.~\ref{fig:robustness_main_text}b-c have been provided in the caption. Fig.~\ref{fig:Simulation_3}:  $\alpha_0 \approx 0.8855,\ \alpha_1 \approx 0.1564 + 0.1307  i,\ \alpha_2 \approx -0.5014 + 0.0499  i,\ \alpha_3 \approx -0.0539 + 0.1047 i,\ 
	 \alpha_4 \approx -0.0023 + 0.0128 i,\
	 \alpha_5 \approx -0.1221 - 0.1279 i,\ 
	 \alpha_6 \approx -0.0009 + 0.0005 i,\ 
	 \alpha_7 \approx 0.4717 + 0.0243 i,\ 
	 \alpha_8 \approx -0.438 - 0.0043 i,\ 
	 \alpha_9 \approx -0.2637 + 0.195 i$, etc.   All conditions had $N=2$ without loss of generality due to the scaling symmetry at the end of Appendix~\ref{apx:Coherent_Regime}.



\begin{thebibliography}{999}
	
	\bibitem{Condensation_2001}
	A. C. Newell, S. Nazarenko, and L. Biven,{\em Wave turbulence and intermittency}, Physica D: Nonlinear Phenomena, {\bf 152}, 520-550 (2001).
	
	\bibitem{Condensation_2004} C. Connaughton, and Y. Pomeau, {\em Kinetic theory and Bose–Einstein condensation}, Comptes Rendus Physique, {\bf 5}, 91-106 (2004).
	
	\bibitem{Condensation_2005_2} V. E. Zakharov, and S. V. Nazarenko, {\em Dynamics of the Bose–Einstein condensation}, Physica D: nonlinear phenomena, {\bf 201}, 203-211  (2005).
		
	\bibitem{Condensation_2005} C. Connaughton, C. Josserand, A. Picozzi, Y. Pomeau, and S. Rica,  {\em Condensation of classical nonlinear waves}, Physical Review Letters, {\bf 95}, 263901 (2005). \arXiv{cond-mat/0502499} [cond-mat.stat-mech].
	
	\bibitem{Condensation_2006} S. Nazarenko, and M. Onorato, {\em Wave turbulence and vortices in Bose–Einstein condensation}, Physica D: Nonlinear Phenomena, {\bf 219}, 1-12 (2006). \arXiv{nlin/0507051} [nlin.CD].
	
	\bibitem{Condensation_2011} P. Aschieri, J. Garnier, C. Michel, V. Doya, and A. Picozzi, {\em Condensation and thermalization of classsical optical waves in a waveguide}, Physical Review A, {\bf 83}, 033838 (2011).
	
	\bibitem{Condensation_2012} C. Sun, S. Jia, C. Barsi, S. Rica, A. Picozzi, and J. W. Fleischer, {\em Observation of the kinetic condensation of classical waves}, Nature Physics, {\bf 8}, 470-474 (2012).
	
	\bibitem{Condensation_2018} N. \v{S}anti\'c, A. Fusaro, S. Salem, J. Garnier, A. Picozzi, and R, Kaiser, {\em Nonequilibrium precondensation of classical waves in two dimensions propagating through atomic vapors}, Physical Review Letters, {\bf 120}, 055301 (2018).
	
	\bibitem{Condensation_2019} A. Fusaro, J. Garnier, K. Krupa, G. Millot, and A. Picozzi, {\em Dramatic acceleration of wave condensation mediated by disorder in multimode fibers}, Physical Review Letters, {\bf 122}, 123902 (2019). \arXiv{2011.05111} [physics.optics].
	
	\bibitem{Condensation_2020} K. Baudin, A. Fusaro, K. Krupa, J. Garnier, S. Rica, G. Millot, and A. Picozzi, {\em Classical Rayleigh-Jeans condensation of light waves: Observation and thermodynamic characterization}, Physical Review Letters, {\bf 125}, 244101 (2020). \arXiv{2007.11950} [physics.optics].
	
	\bibitem{Escobedo} M. Escobedo, and J. J. Vel\'azquez, {\em On the theory of weak turbulence for the nonlinear Schro\"odinger equation}, American Mathematical Soc. (2015). \arXiv{1305.5746} [math-ph].
	
	\bibitem{Velazques} A. H. M. Kierkels, and J. J. L. Vel\'azquez, {\em On the transfer of energy towards infinity in the theory of weak turbulence for the nonlinear Schr\"odinger equation}, Journal of Statistical Physics, {\bf 159}, 668-712  (2015). \arXiv{arXiv:1410.2073} [math-ph].
	
	\bibitem{BEC2} K. Huang,  Statistical Mechanics, Wiley, (1987).
	
	\bibitem{BEC} S.~Pitaevskii, and L.~Stringari, Bose-Einstein Condensation, Oxford Science Publications, Oxford, (2003).
	
	
	\bibitem{Josserand} R.~Jordan, and C.~Josserand,  {\em Self-organization in nonlinear wave turbulence}, Physical Review E, {\bf 61}, 1527 (2000). \arXiv{chao-dyn/9905023}.
	
	\bibitem{Zakharov_1988}
	V. E. Zakharov, A. N. Pushkarev, V. F. Shvetz, and V. V. Yan'kov, {\em Solitonic turbulence}, Pis'ma v Zh. Eksp. Teor. Fiz. {\bf 48}, 79-81 (1988) [JETP Lett. {\bf 48}, 83-85 (1988)].
	
	\bibitem{Zakharov_1989} A. I. Dyachenko, V. E. Zakharov, A. N. Pushkarev, V. F. Shvetz, and V. V. Yan'kov, {\em Solitonic turbulence in nonintegrable wave systems}, Zh. Eksp. Teor. Fiz. {\bf 96}, 2026-2048 (1989) [Sov. Phys. JETP {\bf 69}, 1144-1162 (1989)].
	
	\bibitem{Rumpf} B. Rumpf, and A. C. Newell, {\em Coherent structures and entropy in constrained, modulationally unstable, nonintegrable systems}, Physical Review Letters, {\bf 87}, 054102 (2001).
	

	\bibitem{Rumpf2} B. Rumpf, {\em  Simple statistical explanation for the localization of energy in nonlinear lattices with two conserved quantities}, Physical Review E, {\bf 69}, 016618 (2004). \arXiv{nlin/0312058} [nlin.PS].

	
	\bibitem{NazarenkoBook} S. Nazarenko, {\em Wave Turbulence}, Lectures Notes in Physics, Springer, New York, (2011).
	
	\bibitem{ZakharovBook} V. E. Zakharov, V. S. L'vov, and G. Falkovich, {\em Kolmogorov Spectra of Turbulence I}, Springer, Berlin, (1992).
	
	\bibitem{ZK_Optical_Turbulence} S. Dyachenko, A. C. Newell, A. Pushkarev, and V. Zakharov, {\em Optical turbulence: weak turbulence, condensates and collapsing filaments in the nonlinear Schr\"odinger equation}, Physica D: Nonlinear Phenomena, {\bf 57}, 96-160 (1992).
	
	\bibitem{Self-Similar_2001} R. Lacaze, P. Lallemand, Y. Pomeau, and S. Rica, {\em Dynamical formation of a Bose–Einstein condensate}, Physica D: Nonlinear Phenomena, {\em 152}, 779-786 (2001).
	
	\bibitem{Self-Similar_2009} G. D\"uring, A.  Picozzi, and S. Rica, {\em Breakdown of weak-turbulence and nonlinear wave condensation}, Physica D: Nonlinear Phenomena, {\bf 238}, 1524-1549 (2009).
	
	\bibitem{Self-Similar_2021} B. V. Semisalov, V. N. Grebenev, S. B. Medvedev, and S. V. Nazarenko, {\em Numerical analysis of a self-similar turbulent flow in Bose–Einstein condensates}, Communications in Nonlinear Science and Numerical Simulation, {\bf 102}, 105903 (2021). \arXiv{2104.14591} [physics.flu-dyn].
	
	\bibitem{Condensation_2014} S. Nazarenko, M. Onorato, and D. Proment, {\em Bose-Einstein condensation and Berezinskii-Kosterlitz-Thouless transition in the two-dimensional nonlinear Schr\"odinger model}, Physical Review A, {\bf 90}, 013624 (2014). \arXiv{1305.2737} [physics.flu-dyn].
	
	\bibitem{Condensation_2021} V. Shukla, and S. Nazarenko,  {\em Nonequilibrium Bose-Einstein condensation}, Physical Review A, {\bf 105}, 033305 (2022). \arXiv{2105.07274} [cond-mat.quant-gas].

	\bibitem{ReviewOpticalTurbulence} A. Picozzi, J. Garnier, T. Hansson, P. Suret, S. Randoux, G. Millot,  and D. N. Christodoulides, {\em Optical wave turbulence: Towards a unified nonequilibrium thermodynamic formulation of statistical nonlinear optics}, Physics Reports, {\bf 542}, 1-132 (2014).
	

	
	\bibitem{BBCE1} A. Biasi, P. Bizo\'n, B. Craps, and O. Evnin, {\em Exact lowest-Landau-level solutions for vortex precession in BoseEinstein condensates}, Physical Review A {\bf 96}, 053615 (2017). \arXiv{1705.00867} [cond-mat.quant-gas].
	
	
	\bibitem{BBCE2} A. Biasi, P. Bizo\'n, B. Craps and O. Evnin, {\em Two infinite families of resonant solutions for the Gross-Pitaevskii
	equation}, Physical Review E {\bf 98}, 032222 (2018). \arXiv{1805.01775} [cond-mat.quant-gas].
	
	\bibitem{BEM1} A. Biasi, O. Evnin, and B. A. Malomed, {\em Fermi-PastaUlam phenomena and persistent breathers in the harmonic trap}, Physical Review E, {\bf 104}, 034210 (2021). \arXiv{2106.03870} [nlin.PS].
	
	\bibitem{CEM} B. Craps, M. De Clerck, O. Evnin, and S. Khetrapal, {\em Energy-level splitting for weakly interacting bosons in a harmonic trap}, Physical Review A, {\bf 100}, 023605  (2019). \arXiv{1903.04974} [cond-mat.quant-gas].
	
	\bibitem{BMP} A. F. Biasi, J. Mas, and A. Paredes, {\em Delayed collapses of Bose-Einstein condensates in relation to anti-de Sitter gravity}, Physical Review E, {\bf 95}, 032216 (2017). \arXiv{1610.04866} [nlin.PS].
	
	\bibitem{CEL} B. Craps, O. Evnin and V. Luyten, {\em Maximally rotating waves in AdS and on spheres}, JHEP {\bf 1709} 059 (2017). \arXiv{1707.08501} [hep-th].
	
	\bibitem{BR} P. Bizo\'n, and A. Rostworowski, {\em Weakly turbulent instability of anti–de Sitter spacetime}, Physical Review Letters {\bf 107}, 031102 (2011). \arXiv{1104.3702} [gr-qc].
	
	\bibitem{BMR} P. Bizo\'n, M. Maliborski, and A. Rostworowski, {\em Resonant dynamics and the instability of anti-de Sitter spacetime}, Physical Review Letters {\bf 115}, 081103 (2015). \arXiv{1506.03519} [gr-qc].
	
	\bibitem{CF} P. Bizo\'n, B. Craps, O. Evnin, D. Hunik, V. Luyten and M. Maliborski, {\em Conformal flow on $S^3$ and weak field integrability in AdS4}, Comm. Math. Phys. {\bf 353}, 1179  (2017). \arXiv{1608.07227} [math.AP].
	
	\bibitem{DR} D. Hunik-Kostyra, and A. Rostworowski, {\em AdS instability: resonant system for gravitational perturbations of AdS5 in the cohomogeneity-two biaxial Bianchi IX ansatz}, JHEP, 1-40 (2020). \arXiv{2002.08393} [gr-qc].
	
	\bibitem{MBox} J. Kurzweil, and M. Maliborski, {\em Resonant dynamics and the instability of the box Minkowski model}, Physical Review D, {\bf 106}, 124020  (2022). \arXiv{2209.05608} [gr-qc].
	
	\bibitem{BEF} P. Bizo\'n, O. Evnin and F. Ficek, {\em A nonrelativistic limit for AdS perturbations}, JHEP {\bf 12}, 113 (2018). \arXiv{1810.10574} [gr-qc].
	
	\bibitem{CE} B.~Craps, and O.~Evnin, {\em AdS (in) stability: an analytic approach}, Fortschr. Phys. {\bf 64}, 336–344 (2016). \arXiv{1510.07836} [gr-qc].
	
	\bibitem{Balasubramanian} V. Balasubramanian, A. Buchel, S. R. Green, L. Lehner, and S. L. Liebling, {\em Holographic thermalization, stability of anti–de sitter space, and the fermi-pasta-ulam paradox}, Physical Review Letters, {\bf 113}, 071601 (2014). \arXiv{1403.6471} [hep-th].
	
	\bibitem{E2} O.~Evnin, {\em Resonant Hamiltonian systems and weakly nonlinear dynamics in AdS spacetimes}, Classical and Quantum Gravity, {\bf 38}, 203001 (2021). \arXiv{2104.09797} [gr-qc].
	
	\bibitem{BCE} A. Biasi, B. Craps and O. Evnin, {\em Energy returns
		in global AdS4}, Physical Review D {\bf 100}, 024008 (2019). \arXiv{1810.04753} [hep-th].
	

	
	\bibitem{CEV1} B. Craps, O. Evnin and J. Vanhoof, {\em Renormalization group, secular term resummation and AdS (in)stability}, JHEP {\em 1410}, 48 (2014). \arXiv{1407.6273} [gr-qc].


	\bibitem{CEV2} B. Craps, O. Evnin and J. Vanhoof, {\em Renormalization, averaging, conservation laws and AdS (in)stability}, JHEP {\bf 1501}, 108 (2015). \arXiv{1412.3249} [gr-qc].
	
	\bibitem{GGT} P. G\'erard, P. Germain and L. Thomann, {\em On the cubic
	lowest Landau level equation}, Arch. Rat. Mech. Anal.
	{\bf 231}, 1073 (2019). \arXiv{1709.04276} [math.AP].
	
	\bibitem{GHT} P. Germain, Z. Hani and L. Thomann, {\em On the continuous resonant equation for NLS: I. Deterministic analysis}, J. Math. Pur. App. {\em 105}, 131  (2016). \arXiv{1501.03760} [math.AP].


	
	\bibitem{E} O.~Evnin, {\em Breathing modes, quartic nonlinearities and effective resonant systems}, SIGMA. Symmetry, Integrability and Geometry: Methods and Applications, {\bf 16}, 034 (2020). \arXiv{1912.07952} [math-ph].
	
	\bibitem{BEM2} A. Biasi, O. Evnin, and B. A. Malomed, {\em Obstruction to ergodicity in nonlinear Schr\"odinger equations with resonant potentials}, 	Physical Review E {\bf 108}, 034204 (2023). \arXiv{2304.10308} [nlin.PS].
	
	\bibitem{DWT1} V. S. L'vov, and S. Nazarenko, {\em Discrete and mesoscopic regimes of finite-size wave turbulence}, Physical Review E, {\bf 82}, 056322 (2010). \arXiv{1006.3631} [physics.flu-dyn].
	
	\bibitem{BE} A. Biasi, and O. Evnin, {\em Turbulent cascades in a truncation of the cubic Szeg\H{o} equation and related systems}, Analysis \& PDE, {\bf 15}, 217-243  (2022). \arXiv{2002.07785} [math.AP].
	
	\bibitem{Xu} H. Xu, {\em Large-time blowup for a perturbation of the cubic Szeg\"{o} equation}, Analysis \& PDE {\bf 7}, 717  (2014).
	
	\bibitem{GG} P. G\'erard, and S. Grellier, {\em The cubic Szeg\H{o} equation}, In Annales scientifiques de l'\'ecole Normale Sup\'erieure {\bf 43}, 761-810  (2010). \arXiv{0906.4540} [math.CV].
	
	\bibitem{Jalmuzna} P. Bizon, and J. Jalmuzna, {\em Globally regular instability of AdS3}, Physical Review Letters, {\bf 111}, 1306-0317 (2013).
	
	\bibitem{BBE2} A. Biasi, P. Bizo\'n, and O. Evnin, {\em Complex plane representations and stationary states in cubic and quintic resonant systems}, Journal of Physics A: Mathematical and Theoretical, {\bf 52}, 435201 (2019). \arXiv{1904.09575} [math-ph].
	
	\bibitem{CMEH} B.~Craps, M.~De Clerck, O.~Evnin, P.~Hacker, and M.~Pavlov, {\em Bounds on quantum evolution complexity via lattice cryptography}, SciPost Physics, {\bf 13}, 090  (2022). \arXiv{2202.13924} [quant-ph].
	
	\bibitem{BBE1} A. Biasi, P. Bizo\'n and O. Evnin, {\em Solvable cubic resonant systems}, Comm. Math. Phys. {\bf 369}, 433 (2019). \arXiv{1805.03634} [nlin.SI].

	\bibitem{EP} O.~Evnin, and W.~Piensuk, {\em Quantum resonant systems, integrable and chaotic}, Journal of Physics A: Mathematical and Theoretical, {\bf 52}, 025102 (2018). \arXiv{1808.09173} [math-ph].
	
	\bibitem{CE_2020} M.~De Clerck, and O.~Evnin, {\em Time-periodic quantum states of weakly interacting bosons in a harmonic trap}, Physics Letters A, {\bf 384}, 126930  (2020). \arXiv{2003.03684} [cond-mat.quant-gas].
	
	\bibitem{Catalan} S. Roman, {\em An introduction to Catalan numbers}, Springer International Publishing  (2015).
	
	\bibitem{Cascade_1}	R. H. Kraichnan, and D. Montgomery, {\em Two-dimensional turbulence}, Reports on Progress in Physics, {\bf 43}, 547 (1980).
	
	\bibitem{Nz_2023} Y. Zhu, B. Semisalov, G. Krstulovic, and S. Nazarenko,  {\em Direct and Inverse Cascades in Turbulent Bose-Einstein Condensates}, Physical Review Letters, {\bf 130}, 133001 (2023). \arXiv{2208.09279} [cond-mat.quant-gas].
	
		
	\bibitem{Bourgain} J.~Bourgain, {\em Problems in Hamiltonian PDE's}, In Visions in Mathematics: GAFA 2000 Special Volume, Part I. Basel: Birkh\"auser Basel (2010).
	
	\bibitem{Staffilani2010} J.~Colliander, M.~Keel, G.~Staffilani, H.~Takaoka, and T.~Tao, {\em Transfer of energy to high frequencies in the cubic defocusing nonlinear Schr\"odinger equation}, Inventiones mathematicae, {\bf 181}, 39-113 (2010).

		
	\bibitem{Bourgain2} J. Bourgain, {\em On the growth in time of higher Sobolev norms of smooth solutions of Hamiltonian PDE}, International Mathematics Research Notices, {\bf 1996}, 277-304  (1996).
	
	\bibitem{Kuksin}  S. B. Kuksin, {\em Oscillations in space-periodic nonlinear Schr\"odinger equations}, Geometric \& Functional Analysis GAFA, {\bf 7}, 338-363 (1997).
	
	\bibitem{Maspero} M.~Guardia, Z.~Hani, E.~Haus, A.~Maspero, and M.~Procesi, {\em Strong nonlinear instability and growth of Sobolev norms near quasiperiodic finite gap tori for the 2D cubic NLS equation}, Journal of the European Mathematical Society, {\bf 25}, 1497-1551  (2022). \arXiv{1810.03694} [math.AP].
	
	\bibitem{Hani} Z.~Hani, {\em Long-time instability and unbounded Sobolev orbits for some periodic nonlinear Schr\"odinger equations}, Archive for Rational Mechanics and Analysis, {\bf 211}, 929-964 (2014). \arXiv{1210.7509} [math.AP].
	
	\bibitem{GG2} P. G\'erard, and S. Grellier, {\em On a damped Szego equation (with an appendix in collaboration with Christian Klein)}, SIAM Journal on Mathematical Analysis, {\bf 52}, 4391-4420  (2020). \arXiv{1912.10933} [math.AP].
	
	\bibitem{GGH} P. G\'erard, S. Grellier, and Z. He, {\em Turbulent cascades for a family of damped Szeg\H{o} equations}, Nonlinearity, {\em 35}, 4820  (2022). \arXiv{2111.05247} [math.AP].
	
	\bibitem{Murdock} J. A. Murdock, Perturbations: Theory and Methods, SIAM (1987).
	
	\bibitem{Polylogarithms} D. C. Wood, {\em The computation of polylogarithm}, Tech. Report Technical Report 15-92, University of Kent, 1992.
	
	\bibitem{Sulem} C. Sulem, P. L. Sulem, and H. Frisch, {\em Tracing complex singularities with spectral methods}, Journal of Computational Physics, {\bf 50}, 138-161 (1983).
	
	\bibitem{Shelley} M. J. Shelley, {\em A study of singularity formation in vortex-sheet motion by a spectrally accurate vortex method}, Journal of Fluid Mechanics, {\bf 244}, 493-526 (1992).
	
\end{thebibliography}
\end{document}